\documentclass[reprint,superscriptaddress,showpacs,amsmath,amssymb,aps,prd,nofootinbib]{revtex4-1}
\pdfoutput=1
\usepackage[T1]{fontenc}
\usepackage{hyperref}
\usepackage{graphicx}
\usepackage{dcolumn}
\usepackage{bm}
\usepackage{lineno}
\usepackage{multirow}
\usepackage{color}
\usepackage{soul}
\usepackage{makecell}
\usepackage{xspace}
\usepackage[caption=false]{subfig}

\newcommand{\fluka}{\textsc{fluka}}
\newcommand{\genie}{\textsc{genie}}

\newcommand{\nuwro}{\textsc{nuwro}}
\newcommand{\pipm}{\ensuremath{\pi^{\pm}}\xspace}
\newcommand{\pip}{\ensuremath{\pi^{+}}\xspace}
\newcommand{\pim}{\ensuremath{\pi^{-}}\xspace}

\newcommand{\de}[1]{\left(#1\right)}

\begin{document}

\title{Using world $\pi^{\pm}$--nucleus scattering data to constrain an intranuclear cascade model}

\author{E. S. Pinzon Guerra}
\affiliation{York University, Department of Physics and Astronomy, Toronto, Ontario, Canada}
\author{C.~Wilkinson}
\affiliation{University of Bern, Albert Einstein Center for Fundamental Physics, Laboratory for High Energy Physics (LHEP), Bern, Switzerland}
\author{S.\,Bhadra}
\affiliation{York University, Department of Physics and Astronomy, Toronto, Ontario, Canada}
\author{S.\,Bolognesi}
\affiliation{IRFU, CEA Saclay, Gif-sur-Yvette, France}
\author{J.~Calcutt}
\affiliation{Michigan State University, Department of Physics and Astronomy, East Lansing, Michigan, U.S.A.}
\author{P.~de~Perio}
\altaffiliation[Present address: ]{Columbia University, Physics Department, New York, New York 10027, USA}
\affiliation{University of Toronto, Department of Physics, Toronto, Ontario, Canada}
\author{S.\,Dolan}
\affiliation{IRFU, CEA Saclay, Gif-sur-Yvette, France}
\affiliation{Ecole Polytechnique, IN2P3-CNRS, Laboratoire Leprince-Ringuet, Palaiseau, France}
\author{T.\,Feusels}
\affiliation{University of British Columbia, Department of Physics and Astronomy, Vancouver, British Columbia, Canada}
\affiliation{TRIUMF, Vancouver, British Columbia, Canada}
\author{G.A.\,Fiorentini}
\affiliation{York University, Department of Physics and Astronomy, Toronto, Ontario, Canada}
\author{Y.~Hayato}
\affiliation{Kavli Institute for the Physics and Mathematics of the Universe (WPI),Todai Institutes for Advanced Study, University of Tokyo, Kashiwa, Chiba, Japan}
\affiliation{University of Tokyo, Institute for Cosmic Ray Research, Kamioka Observatory, Kamioka, Japan}
\author{K. Ieki}
\altaffiliation[Present address: ]{University of Tokyo, Department of Physics, Tokyo, Japan}
\affiliation{Kyoto University, Department of Physics, Kyoto, Japan}
\author{K.~Mahn}
\affiliation{Michigan State University, Department of Physics and Astronomy, East Lansing, Michigan, U.S.A.}
\author{K.S.~McFarland}
\affiliation{University of Rochester, Department of Physics and Astronomy, Rochester, New York, USA}
\author{V.\,Paolone}
\affiliation{University of Pittsburgh, Department of Physics and Astronomy, Pittsburgh, Pennsylvania, U.S.A.}
\author{L.~Pickering}
\affiliation{Michigan State University, Department of Physics and Astronomy, East Lansing, Michigan, U.S.A.}
\author{R.~Tacik}
\affiliation{TRIUMF, Vancouver, British Columbia, Canada}
\affiliation{University of Regina, Department of Physics, Regina, Saskatchewan, Canada}
\author{H.A.\,Tanaka}
\affiliation{University of Toronto, Department of Physics, Toronto, Ontario, Canada}
\affiliation{SLAC National Accelerator Laboratory, Stanford University, Menlo Park, California, USA}
\author{R.~Terri}
\affiliation{Queen Mary University of London, School of Physics and Astronomy, London, United Kingdom}
\author{M.O.~Wascko}
\affiliation{Imperial College London, Department of Physics, London, United Kingdom}
\author{M.J.\,Wilking}
\affiliation{State University of New York at Stony Brook, Department of Physics and Astronomy, Stony Brook, New York, U.S.A.}
\author{C.~Wret}
\affiliation{University of Rochester, Department of Physics and Astronomy, Rochester, New York, USA}
\author{M.\,Yu}
\affiliation{York University, Department of Physics and Astronomy, Toronto, Ontario, Canada}

\date{\today}

\begin{abstract}
The NEUT intranuclear cascade model is described and fit to a large body of \pipm--nucleus scattering data. Methods are developed to deal with deficiencies in the available historical data, and robust uncertainty estimates are produced. The results are compared to a variety of simulation packages, and the data itself. This work provides a method for tuning Final State Interaction models, which are of particular interest to neutrino experiments that operate in the few-GeV energy region, and provides results which can be used directly by the T2K and Super-Kamiokande collaborations, for whom NEUT is the primary simulation package.
\end{abstract}

\maketitle

\section{Introduction}
\label{sec:intro}
There is ongoing work dedicated to understanding or measuring neutrino interactions in the 0.1--10 GeV energy range aimed to reduce systematic uncertainties for current and future neutrino oscillation experiments (see, for example, recent reviews in Refs.~\cite{Alvarez-Ruso:2014bla,Garvey:2014exa,Mosel:2016cwa,Katori:2016yel,Alvarez-Ruso:2017oui, Mahn:2018mai, Betancourt:2018bpu}). The key problem is providing the best estimate of the incoming neutrino energy (as their oscillations are energy dependent) using measurable quantities in a detector: namely the outgoing particle content and kinematics. After a neutrino interacts with nucleon(s) within the nucleus, or quark(s) within a nucleon, the interaction products are propagated out of the nuclear medium before being measurable in a detector. Nucleons and pions have a considerable re-interaction probability when propagated through the nucleus, which can change the outgoing particle type, number and kinematics. These effects are known in the neutrino scattering community as Final State Interactions (FSI).

The effect of FSI is particularly important for oscillation analyses that use kinematic neutrino energy reconstruction, which selects neutrino events with a charged lepton and no pions in the final state (CC0$\pi$), such as T2K~\cite{Abe:2011ks} and the planned Hyper-Kamiokande~\cite{Hyper-Kamiokande:2016dsw}. CC0$\pi$ events are dominated by charged-current quasi-elastic (CCQE) interactions $\nu_{l} + n \rightarrow l^{-} + p$ and $\bar{\nu}_{l} + p \rightarrow l^{+} + n$. For such events, one can reconstruct the incoming neutrino energy perfectly, provided that the incoming neutrino beam direction is known, and the initial state struck nucleon is at rest~\cite{long_prd}. In a nuclear environment, the neutrino energy reconstruction is smeared by the initial state nucleon momentum, as they are not at rest inside the nucleus. Additionally, biases are introduced to the reconstructed neutrino energy by introducing events in the CC0$\pi$ sample that are not CCQE, and so do not have the simple kinematic relationship to the neutrino energy. Events where a pion is produced at the primary neutrino interaction vertex but is later absorbed by the nucleus through FSI are a large contribution to this source of bias, thus FSI in carbon and oxygen are of great importance to the T2K oscillation analysis. Pions which interact outside the nucleus, known as secondary interactions (SI), can result in mis-reconstructing the event, and can migrate events from one topology to another. Hadronic interactions inside (FSI) and outside (SI) the target nucleus are related, as discussed in Refs~\cite{intranuke, NuWro, salcedo:pionfsi, Oset:1987re}, although the relationship is not trivial. FSI and SI are among the nuclear effects that dominate systematic uncertainties for long-baseline neutrino oscillation analyses.

In this work, we describe the tuning of the $\pi^{\pm}$--A scattering model used in NEUT~\cite{neut}---the primary interaction simulation package used by the T2K and Super-Kamiokande experiments---to the world dataset and describe its relation to the intranuclear semi-classical cascade models for FSI and SI. Although this work focuses on a single model, as it was motivated by the needs of the T2K experiment, the methods and the general conclusions developed herein are applicable to other cascade models~\cite{fluka1, geant4, genie, NuWro}. Comparisons are made between the results from this tuning and various other models, including the GiBUU simulation package---which goes beyond the semi-classical approximations made by cascade models~\cite{Buss20121}. The procedures used to evaluate the impact of these cascade model uncertainties are described elsewhere~\cite{long_prd,deperio-thesis} and will not be discussed here. The updates to the NEUT cascade model described have been incorporated in recent T2K analyses~\cite{Abe:2018wpn}.

Section \ref{sec:neut-summary} provides an overview of the NEUT cascade model and the parameters used in this tuning work. Section \ref{sec:data} provides a summary of the external scattering data sets. Section \ref{sec:fit-strategy} describes the fit strategy and methods. Section \ref{sec:fit-results} presents the best fit values for the FSI parameters and their correlation matrix. Section \ref{sec:models} provides an overview and set of comparisons to other models for $\pi^\pm$--A interactions. Finally, Section~\ref{sec:outlook} presents our conclusions and outlook.

\section{NEUT Cascade}\label{sec:neut-summary}
\begin{figure}[htbp]
\centering
\includegraphics[width=86mm]{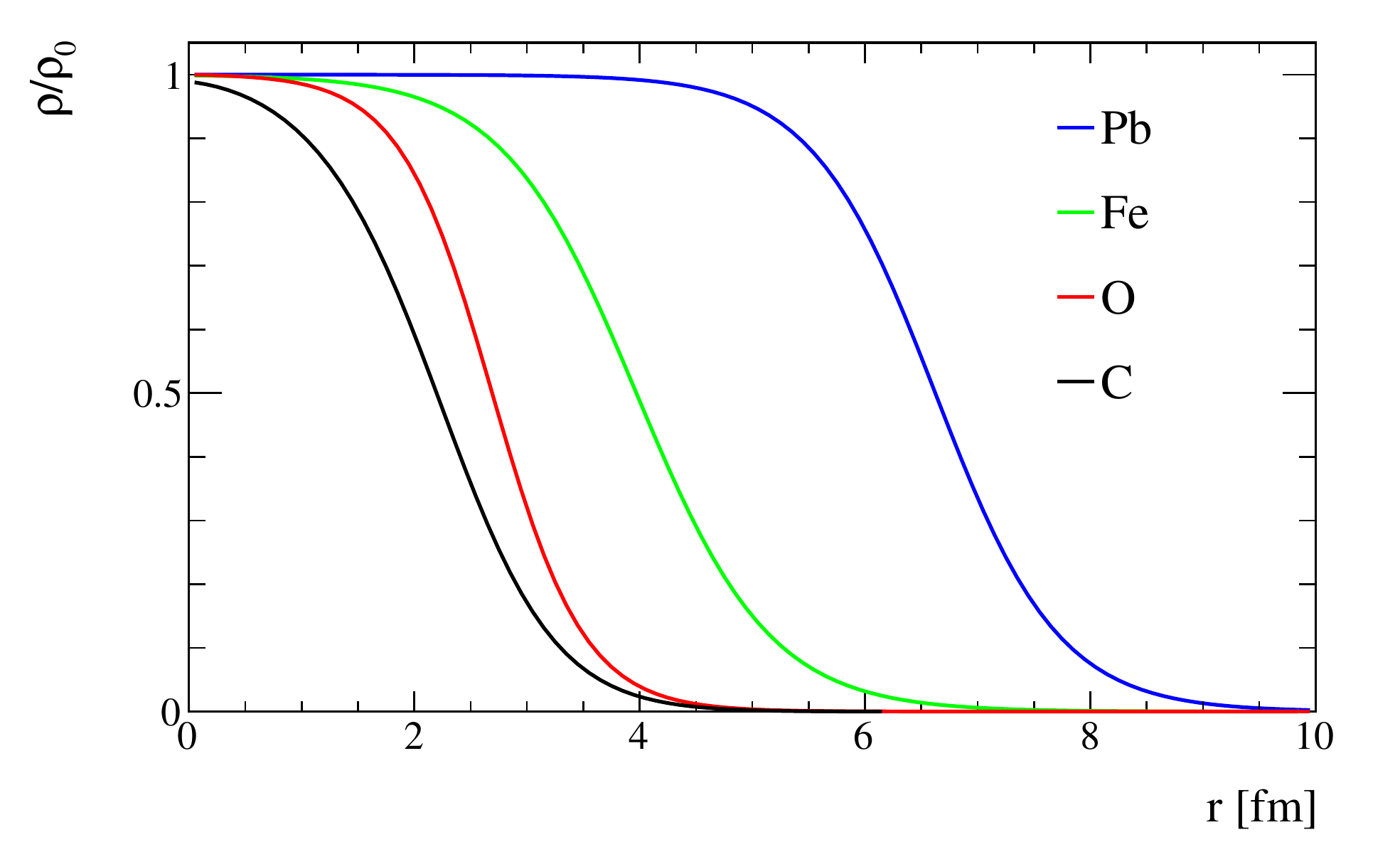} 
\caption{The normalized three-parameter (two-parameter) Fermi model nuclear potentials shown for various nuclei (oxygen). Reproduced from Ref.~\cite{deperio-thesis}.}
\label{fig:fsi-nucl-dens}
\end{figure}

Pion FSI are simulated in NEUT using a semi-classical intranuclear cascade model~\cite{neut, deperio-thesis}. After a neutrino interaction occurs, the starting positions of the interaction products are chosen randomly from a nuclear density profile in the form of a three-parameter Fermi model (Woods-Saxon potential) described in Equation~\ref{eqn:woods} and shown in Figure~\ref{fig:fsi-nucl-dens}:
\begin{equation}\label{eqn:woods}
\frac{\rho(r)}{\rho_0} = \frac{1+w\frac{r^2}{c^2}}{1+\exp\de{{\frac{r-c}{\alpha}}}},
\end{equation}
where $r$ is the distance from the center of the nucleus, $c$ is the nuclear radius, $\alpha$ is the surface thickness, and $w$ modifies the shape of the potential. The parameters $c,\alpha$ and $w$ were determined from an analysis of electron-scattering measurements~\cite{jager:neut}. Note that the mass number (A) dependence of the model is encoded in these parameters. In the case of oxygen, the two-parameter Fermi model is used ($w$ = 0). The initial kinematic information of outgoing particles is taken directly from the neutrino interaction simulation and is fed into the cascade model.

\begin{figure}[htbp]
\centering
\includegraphics[width=86mm]{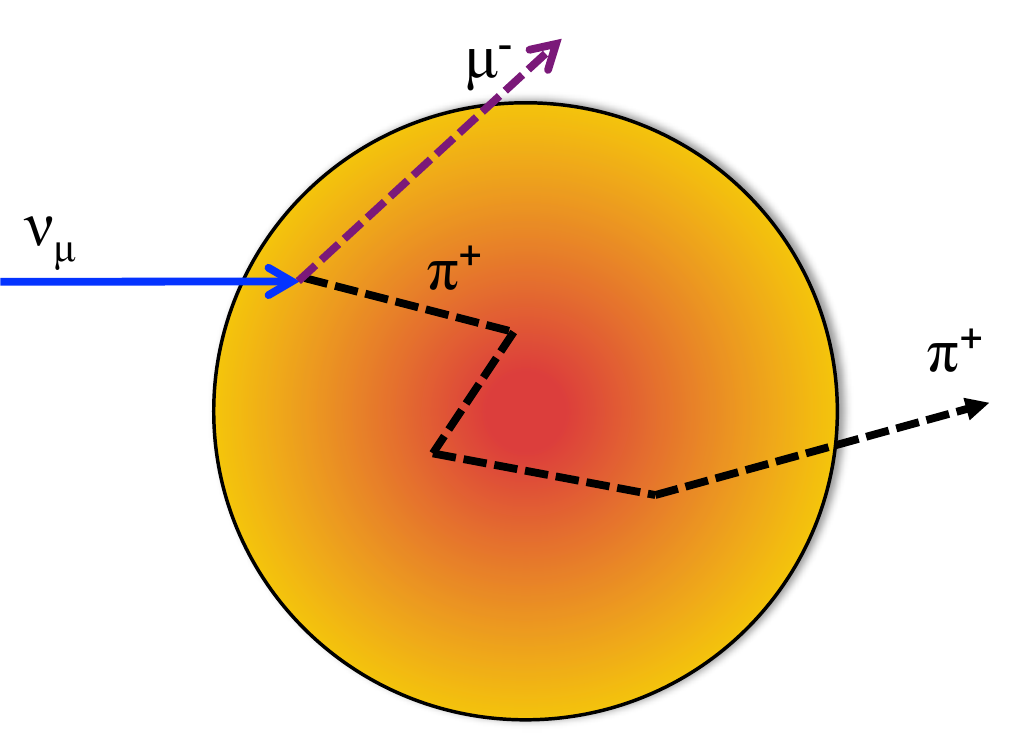} 
\caption{Cartoon illustrating the effect of FSI in the intra-nuclear cascade model.}
\label{fig:fsi-cascade}
\end{figure}

The initial interaction products are propagated ``classically'', in finite steps within the nuclear medium. The steps are in space and were chosen as ${dx=R_{\mathrm{N}}/100}$, where $R_{\mathrm{N}}$ is the size of the nucleus and is defined such that $\rho(R_{\mathrm{N}})/\rho_0 \approx 10^{-4}$, which for carbon is $\sim$2.5 times the nuclear radius from Ref.~\cite{jager:neut}. The step spacing was chosen such that the probability of two or more interactions at each step was negligible. $R_{\mathrm{N}}$ acts to stop the cascade once the nuclear density, and thus the re-interaction probability, becomes negligible. The probabilities for each interaction type is calculated at each step and a random number generator is used to determine which, if any, interaction takes place. The cascade continues until the particle is absorbed or its position exceeds $R_{\mathrm{N}}$. Particles produced through inelastic FSI are also included into the cascade model from the relevant production point(s). The particles are propagated independently and are only subject to the potential from the nucleus, not from other nucleons and pions. Figure \ref{fig:fsi-cascade} shows an example cartoon of the intranuclear cascade mechanism.

For low momentum pions (defined in NEUT as $p_{\pi}<500$ MeV$/c$), tables computed from the Oset et al. model~\cite{salcedo:pionfsi} are used to determine quasi-elastic, single charge exchange, and absorption interaction probabilities inside the nucleus, and relate them to extra-nuclear $\pi^{\pm}$--A measurements. This model involves a many-body calculation in infinite nuclear matter with a local density approximation included. The $\pi^{\pm}$--A scattering is represented as a wave in a complex optical potential. Contributions from individual reaction channels are obtained by separating the real and complex parts of the potential and calculating the corresponding Feynman diagrams. For high momentum pions ($p_{\pi}>500$ MeV$/c$), the interaction probabilities are calculated from $\pi^{\pm}$--A scattering data off free proton and deuterium compiled by the PDG~\cite{bib:pdg}. The two models are blended linearly in the $400<p_{\pi}<500$ MeV$/c$ region to avoid discontinuities.

The most commonly used generators and simulation toolkits, including GENIE~\cite{genie}, NuWro~\cite{NuWro}, FLUKA~\cite{fluka1}, and Geant4~\cite{geant4}, have similar cascade models implemented, although the details vary. The notable exception is GiBUU~\cite{Buss20121}, where the Boltzmann-Uehling-Uhlenbeck transport equations are solved for a more complete description of the nuclear medium. In GiBUU, the particles experience the potential from both the excited nuclear remnant and the other particles produced in the initial interaction. Comparisons between the results of this work, the data, and a variety of these alternate simulations are presented in Section~\ref{sec:models}.

The model is parameterized by the scaling factors summarized in Table~\ref{tab:fsi-parameters}, henceforth referred to simply as ``FSI parameters'' ($f_{\mathrm{FSI}}$). Each parameter scales the corresponding microscopic probability of a $\pi^{\pm}$ interaction at each step of the cascade, except for $f_{\mathrm{CX}}$, which scales the charge exchange fraction of low momentum quasi-elastic (QE) scattering.

A reweighting scheme is used to allow the propagation of variations of these parameters to T2K analyses~\cite{long_prd, deperio-thesis}. The scheme uses the information in the cascade for each event and re-calculates the interaction probabilities for each step with varied parameters to obtain a new value for the event probability. The FSI weight is defined as the ratio between the varied and nominal total event probabilities. However, the reweighting procedure is not exact due to the factorization into individual parameter variations, so was not used in this work.

\begin{table}[htbp]
\small
\centering
\begin{tabular}{c c c}
\hline
Parameter & Description & \thead{Momentum \\ Region (MeV/$c$)} \\
\hline
$f_{\mathrm{ABS}}$  & Absorption                   & $< 500$ \\
$f_{\mathrm{QE}}$   & Quasi-elastic scatter        & $< 500$ \\
$f_{\mathrm{CX}}$   & Single charge exchange       & $< 500$ \\
$f_{\mathrm{QEH}}$  & Quasi-elastic scatter        & $> 400$ \\
$f_{\mathrm{CXH}}$  & Single charge exchange       & $> 400$ \\
$f_{\mathrm{INEL}}$ & Hadron (N+n$\pi$) production & $> 400$ \\ 
\hline
\end{tabular}  
\caption[NEUT FSI Parameter Descriptions]{
Description of the NEUT pion FSI  parameters used in the fits described in this work. The overlap in the momentum regions is due to blending of the high and low energy models in NEUT as described in the text.
}
\label{tab:fsi-parameters}
\end{table}

\section{Summary of External Scattering Data}\label{sec:data}
Although T2K analyses are predominantly interested in light nuclear targets (carbon and oxygen), there are many heavier targets in the T2K detectors~\cite{t2k-nim, ingrid}. As such, this work includes external data taken with carbon, oxygen, aluminium, iron, copper, and lead targets.

We used data from \pip and \pim beams over a momentum range from 60--2000~MeV$/c$. The interaction channels are defined exclusively from the number of pions in the final state, with any number of nucleons. This allows for direct comparisons between the external measurements of cross sections and the NEUT predicted cross sections. The following interaction channels were used:
\begin{itemize}
\item \textbf{Absorption (ABS):} the incident pion is absorbed by the nucleus, and thus no pions are observed in the final state.
\item \textbf{Quasi-elastic Scattering (QE):} only one pion is observed in the final state, and it has the same charge as the incident beam. The interaction is with a nucleon within the nucleus which differentiates QE from {\it elastic} scattering, where the interaction is with the nucleus as a whole. There is no requirement for the struck nucleon to be observed in the final state, as it may undergo FSI itself and not be observable.
\item \textbf{Single Charge Exchange (CX):} the charged pion interacts with the nucleus and a single $\pi^{0}$ is observed in the final state.
\item \textbf{Absorption + Single Charge Exchange (ABS+CX):} sum of ABS and CX (e.g., a charged pion is present in the initial state, but none are observable in the final state).
\item \textbf{Reactive (REAC):} sum of ABS + CX + QE, double charge exchange, and hadron production. Double charge exchange events have a single pion in the final state which has the opposite charge to the incident beam. Hadron production events have final states with more than one pion.
\end{itemize}

These channels do not have a one-to-one correspondence to the parameters defined in Table~\ref{tab:fsi-parameters}, but describe experimentally distinguishable interactions. Elastic scattering is experimentally defined as events with a very small scattering angle, where there is a very small momentum transfer to the target nucleus. Measurements of the elastic and total (reactive + elastic) cross sections are not used for this tuning as NEUT does not simulate elastic $\pi^\pm$--A scattering.

\begin{table*}[htbp]
\centering
    {\renewcommand{\arraystretch}{1.2}
      \begin{tabular}{c c c c c c c c c c c c}
\hline
Reference & Polarity & Targets & $p_{\pi}$ [MeV$/c$] & Channel(s)\\ \hline
B. W. Allardyce {\it et al.}~\cite{allardyce:piscat}&$\pi^{\pm}$ & $^{12}$C, $^{27}$Al, $^{207}$Pb & 710-2000 & REAC \\
A. Saunders {\it et al.}~\cite{saunders:piscat}&$\pi^{\pm}$ & $^{12}$C, $^{27}$Al & 116-149 & REAC \\
C. J. Gelderloos {\it et al.}~\cite{gelderloos:piscat} & $\pi^{-}$ & $^{12}$C, $^{27}$Al, $^{63}$Cu, $^{207}$Pb & 531-615 & REAC \\
F. Binon {\it et al.}~\cite{binon:piscat} & $\pi^{-}$ & $^{12}$C & 219-395 & REAC\\
O. Meirav {\it et al.}~\cite{Meirav:piscat} & $\pi^{\pm}$ & $^{12}$C, $^{16}$O & 128-169 & REAC \\
C. H. Q. Ingram~\cite{ingram:piscat} & $\pi^{+}$ & $^{16}$O & 211-353 & QE \\
S. M. Levenson {\it et al.}~\cite{levenson:piscat} & $\pi^{+}$ & $^{12}$C & 194-416 & QE \\
M. K. Jones {\it et al.}~\cite{jones:piscat} & $\pi^{+}$ & $^{12}$C, $^{208}$Pb & 363-624 & QE, CX\\
D. Ashery {\it et al.}~\cite{ashery:piscat} & $\pi^{\pm}$ & $^{12}$C, $^{27}$Al, $^{56}$Fe & 175-432 & QE, ABS+CX\\
H. Hilscher {\it et al.}~\cite{Hilscher:piscat} & $\pi^{-}$ & $^{12}$C & 156 & CX \\ 
T. J. Bowles~\cite{bowles:piscat} & $\pi^{+}$ & $^{16}$O & 128-194 & CX \\
D. Ashery {\it et al.}~\cite{ashery:pioncx} & $\pi^{\pm}$ & $^{12}$C, $^{16}$O, $^{207}$Pb & 265 & CX \\
K. Nakai {\it et al.}~\cite{nakai:piscat} & $\pi^{\pm}$ & $^{27}$Al, $^{63}$Cu & 83-395 & ABS\\
E. Bellotti {\it et al.}~\cite{Bellotti1973_abs:piscat} & $\pi^{+}$ & $^{12}$C & 230 & ABS \\
E. Bellotti {\it et al.}~\cite{Bellotti1973_cx:piscat} & $\pi^{+}$ & $^{12}$C & 230 & CX \\
I. Navon {\it et al.}~\cite{navon:piscat} & $\pi^{+}$ & $^{12}$C, $^{56}$Fe & 128 & ABS+CX \\
R. H. Miller {\it et al.}~\cite{miller:piscat}&$\pi^{-} $ & $^{12}$C, $^{207}$Pb & 254 & ABS+CX \\
E. S. Pinzon Guerra {\it et al.}~\cite{duet:piscat} & $\pi^{+}$ & $^{12}$C & 206-295 & ABS, CX \\ \hline
\end{tabular}}  
\caption{Summary of $\pi^\pm$--A scattering data used for this tuning, including beam polarity, nuclear target type(s), momentum range and interaction channel(s). Some of these experiments may have measured data on other target nuclei that are not considered here.}
\label{tab:piscatdat}
\end{table*}

\begin{figure}[htbp]
\centering
\includegraphics[width=86mm]{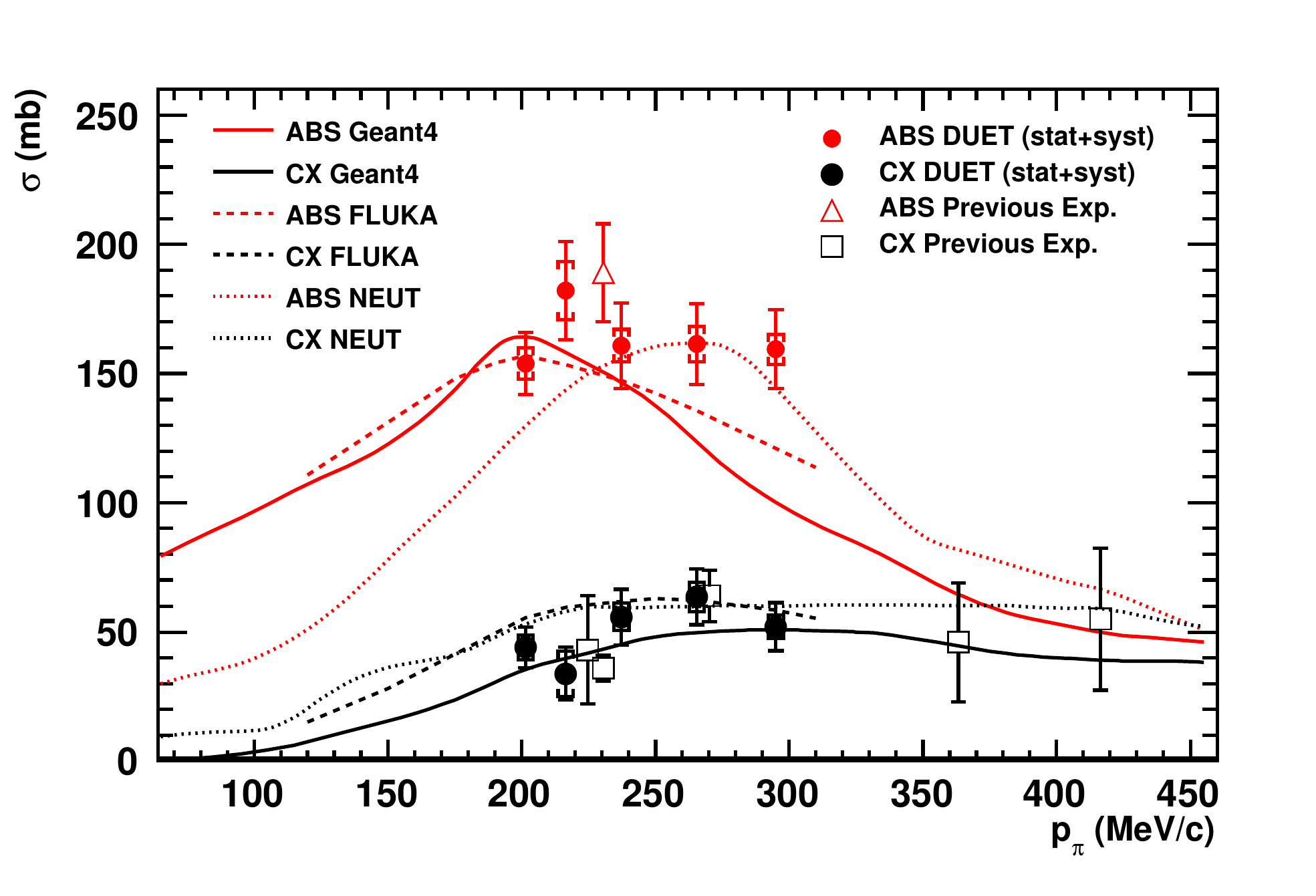} 
\caption{DUET measurements of ABS and CX compared with previous measurements~\cite{jones:piscat,ashery:pioncx,Bellotti1973_abs:piscat,Bellotti1973_cx:piscat} and ABS (red) and CX (black) model predictions from GEANT4 (solid line), FLUKA (dashed line), and NEUT (dotted line). Reproduced from Ref.~\cite{duet:piscat}.}
\label{fig:duet-abs-cx}
\end{figure}

\begin{figure}[htbp]
\centering
\includegraphics[width=86mm]{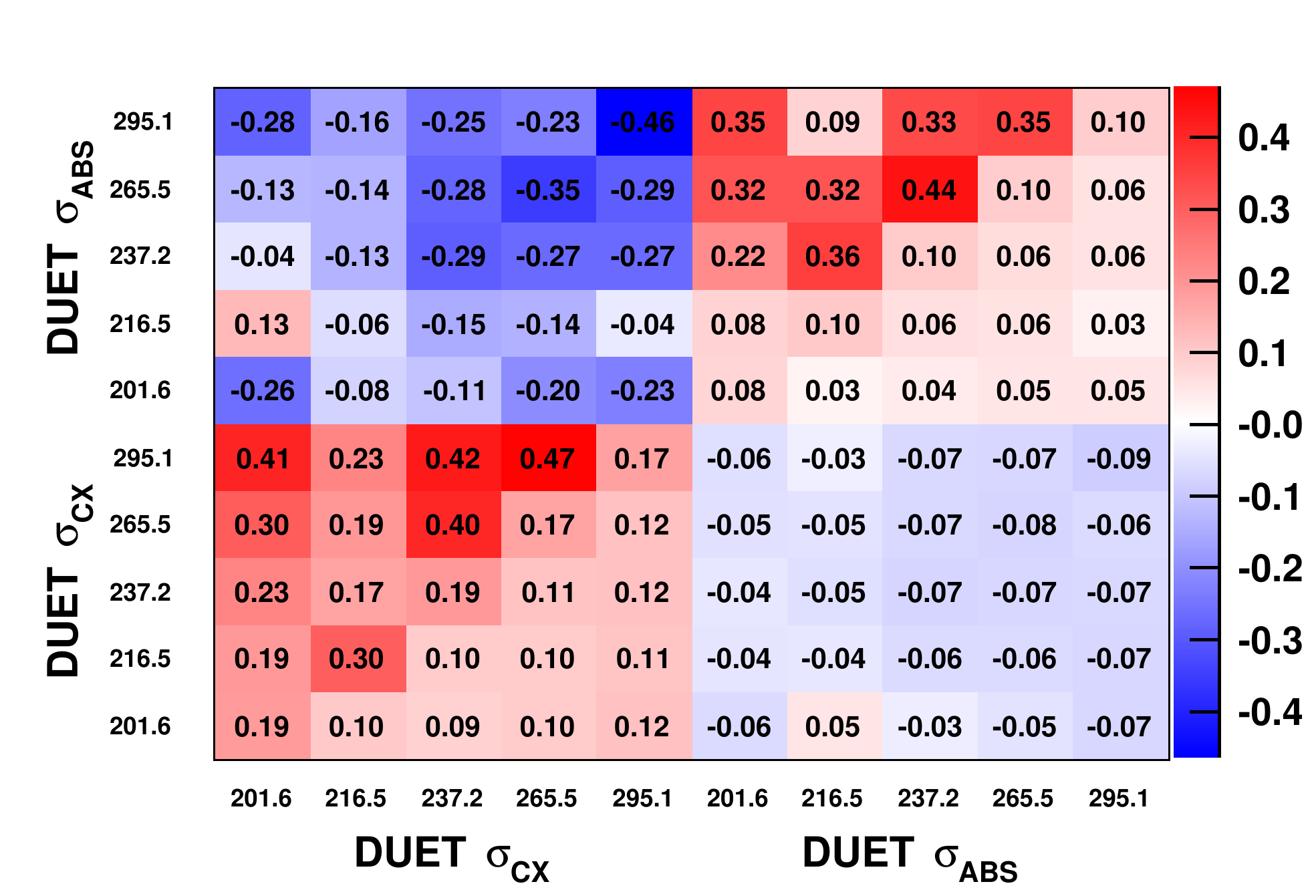} 
\caption{Fractional covariance and correlation for the DUET measurements of ABS and CX. The diagonal and lower triangle show the covariance $(\mathrm{sgn}(V_{ij} )\times \sqrt{V_{ij}} )$, while the upper triangle of the matrix shows the correlation coefficients. Reproduced from Ref.~\cite{duet:piscat}.}
\label{fig:duet-covariance}
\end{figure}

Table~\ref{tab:piscatdat} lists the external data, specifying the channels, targets, and $p_\pi$ range measured by each experiment. Most of the datasets used date from the 1950's--1990's, and contain limited information: in particular, covariances between datapoints in each channel, and between channels, are not available. A notable exception is the recent measurement of the ABS and CX cross sections by the DUET collaboration~\cite{duet:piscat}. These measurements were partially motivated by the need to cover the lack of ABS measurements in the $\Delta$(1232) resonance region, which is of particular importance for T2K. Figure~\ref{fig:duet-abs-cx} shows the cross sections measured by DUET, and Figure~\ref{fig:duet-covariance} shows the covariance matrix provided between all DUET datapoints.

There are some changes to the datasets used, compared to earlier iterations of this work, described in Ref~\cite{long_prd,deperio-thesis}. In addition to the new results from the DUET collaboration, Refs.~\cite{Meirav:piscat,Hilscher:piscat} have been added. Measurements of exclusive pion production~\cite{grion:piscat,rahav:piscat}, double charge exchange~\cite{wood:piondcx} and diffractive production~\cite{cronin:piscat} are not used, as more inclusive channels are preferred here. Ref.~\cite{rowntree:piscat} was taken on nuclear targets other than those identified in Table~\ref{tab:piscatdat}, and so is not included here. Some measurements of ABS~\cite{jones:piscat,giannelli:piscat,ransome:piscat} were removed as those were performed with the goal of understanding multi-nucleon correlations and concentrated on final states with multiple protons instead of the more inclusive cross sections considered here. Measurements of the total cross section were not used~\cite{wilkin:piscat,clough:piscat}, and measurements of reactive cross section where an optical model was used to separate the reactive and the elastic components from the total cross section were neglected~\cite{takahashi:piscat,carroll:piscat, aoki:piscat} as it introduces model dependence. Finally, Ref.~\cite{crozon:piscat} has been neglected as the normalization of the results is inconsistent with other measurements, discussed in Ref.~\cite{allardyce:piscat}.

\section{Fit Strategy}\label{sec:fit-strategy}
The goal of the fit is two-fold:
\begin{enumerate}
\item Find the set of FSI parameters that best fit the external scattering data listed in Section~\ref{sec:data}.
\item Set uncertainties for the FSI parameters that span the errors from the external data, and extract correlations between the parameters
\end{enumerate}
Given the computational cost of the NEUT cascade, predictions for the relevant $\pi^{\pm}$--A cross sections, $\sigma_j^{\mathrm{NEUT}}(f_{\mathrm{FSI}})$, as a function of the FSI parameters, $f_{\mathrm{FSI}}$, were pre-computed for a finite grid of FSI parameters. The FSI parameters are rescaling parameters of the nominal NEUT FSI prediction, where a parameter value of 1 is the nominal NEUT value, and are not fundamental physics parameters in their own right. The minimum and maximum values allowed for the FSI parameters and the step sizes are summarized in Table~\ref{tab:fsi-grid}. The predictions were only calculated for values of p$_\pi$ for which data was available to reduce the computational load. The predictions were calculated by running the NEUT cascade with a mono-energetic pion beam using the \texttt{piscat} simulation of NEUT~\cite{neut, deperio-thesis}. 

\begin{table}[htbp]
\centering
{\renewcommand{\arraystretch}{1.2}
\begin{tabular}{cccc}
\hline
\textbf{Parameter} & \textbf{Grid min.} & \textbf{Grid max.} & \textbf{Step size}\\ 
\hline
$f_{\mathrm{QE}}$   & 0.1  & 1.7 & 0.1 \\
$f_{\mathrm{ABS}}$  & 0.35 & 1.95 & 0.1 \\
$f_{\mathrm{CX}}$   & 0.1 & 1.6 & 0.1 \\
$f_{\mathrm{INEL}}$ & 0.2 & 2.6 & 0.2\\
$f_{\mathrm{QEH}}$  & 0.8 & 2.8 & 0.2 \\
$f_{\mathrm{CXH}}$  & 1.2 & 2.4 & 0.2\\ 
\hline
\end{tabular}}
\caption{
Minimum and maximum values for the NEUT FSI parameters allowed in the finite grid of pre-computed cross section predictions.}
\label{tab:fsi-grid}
\end{table}

The best fit is found by minimizing the $\chi^{2}$-statistic defined by

\begin{widetext}
\begin{align}
\begin{split}
\chi^{2} (f_{\mathrm{FSI}})=
\sum_{i}^{\mathrm{Datasets}}\left(\sum_{j}^{n_i}\frac{1}{n_i}\left(\frac{\sigma_j^{\mathrm{Data}}-\lambda_i^{-1}\sigma_j^{\mathrm{NEUT}}(f_{\mathrm{FSI}})}{\Delta\sigma_j^{\mathrm{Data}}}\right)^{2}+\left(\frac{\lambda_i-1}{\epsilon_i}\right)^{2}\right) \\
+\sum_{i,j}^{10}\left(\sigma_i^{\mathrm{DUET}}-\sigma_i^{\mathrm{NEUT}}(f_{\mathrm{FSI}})\right)(V_{ij}^{-1})^{\mathrm{DUET}}\left(\sigma_j^{\mathrm{DUET}}-\sigma_j^{\mathrm{NEUT}}(f_{\mathrm{FSI}})\right), 
\end{split}
\label{eqn:chi2}
\end{align}

\end{widetext}
\noindent where $n_i$ represents the number of data points in each data set except DUET, $\sigma_j^{\mathrm{Data}}$ and $\Delta\sigma_j^{\mathrm{Data}}$ are the external data set cross sections and their respective uncertainties, $\sigma_i^{\mathrm{DUET}}$ are the DUET data, and $(V_{ij})^{\mathrm{DUET}}$ is the DUET covariance matrix shown in Figure~\ref{fig:duet-covariance}. Note that every channel measured by the uncorrelated datasets ($n_i$) is treated as a single, uncorrelated point, in the fit. The normalization parameters $\lambda_i$ were included as penalty terms for each data set (except DUET). The uncertainty for the normalization parameters ($\epsilon_i$) was assigned to be 40\% following the representative correlations in the DUET data sets as seen in Figure~\ref{fig:duet-covariance}, and was an ad-hoc choice made for this analysis. These $\lambda_i$ normalization parameters include a fully correlated component between the datapoints of each dataset, to allow the effect of no correlations to be investigated. Note that they are not used as fit parameters for all fits described in this work.

The dependence of the $\chi^2$ in Equation~\ref{eqn:chi2} to the $f_{\mathrm{CXH}}$ parameter was found to be very weak, so it was decided to fix this parameter to its nominal value for the fit. An additional parameter designed to uniformly scale all the microscopic interaction probabilities ($f_{\mathrm{ALL}}$) was also investigated, but similarly, it was found to have little impact on the fit and was consequently not included.

The minimization was performed using the MIGRAD algorithm of the MINUIT package~\cite{minuit}. The advantage of using this algorithm is that it provides both best fit parameters and their correlations. The difficulty is that the algorithm requires a $\chi^2$ surface with smooth first and second order derivatives. Two different interpolation methods were used to smooth the finite pre-computed grid, to determine if biases were introduced by either.
\begin{enumerate}
\item \textbf{TMultiDimFit} The TMultiDimFit~\cite{tmultidimfit}. class of ROOT~\cite{Brun:root} was used to obtain a polynomial expression for the $\chi^2$ grid in terms of the FSI parameters. The best-fit polynomial function contained up to fourth-degree polynomials with 53 terms in total, including cross-terms. A comparison of the best-fit polynomial function to the finite grid reported a $\chi^2$/n.d.o.f of 0.29.

\item \textbf{GNU-Octave n-dim splines} The \texttt{interpn} function of GNU-Octave~\cite{octave} was used to obtain a multi-dimensional spline interpolation of the $\chi^2$ grid. Cubic splines are evaluated around the requested point.
\end{enumerate}

\begin{figure*}[htbp]
  \centering
  \subfloat[Absorption ($p_{\pi} \leq 500$ MeV/c)]{\includegraphics[width=0.33\linewidth]{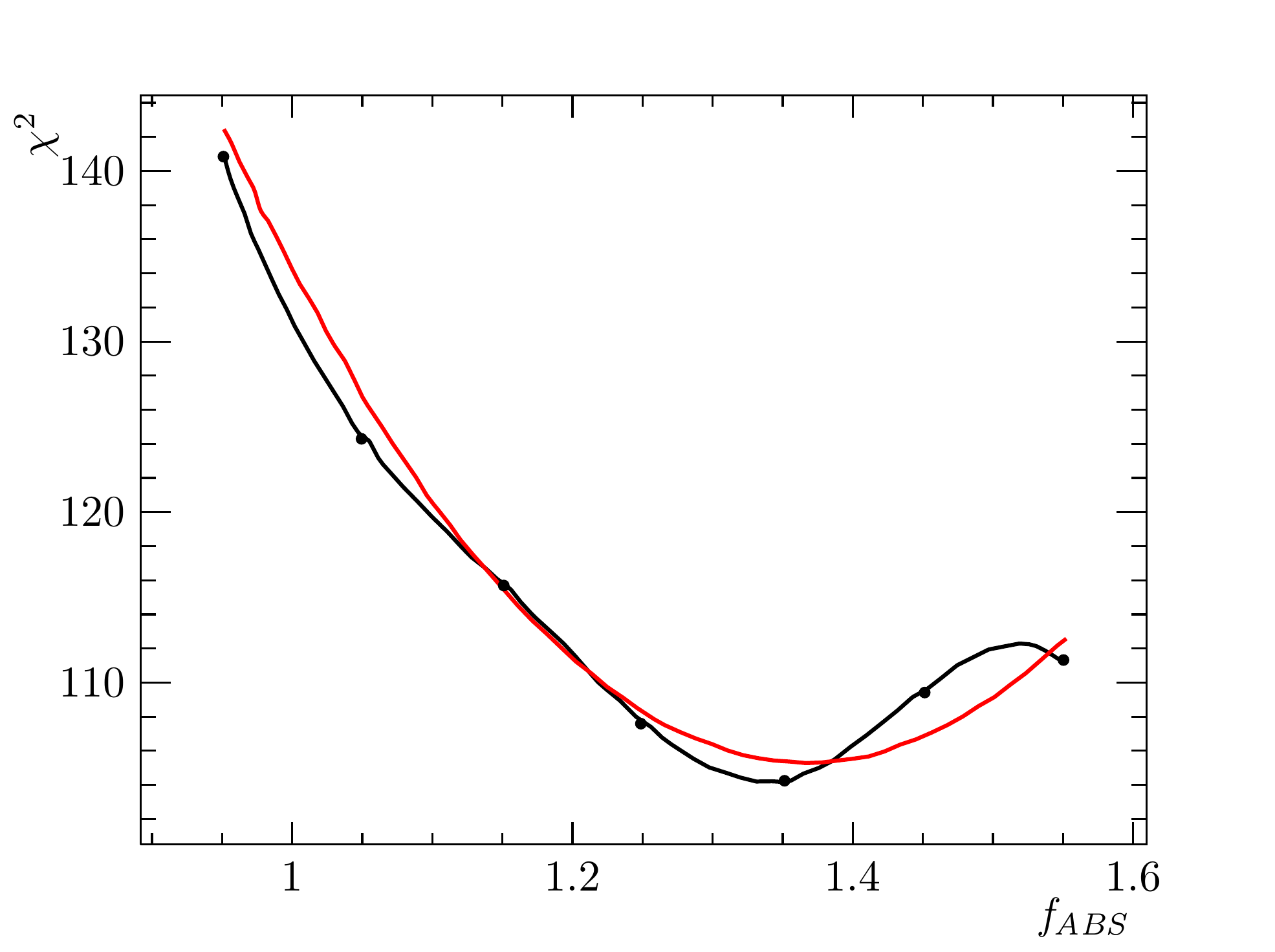}}
  \subfloat[Quasi-elastic ($p_{\pi} \leq 400$ MeV/c)]{\includegraphics[width=0.33\linewidth]{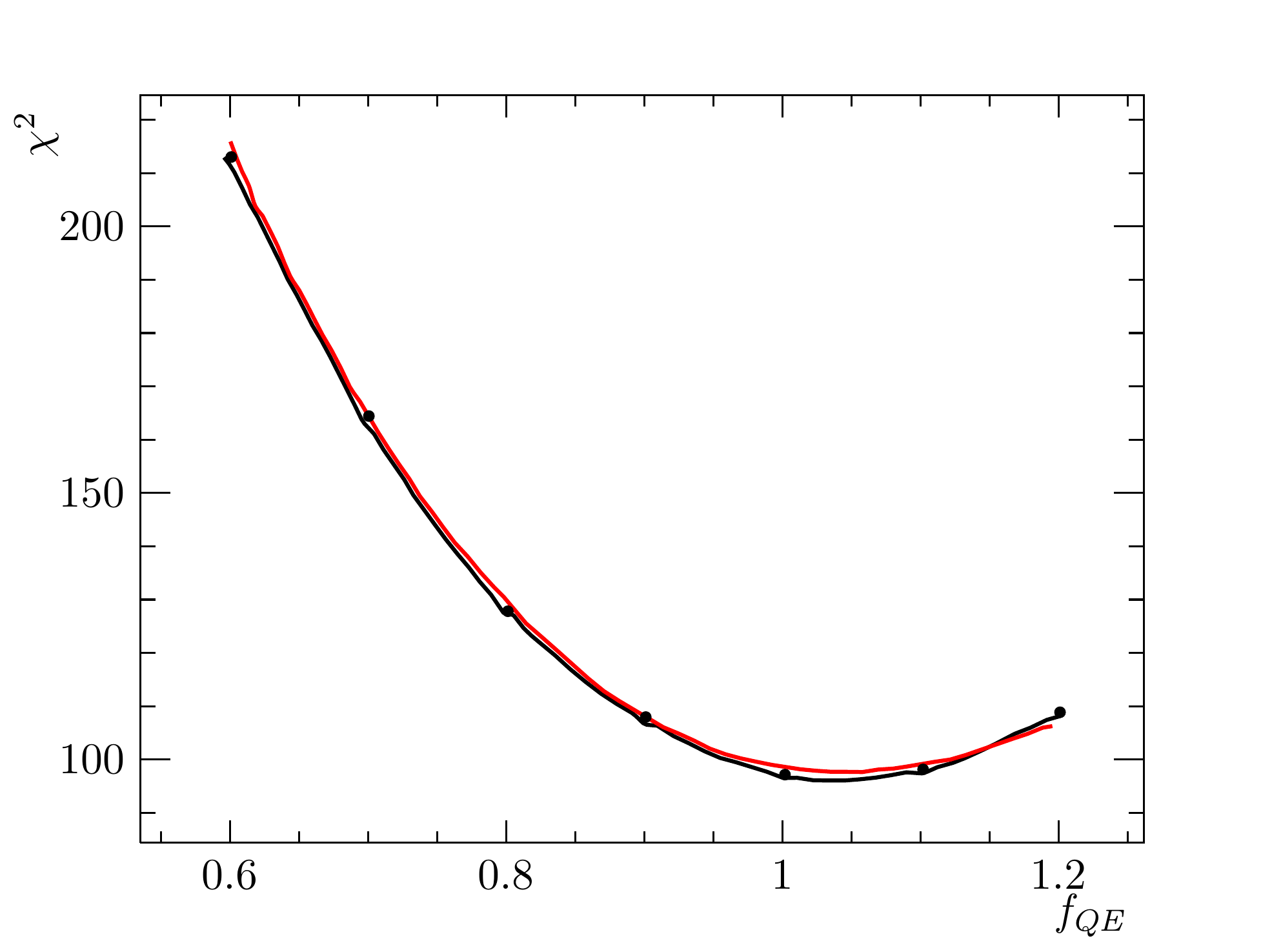}}  
  \subfloat[Quasi-elastic ($p_{\pi} \geq 400$ MeV/c)]{\includegraphics[width=0.33\linewidth]{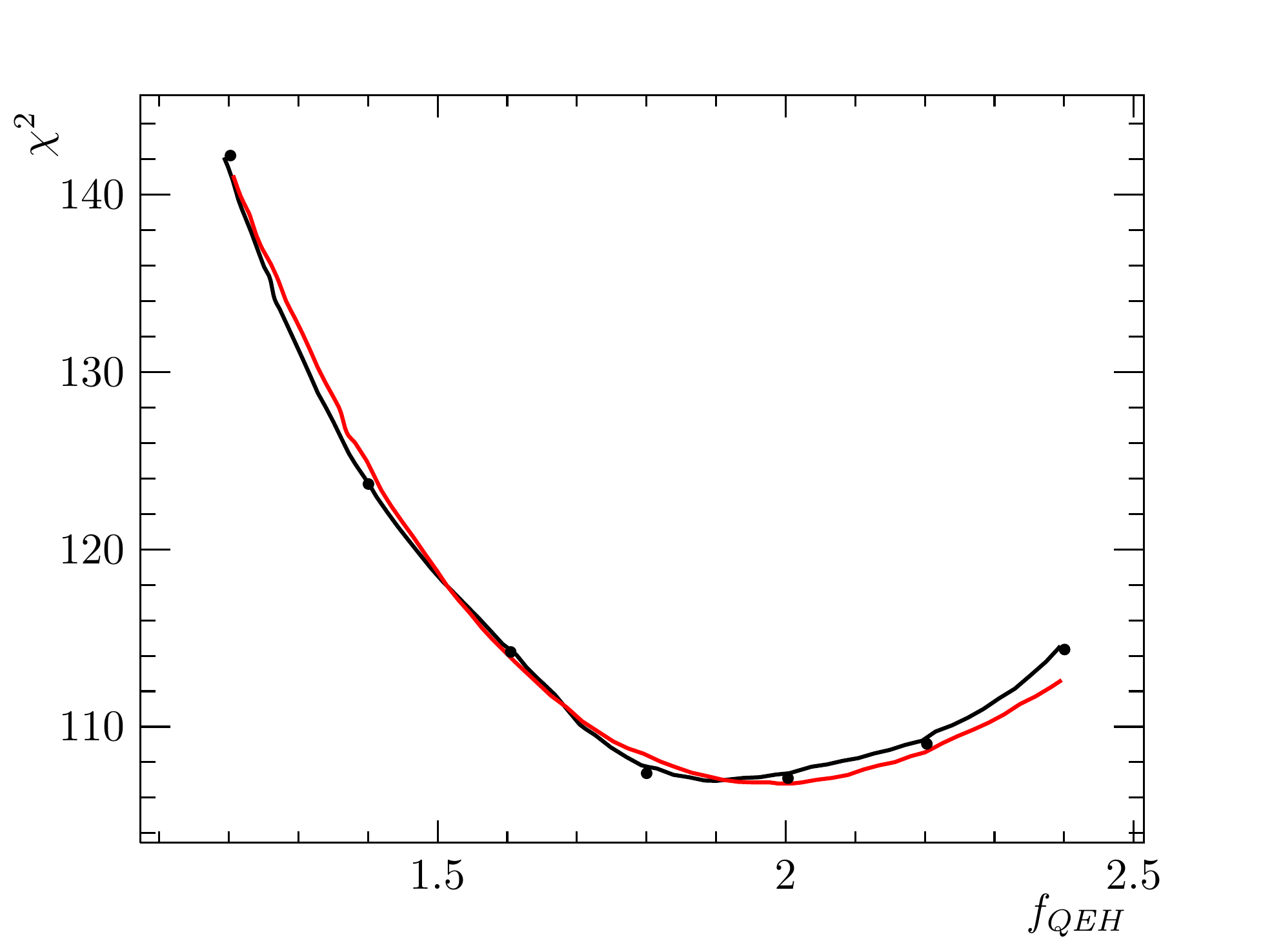}}\\
  \subfloat[Charge exchange ($p_{\pi} \leq 400$ MeV/c)]{\includegraphics[width=0.33\linewidth]{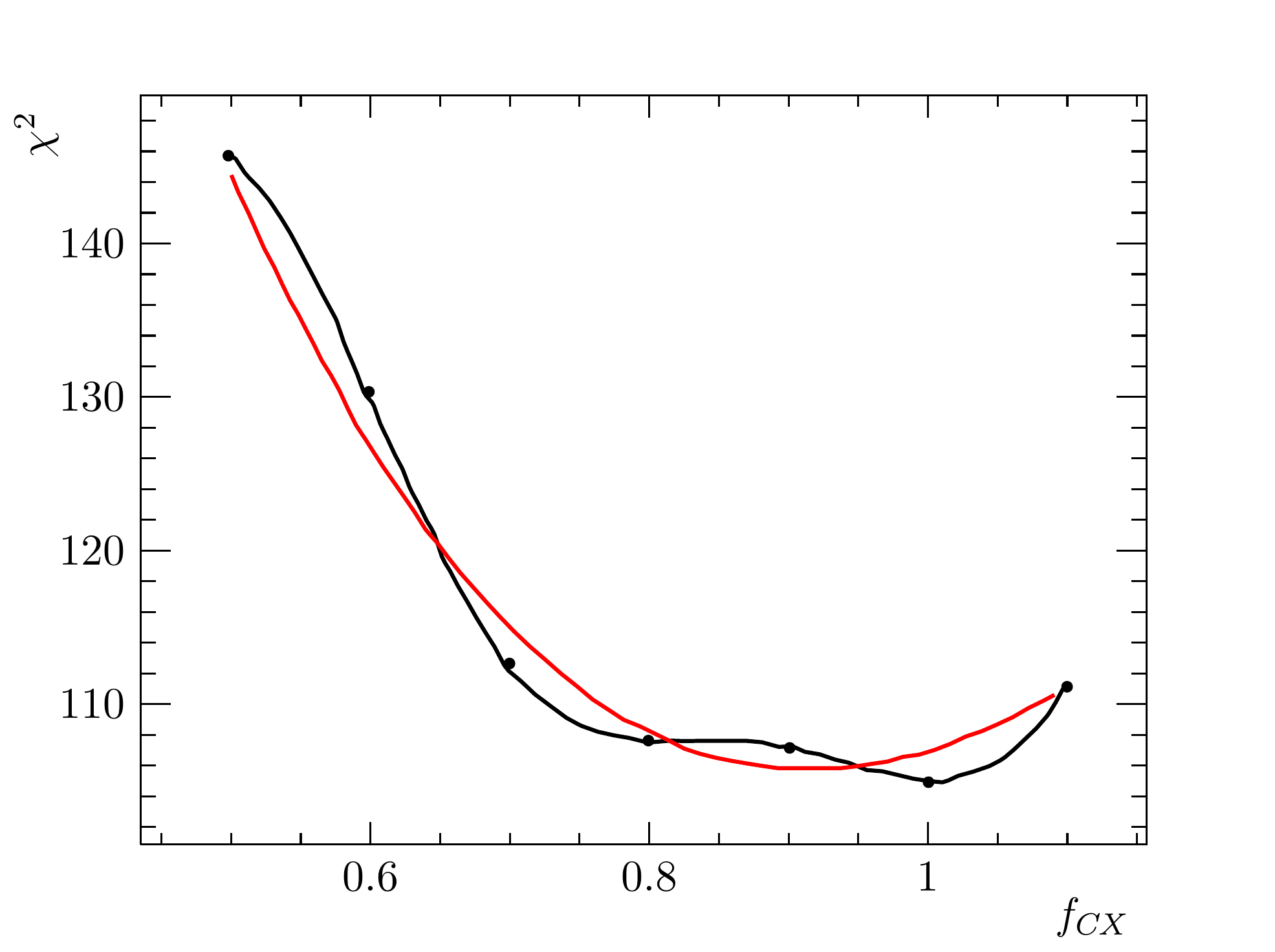}} 
  \subfloat[Charge exchange ($p_{\pi} \geq 400$ MeV/c)]{\includegraphics[width=0.33\linewidth]{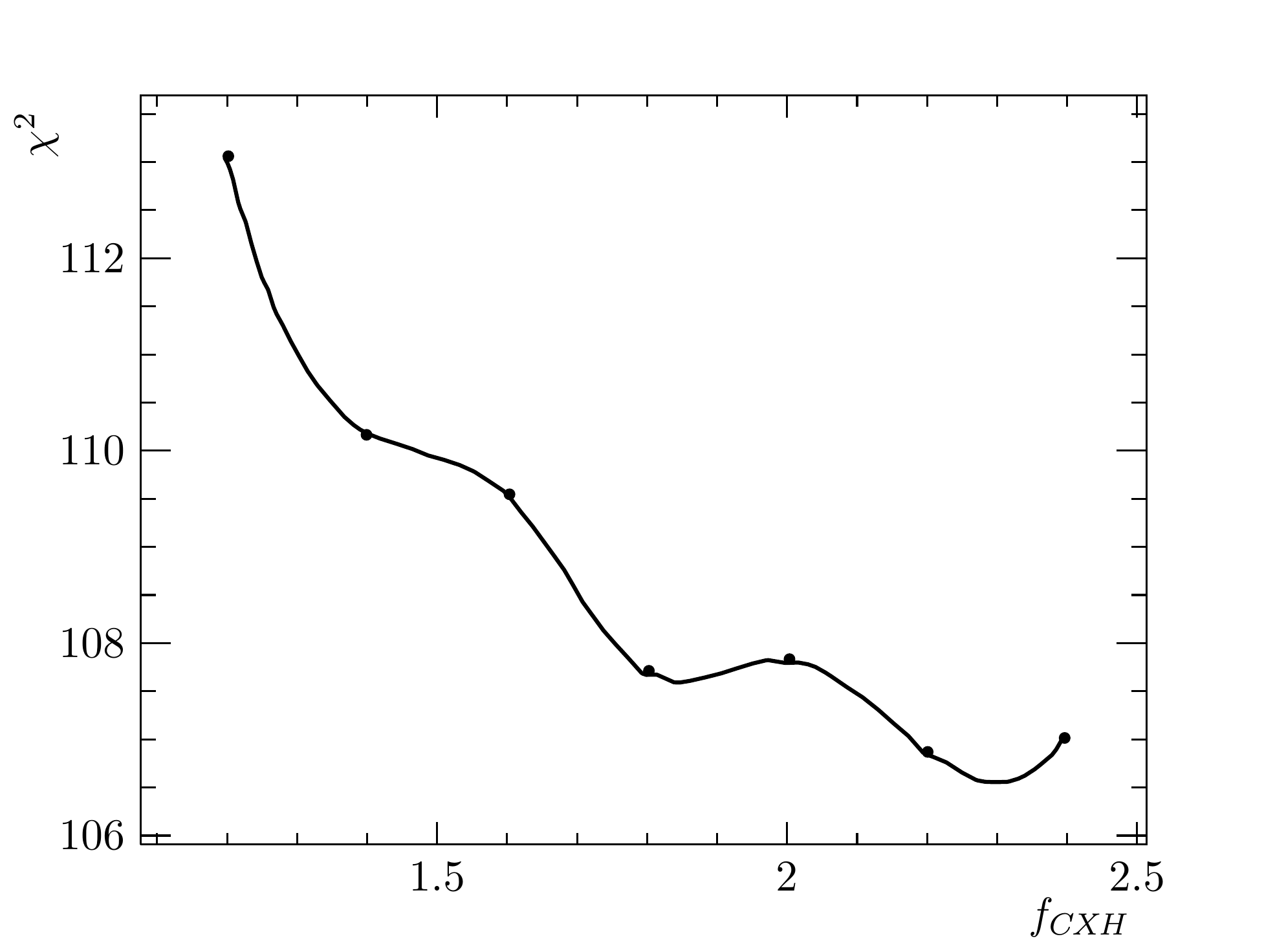}}
  \subfloat[Hadron production ($p_{\pi} \geq 400$ MeV/c)]{\includegraphics[width=0.33\linewidth]{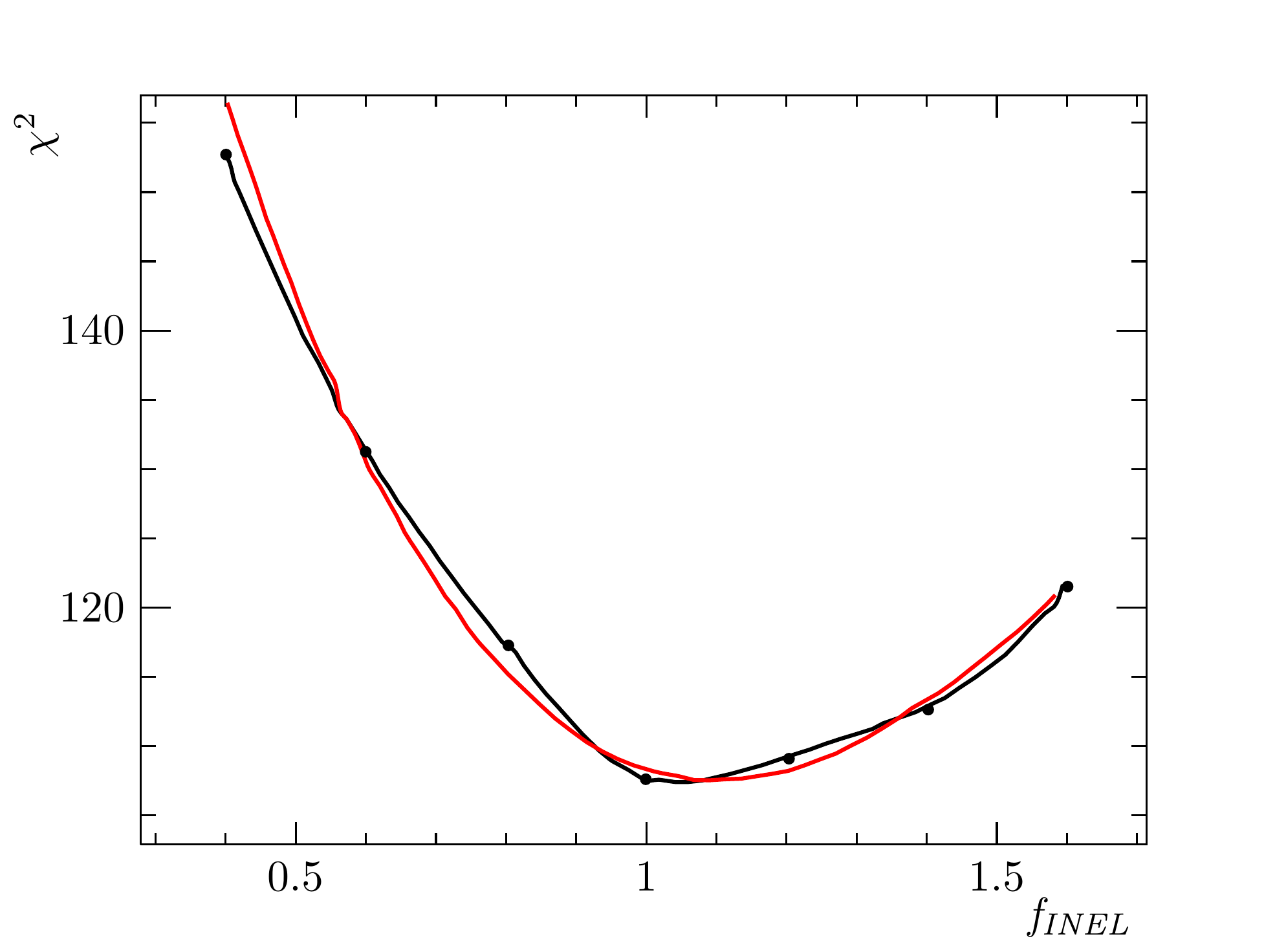}}
  \caption{One-dimensional projections of the interpolated $\chi^2$ grid for a point around the center of the grid using GNU-Octave (black) and TMultiDimFit (red). The points are given around the point: $f_{\mathrm{ABS}} = 1.25$, $f_{\mathrm{QE}} = 0.90$, $f_{\mathrm{QEH}} = 1.80$, $f_{\mathrm{CX}} = 0.80$, $f_{\mathrm{CXH}} = 1.80$, $f_{\mathrm{INEL}} = 1.00$. The points are the $\chi^2$ values in the pre-computed finite grid which is being interpolated.}
  \label{fig:interp-comp-1}
\end{figure*}

In general, it is difficult to compare multi-dimensional distributions and the two interpolation methods. Their one-dimensional projections of each is shown for illustrative purposes in Figure~\ref{fig:interp-comp-1}. The $f_{\mathrm{CXH}}$ parameter caused problems with the convergence of the TMultiDimFit parameterization due to its low constraining power (there is very little available relevant data), and thus was not included for this comparison. The interpolation methods are found to be consistent and no significant biases are expected.

\section{Fit Results}\label{sec:fit-results}
The fit was carried out in two ways. Firstly, the normalization parameters for each dataset, $\lambda_i$ from Equation~\ref{eqn:chi2}, were fixed to their nominal value of 1.0, with results shown in Section~\ref{sec:fixed-lambda}. Secondly, the $\lambda_i$'s were treated as constrained parameters in the fit, as described in Section~\ref{sec:float-lambda}. The former strategy forms the main result of this work, whereas the latter was used as a cross check.

\subsection{Fixed Normalization Parameters}
\label{sec:fixed-lambda}
The best fit FSI parameters, while keeping the normalization parameters fixed, are presented in Table~\ref{tab:fsifitter-bestfit} for both interpolation methods. The spread in the results from the methods was covered by the uncertainties on the fitted parameters. The $\chi^{2}$ between the two methods, using the covariance matrix calculated for the GNU-Octave fit, was 4.14 for 5 degrees of freedom. The minimum $\chi^2$ values from each method are in agreement and are shown in the last row of Table~\ref{tab:fsifitter-bestfit}, along with the number of degrees of freedom in the fit. This confirms that the interpolation methods are not introducing biases, and indicates that the $\chi^2$ did not have a local minimum that would affect the minimization process. Thus, no additional uncertainties due to the interpolation method choice were deemed necessary.

\begin{table}[htbp]
  \centering  
{\renewcommand{\arraystretch}{1.2}
\begin{tabular}{c c c}
\hline
\multirow{2}{*}{\textbf{Parameter}} & \multicolumn{2}{c}{\textbf{Best fit $\pm$ 1$\sigma$}} \\
& \textbf{TMultiDimFit} & \textbf{GNU-Octave} \\
\hline
$f_{\mathrm{QE}}$  & 1.07 $\pm$ 0.06 & 1.07 $\pm$ 0.04 \\
$f_{\mathrm{ABS}}$ & 1.50 $\pm$ 0.08 & 1.40 $\pm$ 0.06 \\
$f_{\mathrm{CX}}$  & 0.69 $\pm$ 0.05 & 0.70 $\pm$ 0.04 \\
$f_{\mathrm{INEL}}$ & 0.89 $\pm$ 0.20 & 1.00 $\pm$ 0.15 \\
$f_{\mathrm{QEH}}$ & 1.90 $\pm$ 0.25 & 1.82 $\pm$ 0.11 \\\hline
$\chi^2$/n.d.o.f & 150.74/59 & 149.03/59 \\ \hline 
\end{tabular}}  
\caption{Post-fit FSI parameters and the minimum $\chi^2$ value with fixed normalization parameters using the TMultiDimFit and GNU-Octave interpolation methods.}
\label{tab:fsifitter-bestfit}
\end{table}

\begin{figure}[htbp]
\centering
\includegraphics[width=86mm]{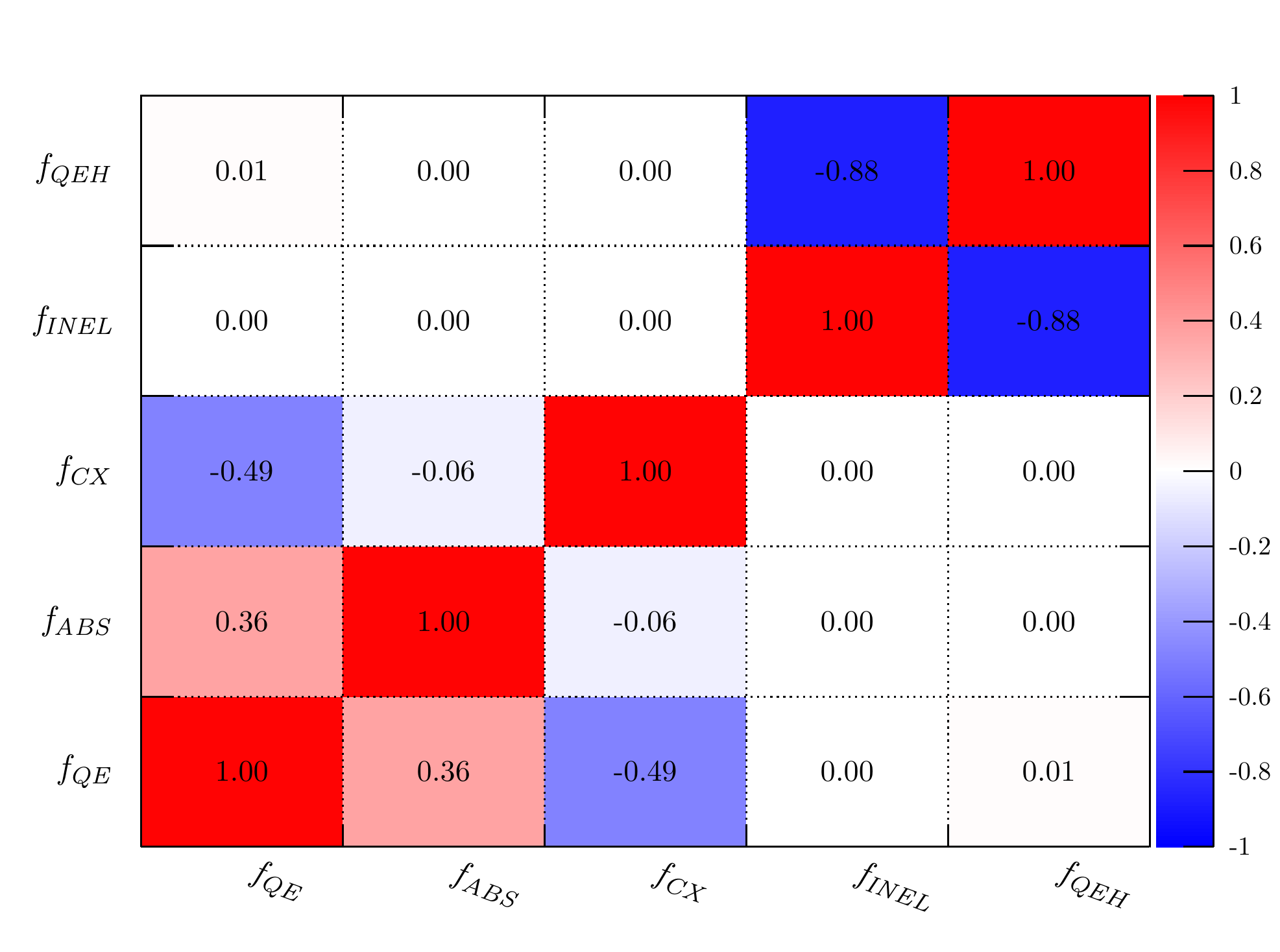} 
\caption{Correlation matrix using the TMultiDimFit interpolation for the fit with fixed normalization parameters.}
\label{fig:fsifitter-covariance-multidimfit}
\end{figure}

\begin{figure}[htbp]
\centering
\includegraphics[width=86mm]{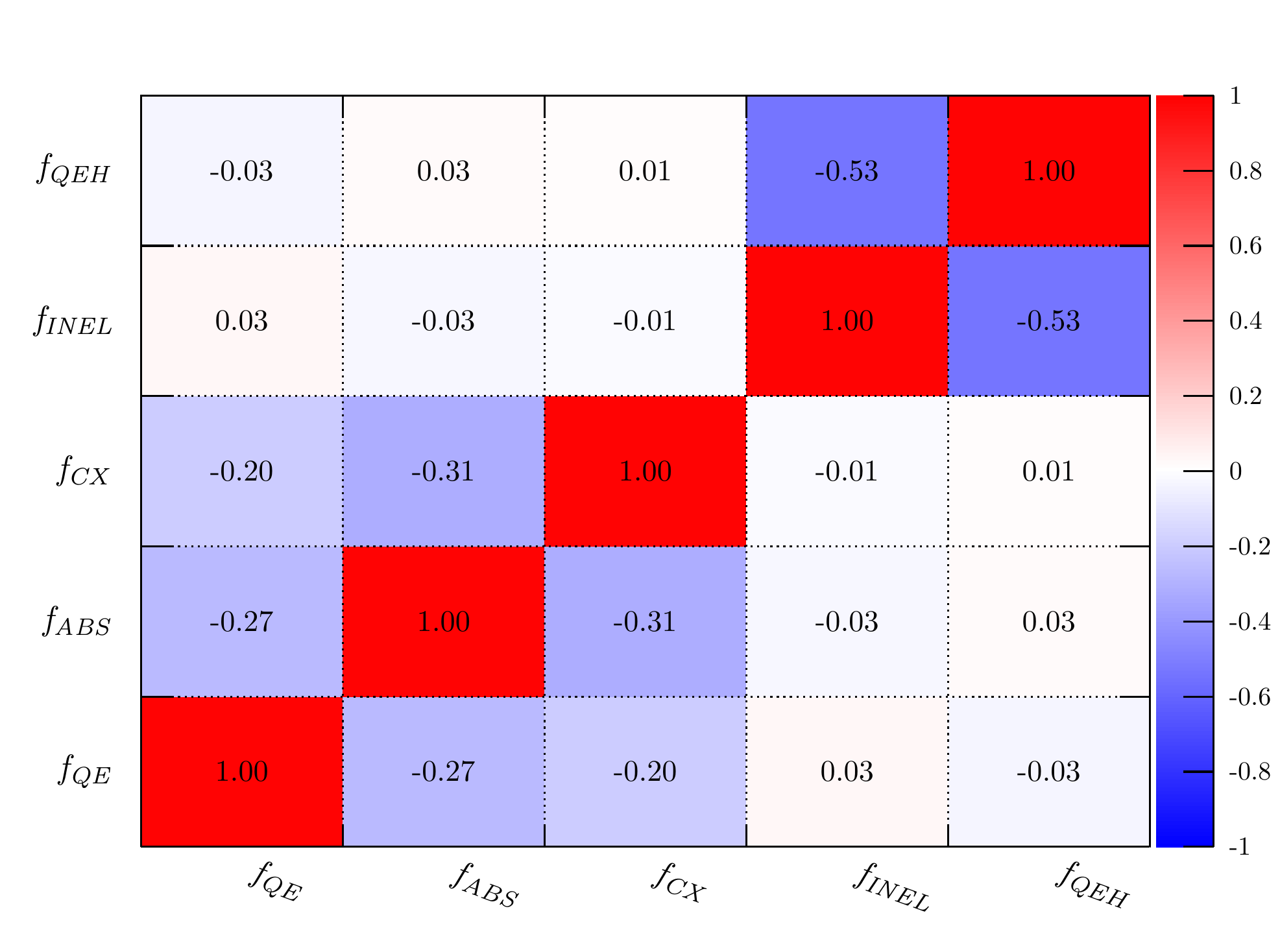} 
\caption{Correlation matrix using the GNU-Octave interpolation for the fit with fixed normalization parameters.}
\label{fig:fsifitter-covariance-octave}
\end{figure}

Figures ~\ref{fig:fsifitter-covariance-multidimfit} and~\ref{fig:fsifitter-covariance-octave} show the covariance matrices obtained from MINUIT using each interpolation method. Stronger correlations across the FSI parameters are observed when using the TMultiDimFit interpolation. This can be understood as the effect of the polynomial parameterization, which inherently carries strong correlations from the large number of cross-terms. For this reason, it was decided to use the GNU-Octave interpolation for the final results.

\subsection{Varying Normalization Parameters}
\label{sec:float-lambda}
To investigate the dependence of the $\chi^2$ and fit results on each data set, the normalization parameters ($\lambda_{i}$ in Equation~\ref{eqn:chi2}) were allowed to vary. This keeps the n.d.o.f the same, but the $\chi^{2}$ is expected to reduce as correlated changes in the datasets from a single measurement introduce a smaller $\chi^{2}$ penalty (e.g., if there is a flux modeling issue with a dataset). This approach has been used elsewhere when fitting to datasets with missing covariance information~\cite{niwg_paper,pumplin2000_pdfs}. Including these parameters as pull terms, rather than adding correlations between datapoints in a covariance matrix, gives greater insight into the behaviour of the fitter. For instance, a dataset with datapoints that all pull strongly on the fitted FSI parameters in the same direction in Section~\ref{sec:fixed-lambda} would have a normalization parameter largely deviating from 1.0.

The fit was performed for three selections of external data sets: data on carbon nuclei only; data on light nuclei (carbon, oxygen, aluminum); and data on light and heavy nuclei (carbon, oxygen, aluminum, iron, copper, and lead). This was done to investigate how the cascade model scales with increasing nuclear size, and the effect on the FSI parameters. Table~\ref{tab:fsifitter-bestfit-floating} shows the best fit FSI parameters for each case using the GNU-Octave method. The agreement across the three cases indicates that the model is able to consistently describe all the data, and that the fit is well behaved.

\begin{figure*}[htbp]
  \centering
  \subfloat[$^{12}$C-only datasets] {\includegraphics[height=0.30\textheight]{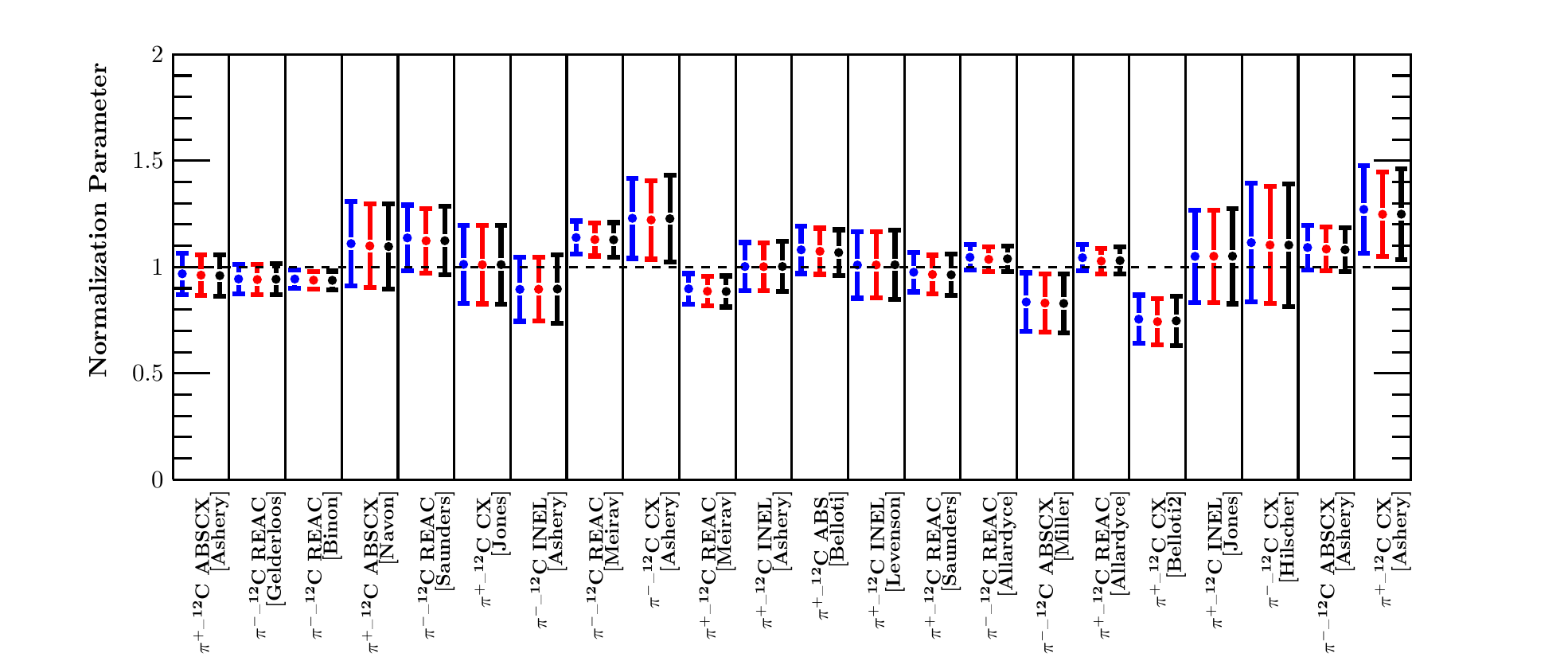}}\\\vspace{-12pt}
  \subfloat[Other light nuclei]    {\includegraphics[height=0.30\textheight]{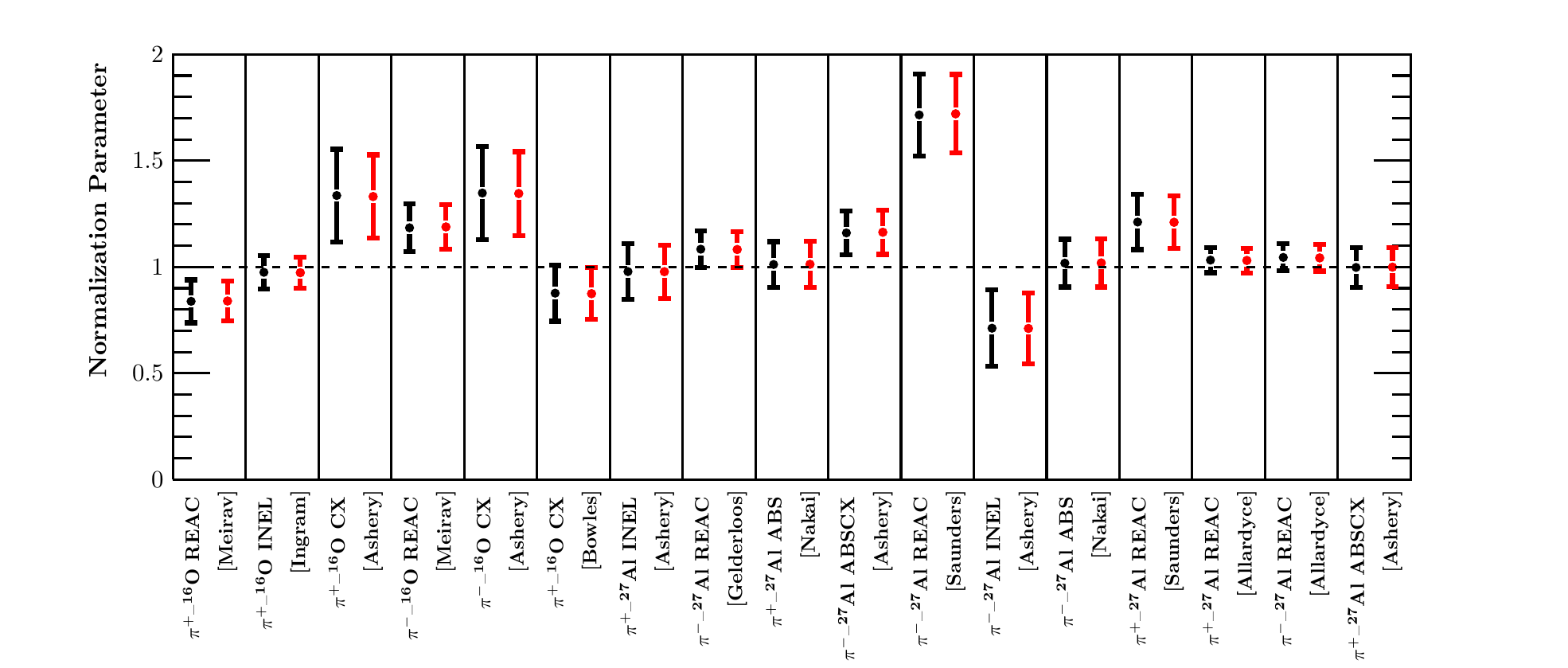}}\\\vspace{-12pt}
  \subfloat[Heavy nuclei]          {\includegraphics[height=0.30\textheight]{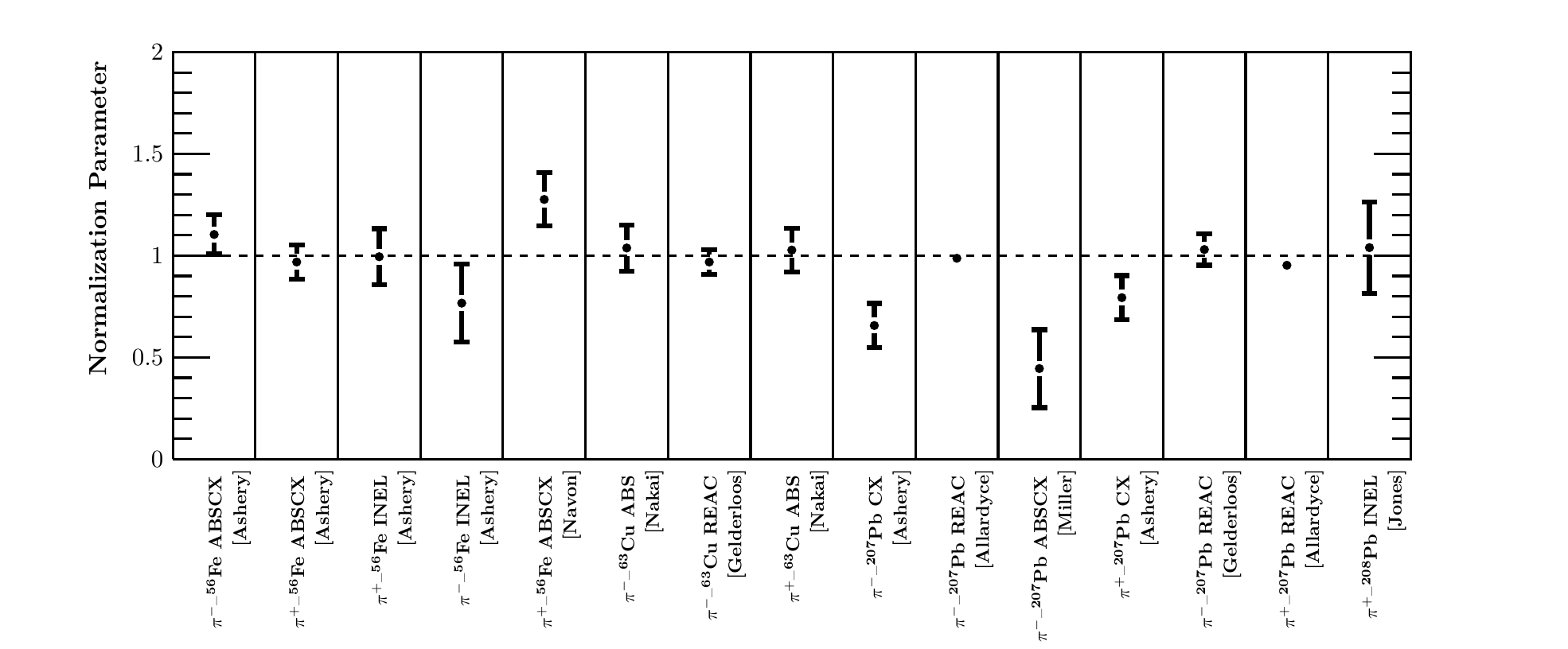}}
\caption{Post-fit normalization parameters for each data set for carbon-only (blue), light nuclei (red) and all-nuclei (black) fits. Some points overlap each other.}
\label{fig:fsifitter-normalization}
\end{figure*}

Figure~\ref{fig:fsifitter-normalization} shows the fitted normalization parameters for the three selections of external data. The normalization parameters roughly follow a Gaussian distribution, and the results from each of the three cases essentially overlap each other. The varying sizes of the post-fit errors is understood to be a consequence of assigning the same 40\% pre-fit error to all the parameters, as different channels have been experimentally measured with different levels of precision. For instance, inelastic and reactive processes tend to have much smaller uncertainties and the model does much better there.

\begin{table}[htbp]
\centering
{\renewcommand{\arraystretch}{1.2}
\begin{tabular}{c c c c}
\hline
\multirow{2}{*}{\textbf{Parameter}} & \multicolumn{3}{c}{\textbf{Best fit $\pm$ 1$\sigma$}} \\
& \textbf{Carbon-only} & \textbf{Light nuclei} & \textbf{All nuclei} \\
\hline
$f_{\mathrm{QE}}$   & 1.07 $\pm$ 0.07 & 1.08 $\pm$ 0.07 & 1.08 $\pm$ 0.07 \\
$f_{\mathrm{ABS}}$  & 1.24 $\pm$ 0.05 & 1.25 $\pm$ 0.05 & 1.26 $\pm$ 0.05 \\
$f_{\mathrm{CX}}$   & 0.79 $\pm$ 0.05 & 0.80 $\pm$ 0.04 & 0.80 $\pm$ 0.04 \\
$f_{\mathrm{INEL}}$ & 0.63 $\pm$ 0.27 & 0.71 $\pm$ 0.21 & 0.70 $\pm$ 0.20 \\
$f_{\mathrm{QEH}}$  & 2.16 $\pm$ 0.34 & 2.14 $\pm$ 0.24 & 2.13 $\pm$ 0.22 \\ \hline
$\chi^2$/n.d.o.f & 18.36/27 & 40.14/44 & 53.48/59 \\ \hline 
\end{tabular}}
\caption{Post-fit NEUT FSI parameters and the minimum $\chi^2$/n.d.o.f values obtained for fits with floating normalization parameters and the specified external data selections.}
\label{tab:fsifitter-bestfit-floating}
\end{table}

As a further cross-check, the fit was repeated after removing the five data sets whose post-fit normalization parameters were more than 1$\sigma$ away from the nominal value of 1.0 in Figure~\ref{fig:fsifitter-normalization}. Table~\ref{tab:fsifitter-bestfit-floating-removed} shows the post-fit FSI parameters for that study. The effect on all parameters and the remaining normalizations was found to be negligible.

\begin{table}[htbp]
\centering
{\renewcommand{\arraystretch}{1.2}
\begin{tabular}{c c c c}
\hline
\multirow{2}{*}{\textbf{Parameter}} & \multicolumn{3}{c}{\textbf{Best fit $\pm$ 1$\sigma$}} \\
& \textbf{Carbon-only} & \textbf{Light nuclei} & \textbf{All nuclei} \\
\hline
$f_{\mathrm{QE}}$  & 1.07 $\pm$ 0.08 & 1.09 $\pm$ 0.06 & 1.09 $\pm$ 0.06 \\
$f_{\mathrm{ABS}}$ & 1.23 $\pm$ 0.05 & 1.25 $\pm$ 0.04 & 1.26 $\pm$ 0.04 \\
$f_{\mathrm{CX}}$  & 0.79 $\pm$ 0.05 & 0.81 $\pm$ 0.04 & 0.80 $\pm$ 0.04 \\
$f_{\mathrm{INEL}}$ & 0.63 $\pm$ 0.27 & 0.71 $\pm$ 0.21 & 0.71 $\pm$ 0.19 \\
$f_{\mathrm{QEH}}$  & 2.15 $\pm$ 0.34 & 2.14 $\pm$ 0.24 & 2.13 $\pm$ 0.21 \\ \hline
$\chi^2$/n.d.o.f & 17.95/26 & 35.55/42 & 45.13/54 \\ \hline 
\end{tabular}}
\caption{Post-fit NEUT FSI parameters and the minimum $\chi^2$/n.d.o.f values obtained for fits with floating normalization parameters and the five data sets with strongest pulls in Figure~\ref{fig:fsifitter-normalization} removed.}
\label{tab:fsifitter-bestfit-floating-removed}
\end{table}

\subsection{Drawing Error Envelopes}\label{sec:drawing-envelopes}
\begin{figure}[htbp]
\centering
\includegraphics[width=86mm]{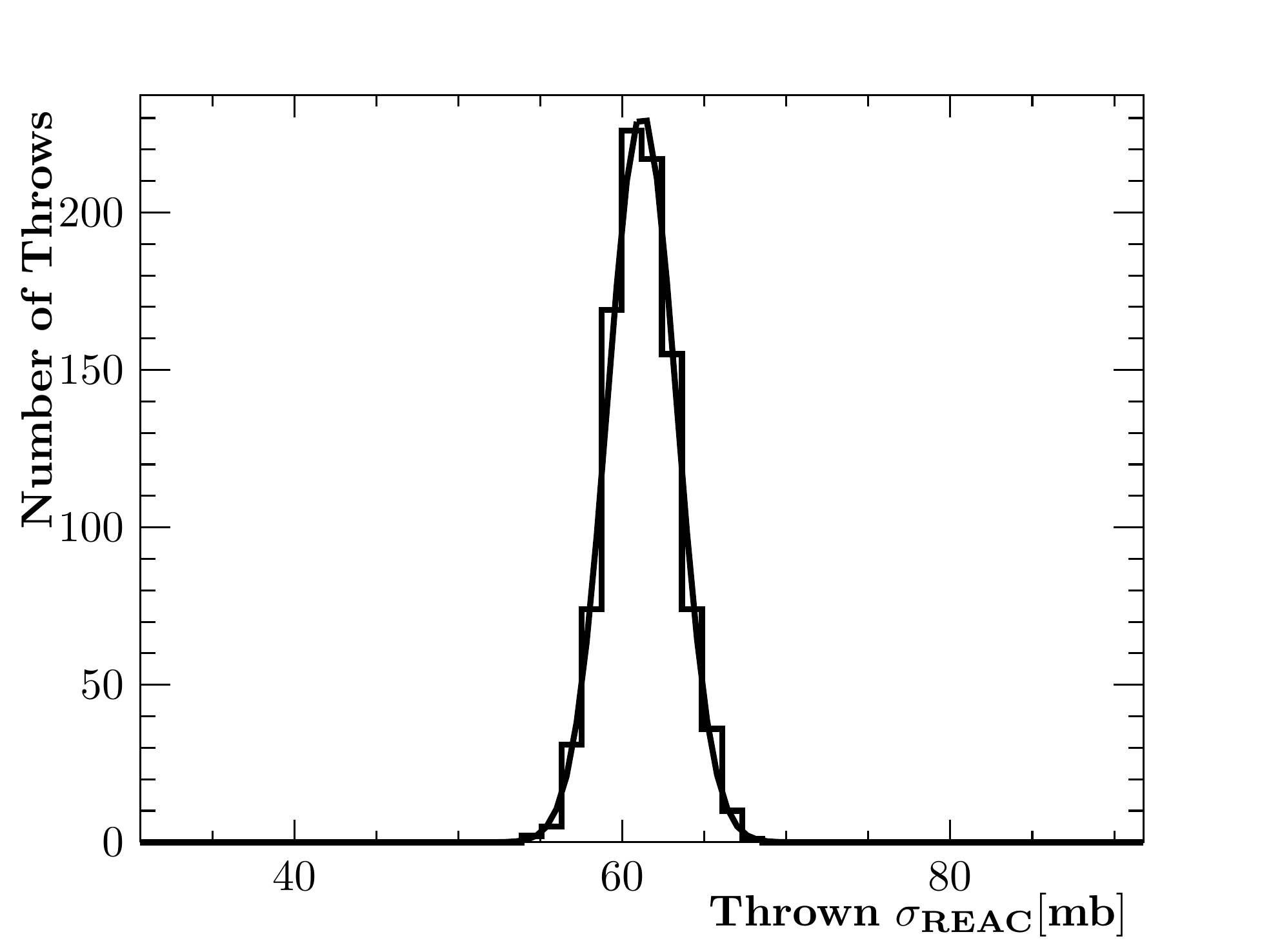} 
\caption{PDF for $\pi^{+}$--$^{12}$C $\sigma_{\mathrm{REAC}}$ for $p_\pi = 20.0$ MeV$/c$.}
\label{fig:fsifitter-pdf-1}
\end{figure}

\begin{figure}[htbp]
\centering
\includegraphics[width=86mm]{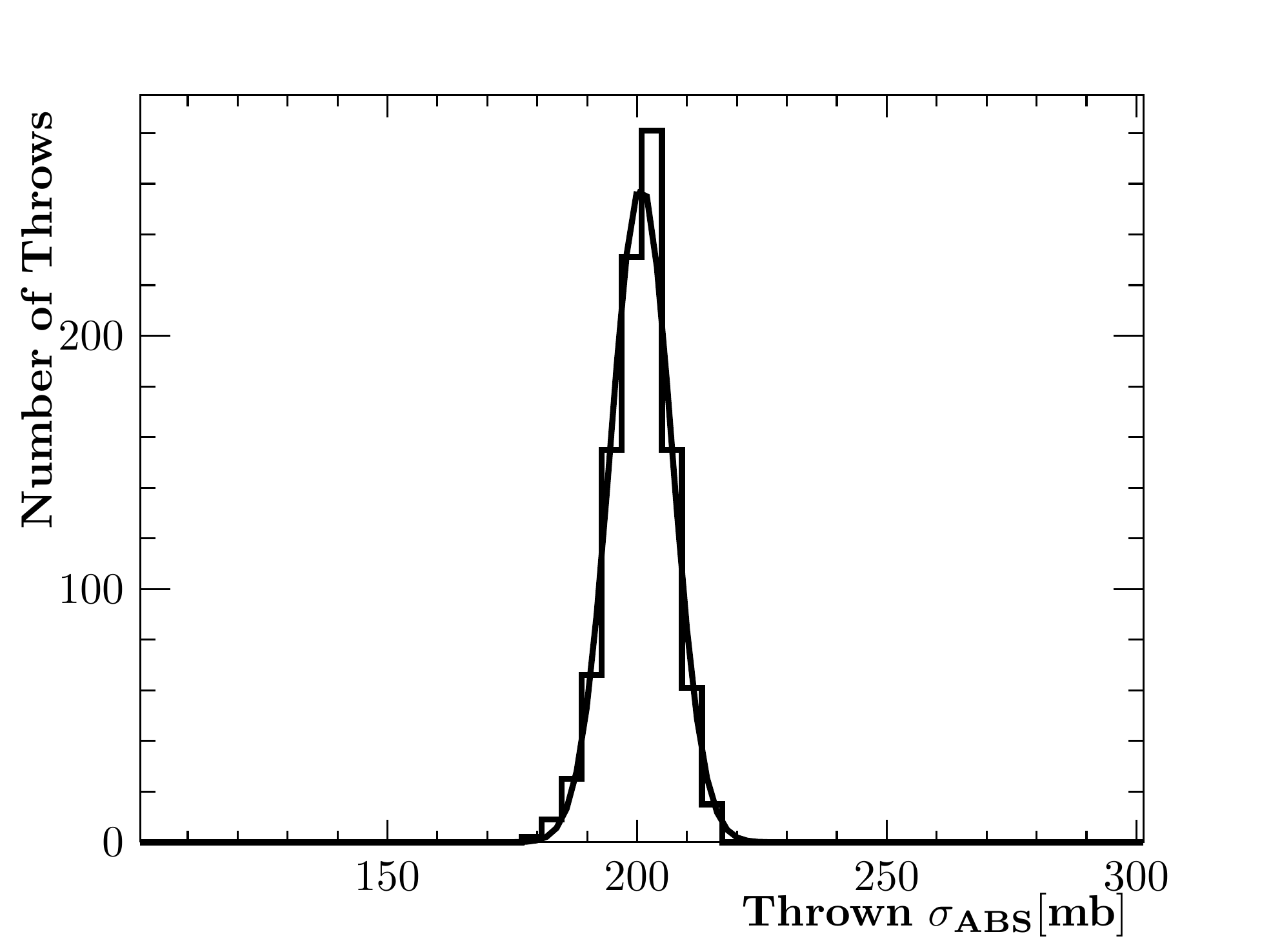} 
\caption{PDF for $\pi^{+}$--$^{12}$C $\sigma_{\mathrm{ABS}}$ for $p_\pi = 265.07$ MeV$/c$.}
\label{fig:fsifitter-pdf-2}
\end{figure}

The constraints of the FSI parameters from the fit were used to form the 1$\sigma$ variations of the macroscopic scattering cross sections to allow for comparisons with external data. This was done as:
\begin{enumerate}
\item Generate a random correlated throw from the obtained covariance matrix (e.g., Figure~\ref{fig:fsifitter-covariance-octave}) through Cholesky decomposition. Throws outside the finite grid defined in Table~\ref{tab:fsi-grid} are discarded since the interpolation does not apply.
\item Draw the probability distribution function (PDF) for each momentum value for which $\sigma^{\mathrm{NEUT}}(f_{\mathrm{FSI}})$ has been calculated using the GNU-Octave spline interpolation. Figures~\ref{fig:fsifitter-pdf-1} and~\ref{fig:fsifitter-pdf-2} show examples of these distributions for two combinations of interaction channel and momentum.
\item Fit a Gaussian function to the PDFs.
\item Use the mean and variance from the Gaussian for each value of momentum to draw the 1$\sigma$ error envelopes.
\end{enumerate}

\begin{figure*}[htbp]
  \centering
  \subfloat[Reactive]            {\includegraphics[width=0.33\linewidth]{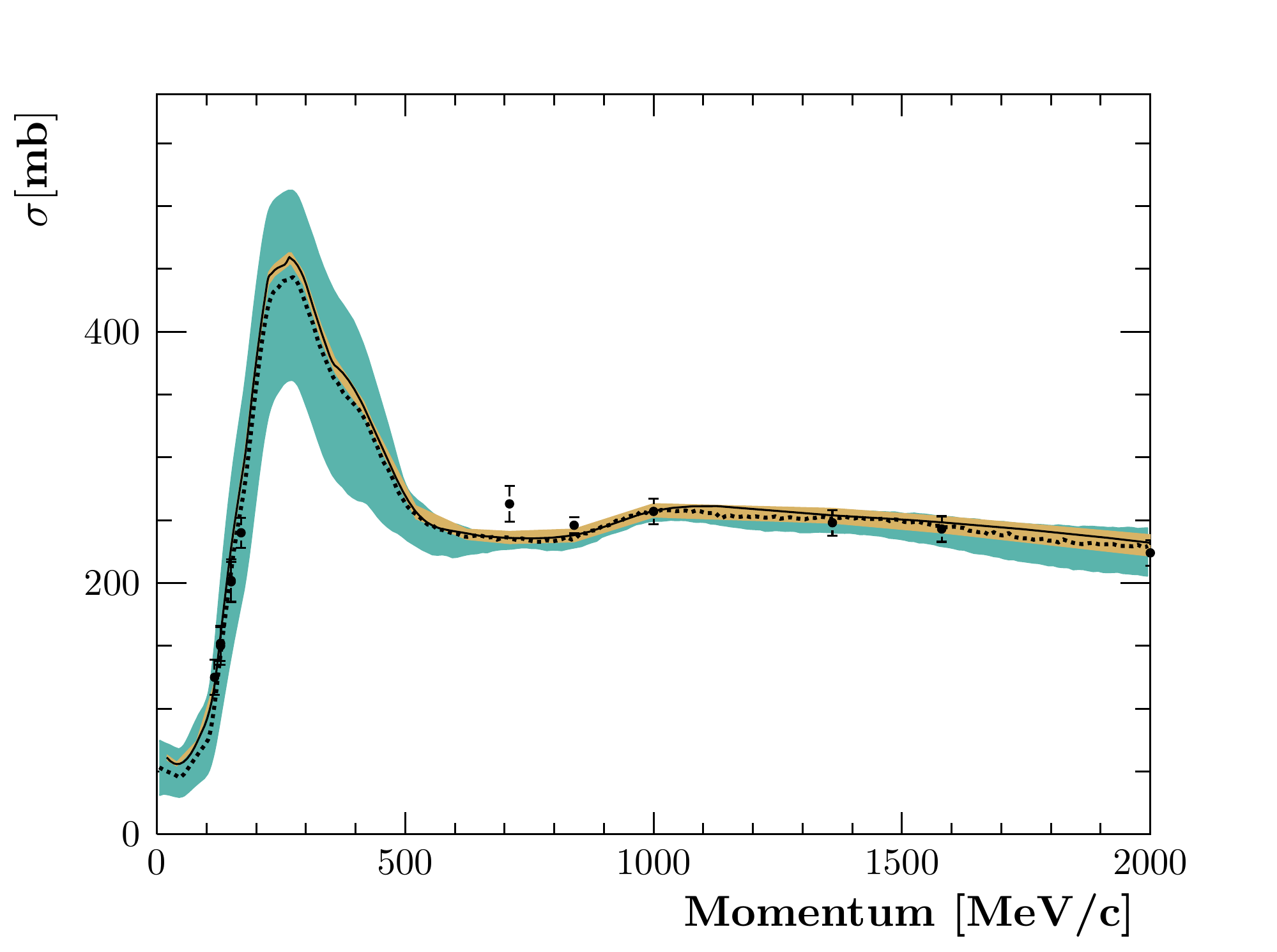}}
  \subfloat[Quasi-elastic]       {\includegraphics[width=0.33\linewidth]{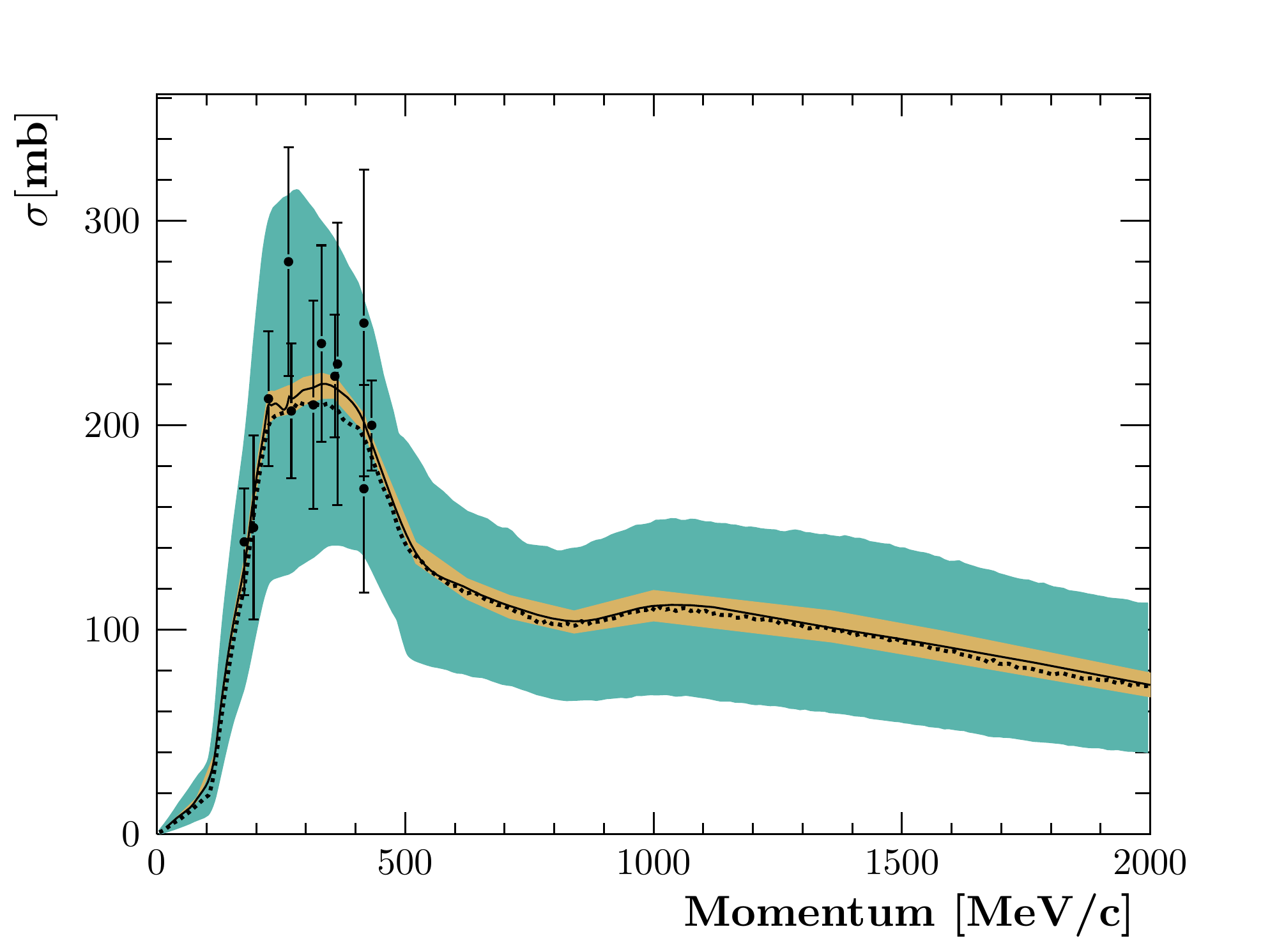}}
  \subfloat[Absorption (ABS)]    {\includegraphics[width=0.33\linewidth]{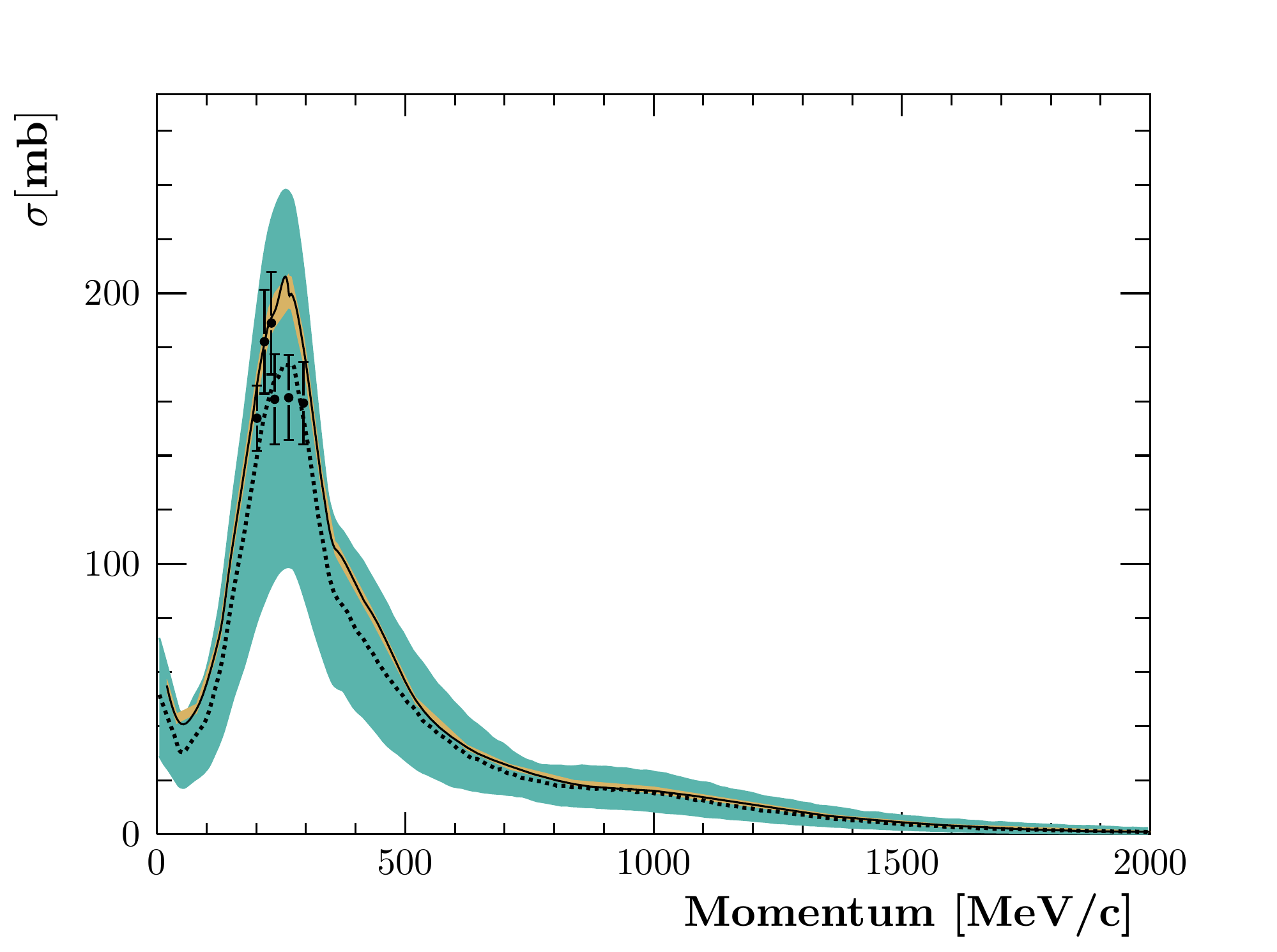}}\\\vspace{-12pt}
  \subfloat[Charge exchange (CX)]{\includegraphics[width=0.33\linewidth]{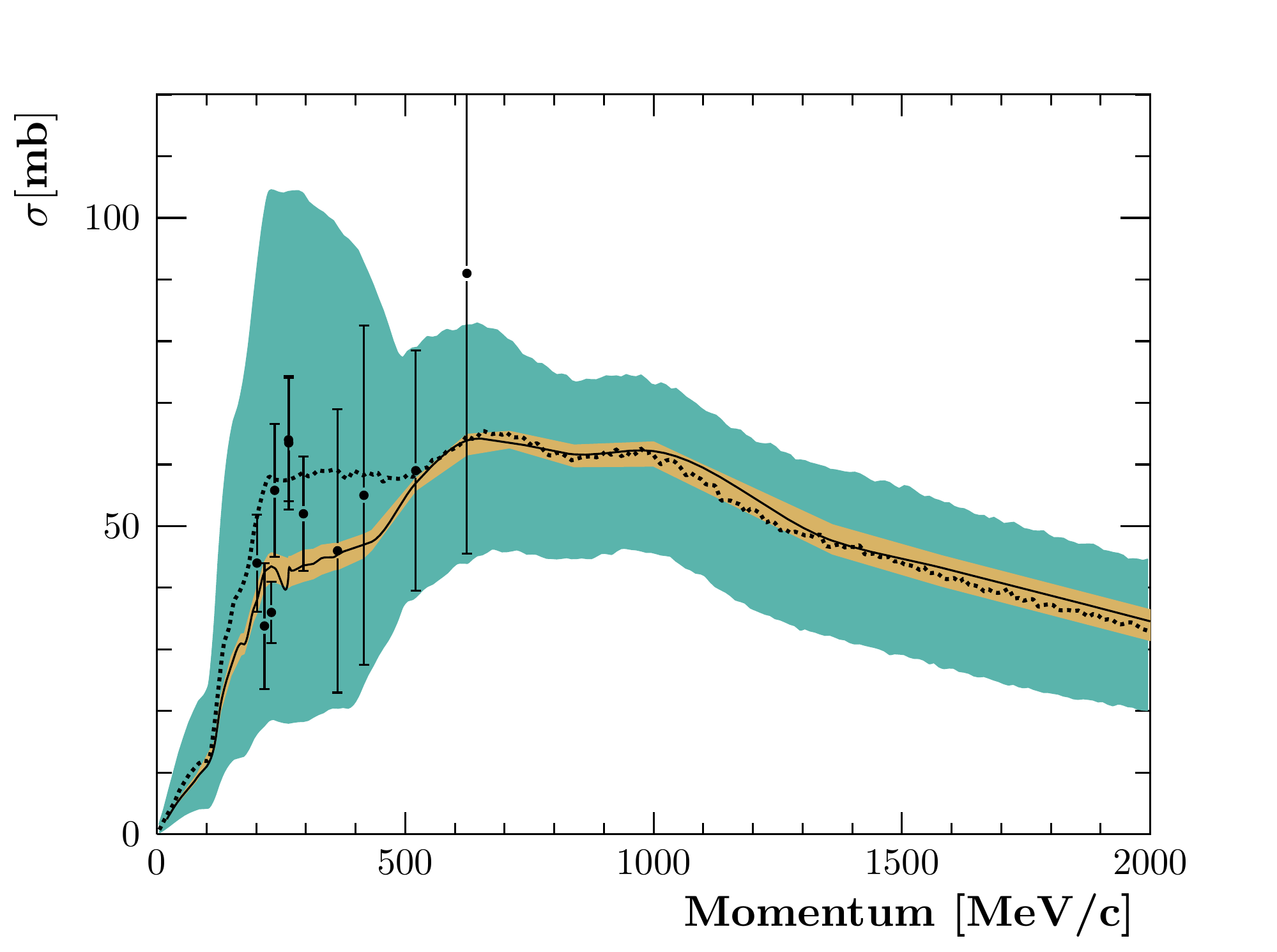}}
  \subfloat[ABS+CX]              {\includegraphics[width=0.33\linewidth]{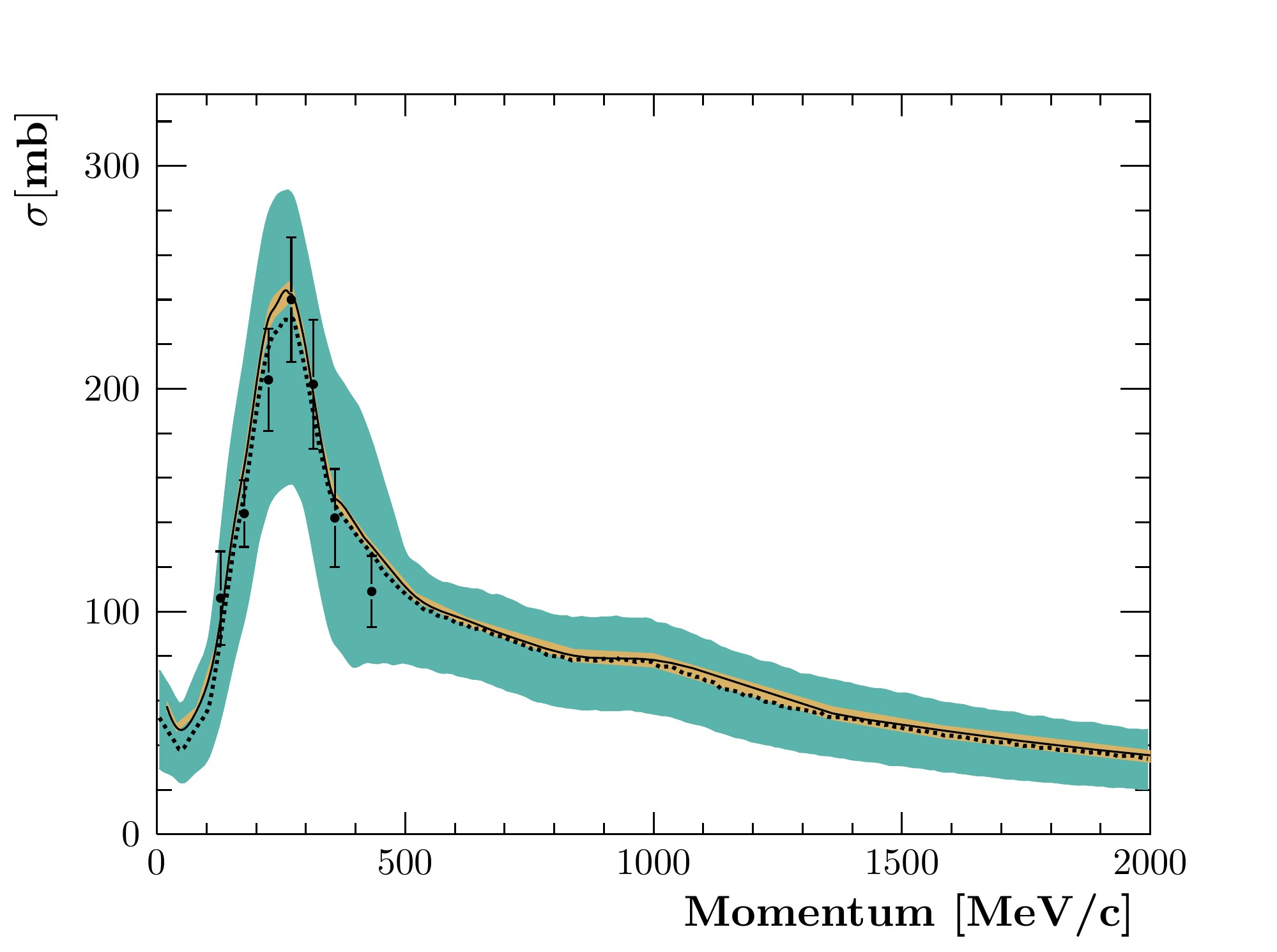}}
  \subfloat                      {\includegraphics[width=0.33\linewidth]{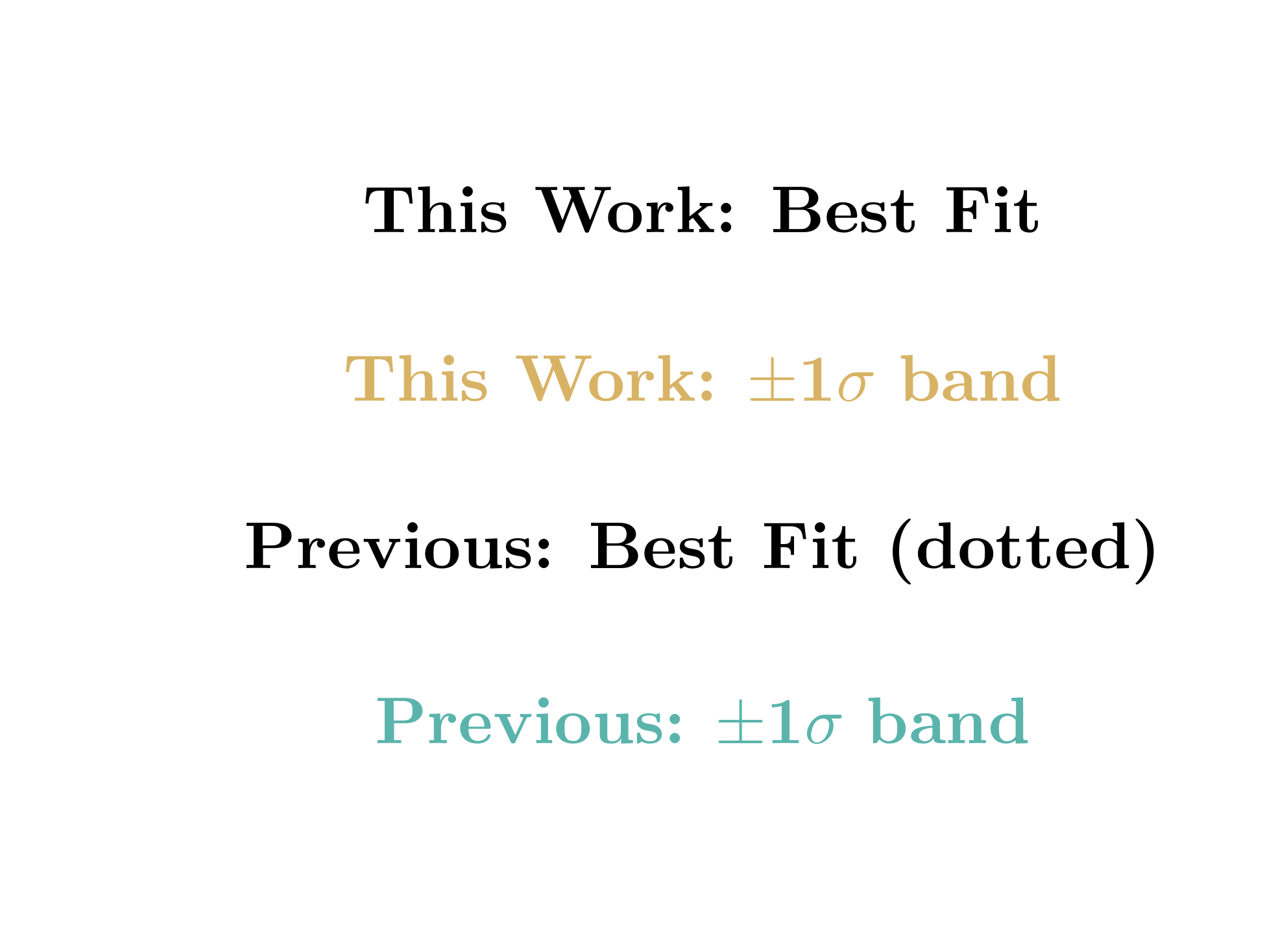}}
\caption{Comparison of the available $\pi^+$--$^{12}$C cross section external data with the NEUT best fit (solid black line) and $1\sigma$ band (red), and the best fit (dashed black line) and 1$\sigma$ band (blue) obtained from a previous iteration of this fit~\cite{long_prd, deperio-thesis}.}
\label{fig:fsifitter-envelopes-scaled-1}
\end{figure*}

Figure~\ref{fig:fsifitter-envelopes-scaled-1} shows the resulting error bands for the $\pi^+$--$^{12}$C cross sections, obtained using the constraints from the correlation matrix in Figure~\ref{fig:fsifitter-covariance-octave} and the procedure described above. The coverage of the error is insufficient when using the constraints from Table~\ref{tab:fsifitter-bestfit}, and was also true for the data on more nuclei. This is dealt with in Section \ref{sec:error-inflation}.

\subsection{Error Inflation}
\label{sec:error-inflation}
The problem of lack of coverage arises from fitting to datasets for which no covariances are provided. Here we follow the same procedure as described in Ref.~\cite{niwg_paper}, which builds on an error inflation procedure defined in Ref.~\cite{pumplin2000_pdfs}. We inflate the fit uncertainties such that 68\% of the data is indeed covered at 1$\sigma$, as one would expect from the fit, whilst keeping the post-fit central values and correlations of the FSI parameters. Explicitly, we scale the cross section uncertainties obtained from the fit, $\Delta\sigma_j^{\mathrm{MC}}(f_{\mathrm{FSI}})$, by some scaling factor $\xi$, and calculate the level of agreement between the pion-nucleus cross section corresponding to the best fit FSI parameters, $\sigma_j^{\mathrm{MC}}(f_{\mathrm{FSI}})$, and the data, $\sigma_j^{\mathrm{Data}}$, given the inflated uncertainty on the pion-nucleus cross section, $\Delta\sigma_j^{\mathrm{MC}}(\xi\Delta f_{\mathrm{FSI}})$, using the figure of merit $\Psi$,
\begin{equation}\label{eqn:psi}
\Psi = \frac{\sigma_j^{\mathrm{Best\;fit}}(f_{\mathrm{FSI}}) - \sigma_j^{\mathrm{Data}}}{\Delta\sigma_j^{\mathrm{MC}}(\xi \Delta f_{\mathrm{FSI}})}.
\end{equation}

By increasing the scaling factor $\xi$ until the distribution of $\Psi$ plotted for all data points has an RMS of 1, we enforce the expected coverage in a naive way---e.g. 68\% of the data are covered by the post-fit uncertainties at 1$\sigma$, neglecting correlations between data points, and between different points in the MC. This approximation almost certainly results in over coverage, and very conservative inflated uncertainty bands, but given the lack of information available in the fit, rigorous definitions of coverage are not possible and we prefer over- to under-coverage. Table~\ref{tab:Psi_scaling} shows the $\Psi$ distribution mean and RMS for various values of $\xi$. A linear fit to these RMS values was performed, and the scaling value for which the RMS was equal to 1, $\xi$ = 57.0, was found. Such a large scaling factor is not very surprising given the large number of datasets without correlations and is consistent with the results found elsewhere~\cite{pumplin2000_pdfs}.

\begin{table}[htbp]
\centering
{\renewcommand{\arraystretch}{1.2}
\begin{tabular}{c c c}
\hline
\textbf{Scale} & \textbf{Mean} & \textbf{RMS} \\
\hline
$1/1$  &  0.39 & 2.86 \\
$1/9$  &  0.00 & 2.02 \\
$1/25$ & -0.11 & 1.28 \\
$1/49$ & -0.14 & 1.09 \\
$1/81$ & -0.09 & 1.00 \\
\hline
\end{tabular}}
\caption{$\Psi$ for various scaling factors.}
\label{tab:Psi_scaling}
\end{table}

\begin{figure*}[htbp]
  \centering
  \subfloat[Reactive]            {\includegraphics[width=0.33\linewidth]{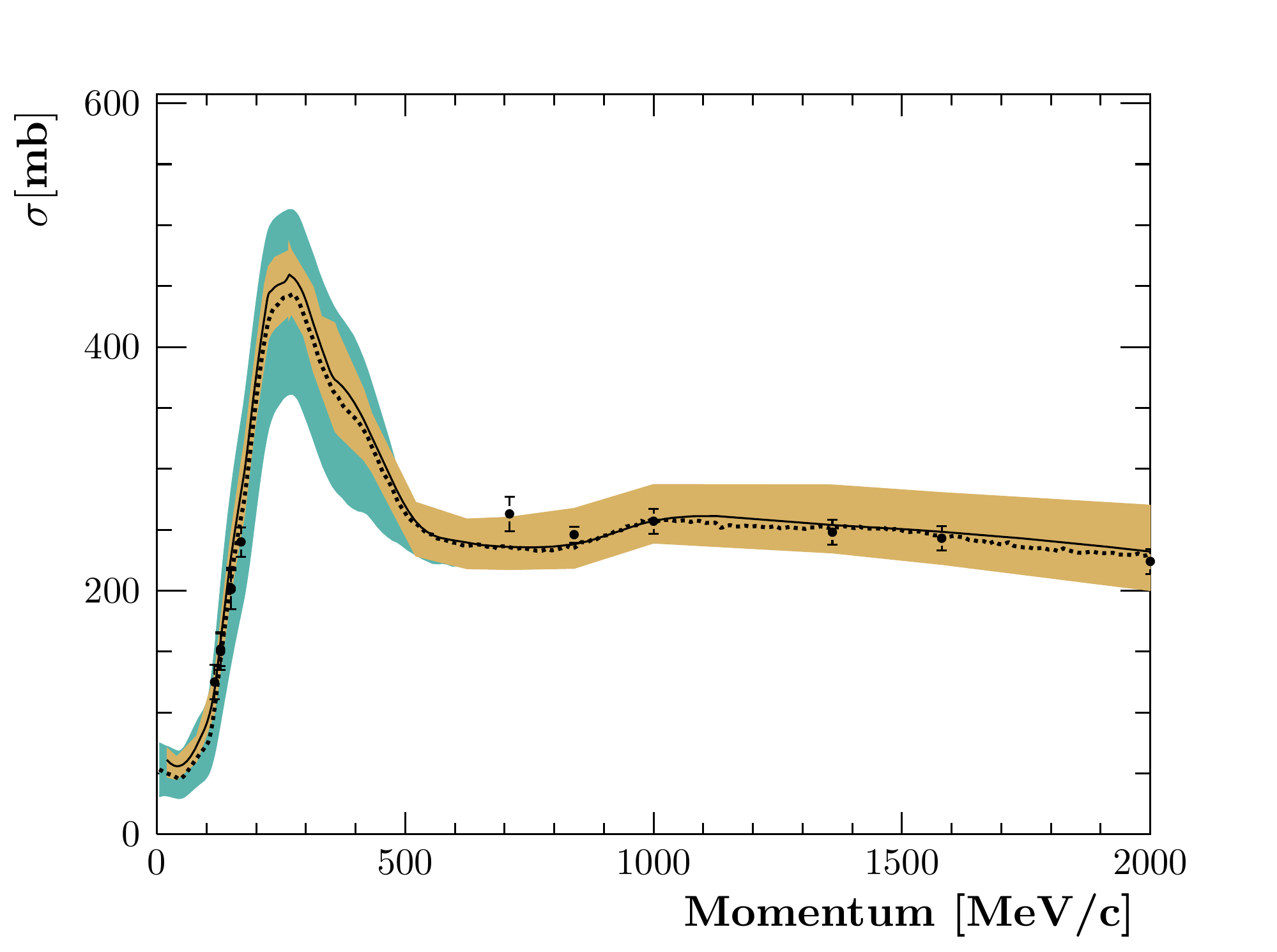}}
  \subfloat[Quasi-elastic]       {\includegraphics[width=0.33\linewidth]{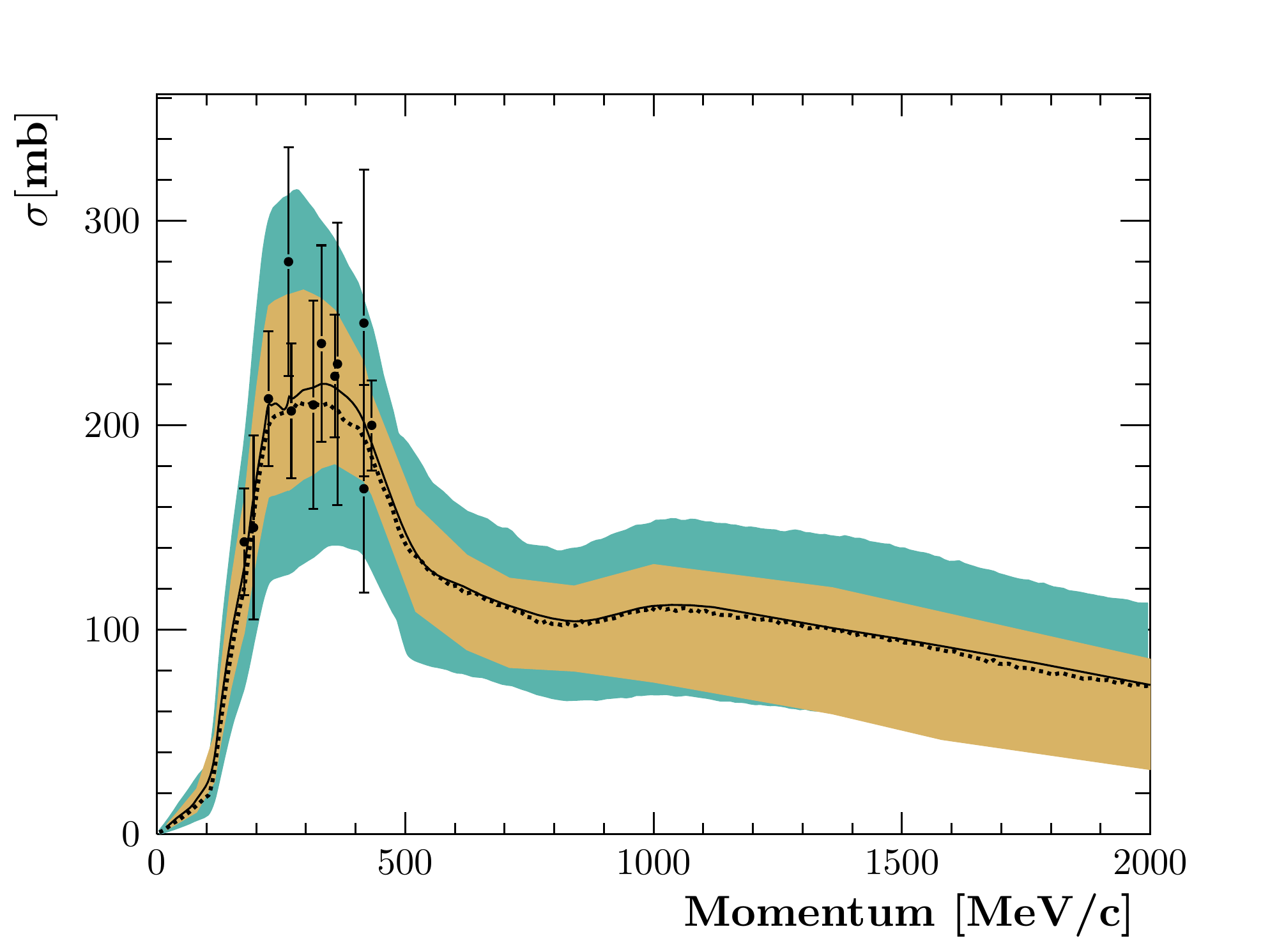}}
  \subfloat[Absorption (ABS)]    {\includegraphics[width=0.33\linewidth]{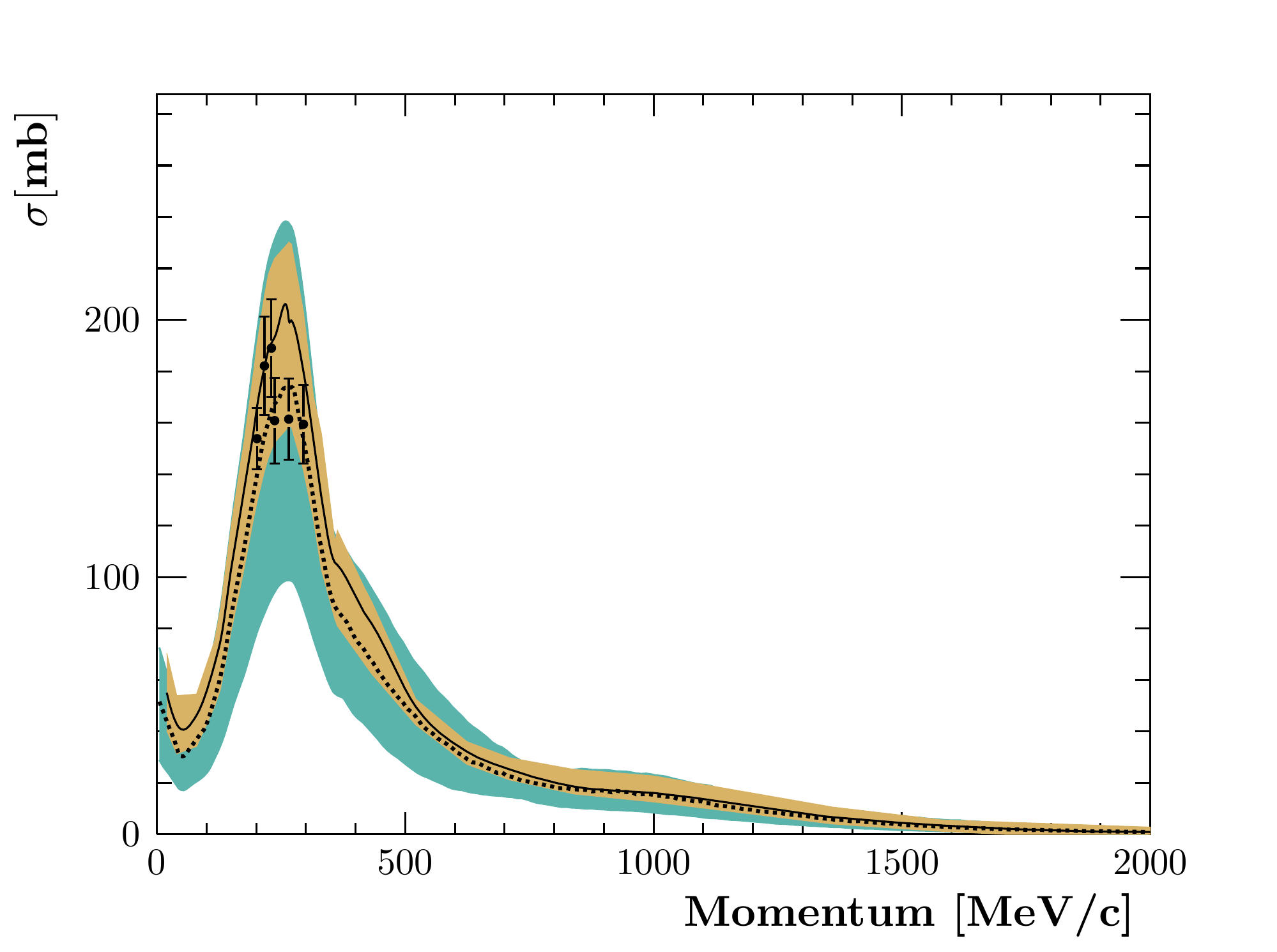}}\\\vspace{-12pt}
  \subfloat[Charge exchange (CX)]{\includegraphics[width=0.33\linewidth]{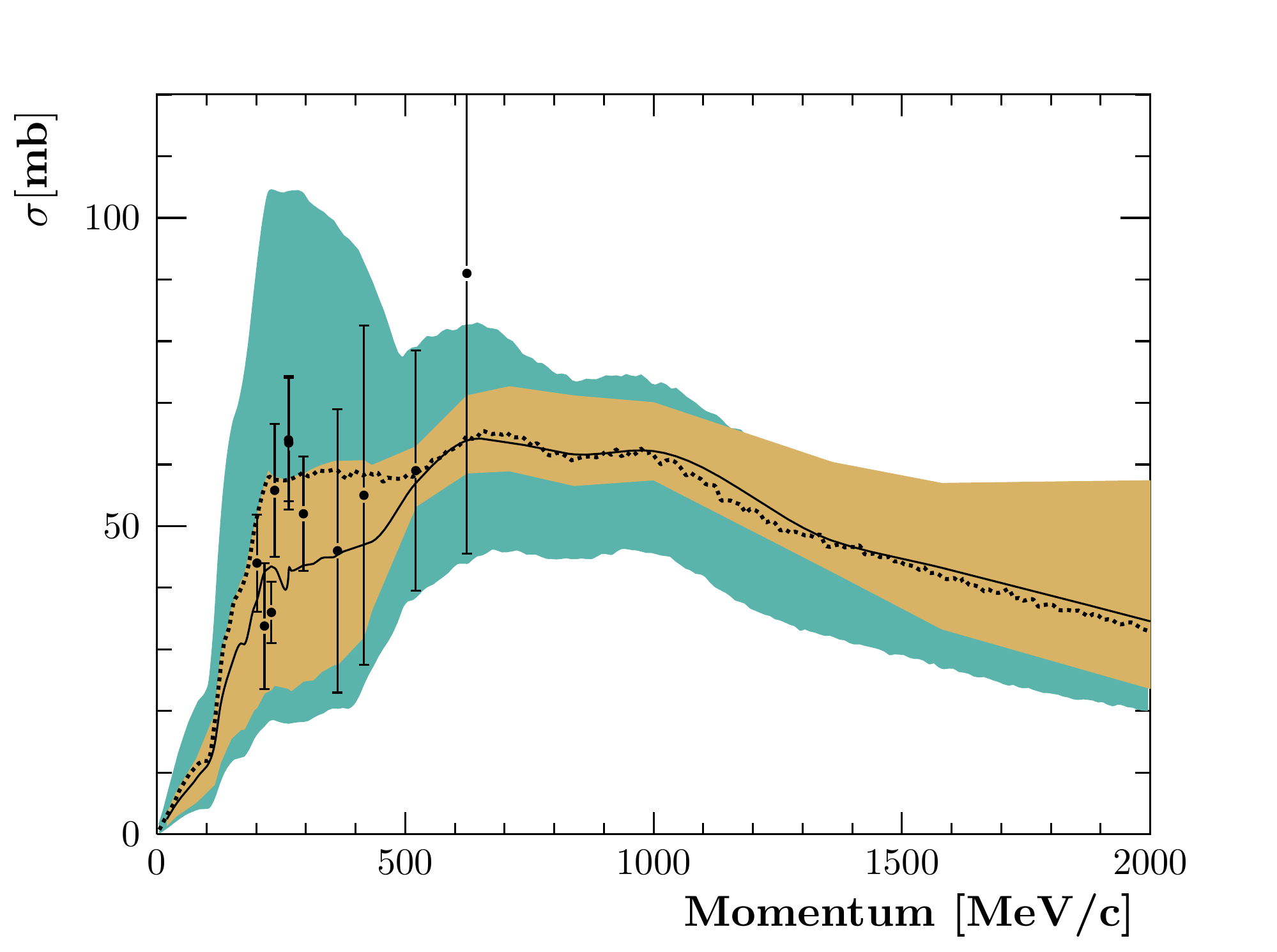}}
  \subfloat[ABS+CX]              {\includegraphics[width=0.33\linewidth]{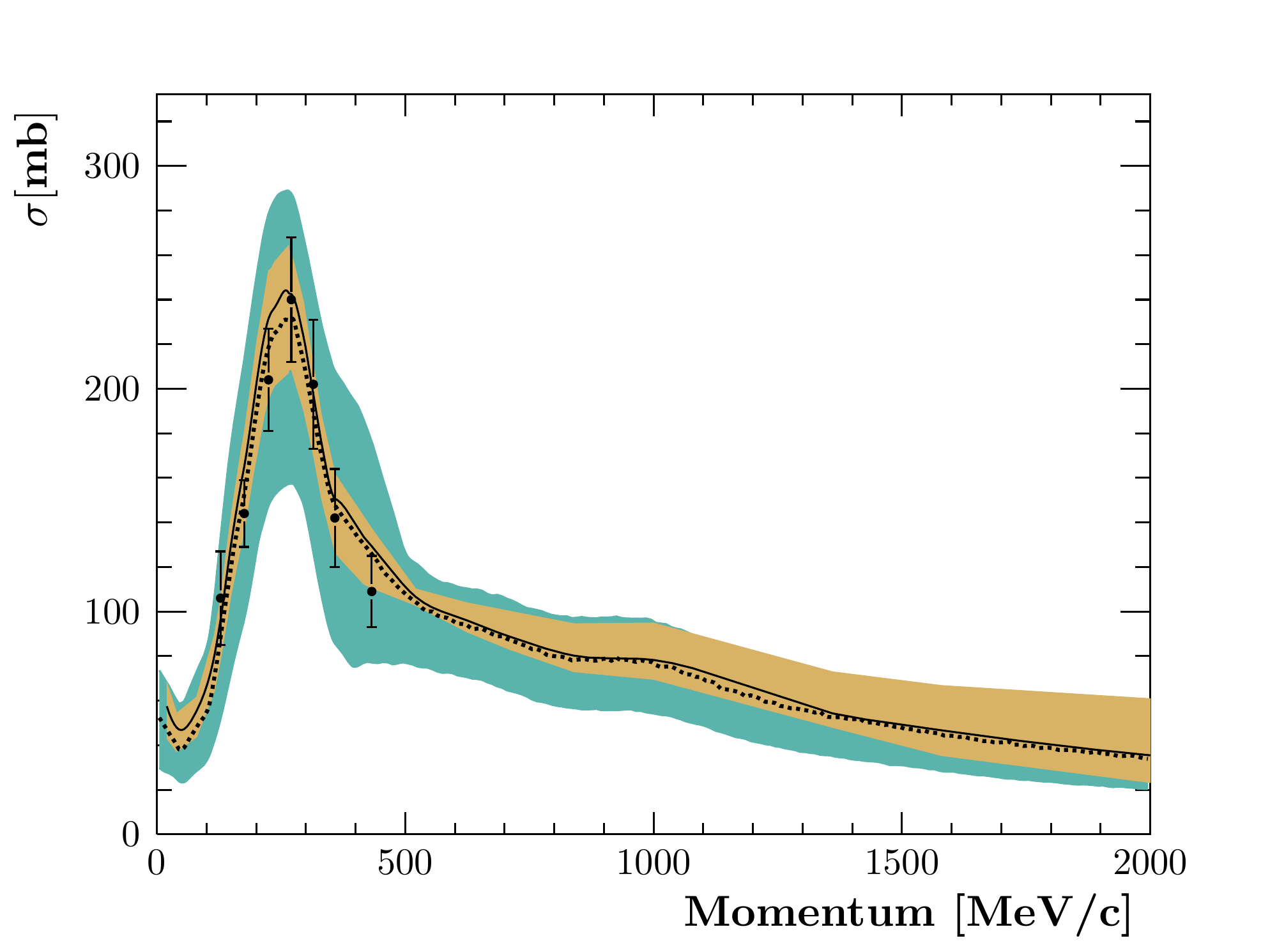}}
  \subfloat                      {\includegraphics[width=0.33\linewidth]{figures/fit_results_pave.pdf}}
\caption{Comparison of the available $\pi^+-^{12}$C cross section external data with the NEUT best fit (solid black line) and $1\sigma$ band after scaling (red), and the best fit (dashed black line) and 1$\sigma$ band (blue) obtained from a previous iteration of this fit~\cite{long_prd, deperio-thesis}.}
\label{fig:fsifitter-c-pip}
\end{figure*}

The final post-fit values of the FSI parameters after applying scaling to the $\chi^2$ surface are shown in Table~\ref{tab:fsifitter-bestfit-scaled}. As the correlation among parameters is not affected by the scaling procedure the correlation matrix presented in Figure~\ref{fig:fsifitter-covariance-octave} remains the same. The scaled uncertainty bands for $\pi^{+}$--$^{12}$C scattering is shown in Figure~\ref{fig:fsifitter-c-pip} and can be compared with the unscaled uncertainty bands shown in Figure~\ref{fig:fsifitter-envelopes-scaled-1}. The results are compared to earlier work~\cite{long_prd, deperio-thesis} dedicated to constraining the FSI parameters to data on T2K. Comparisons of the old and new results with all considered $\pi^{\pm}$--A combinations is found in Ref.~\cite{elder-thesis}. We note that after scaling of the uncertainties, there is good agreement found between the fit results with fixed and varied normalization parameters, shown in Tables~\ref{tab:fsifitter-bestfit-floating} and~\ref{tab:fsifitter-bestfit-scaled}.

\begin{table}[htbp]
\centering
{\renewcommand{\arraystretch}{1.2}
\begin{tabular}{c c}
\hline
\textbf{Parameter} & \textbf{Best fit $\pm$ 1$\sigma$} \\
\hline
$f_{\mathrm{QE}}$  & 1.07 $\pm$ 0.31 \\
$f_{\mathrm{ABS}}$ & 1.40 $\pm$ 0.43 \\
$f_{\mathrm{CX}}$  & 0.70 $\pm$ 0.30 \\
$f_{\mathrm{INEL}}$ & 1.00 $\pm$ 1.10 \\
$f_{\mathrm{QEH}}$  & 1.82 $\pm$ 0.86 \\
\hline
\end{tabular}}
\caption{Post-fit FSI parameters after error scaling}
\label{tab:fsifitter-bestfit-scaled}
\end{table}

\section{Model Comparisons}\label{sec:models}
The modeling of $\pi^\pm$--A interactions is fundamental for a complete description of neutrino-nucleus interactions. All complete neutrino event generators include a model for these Final State Interactions. In this section, a variety of available models are compared to the tuned NEUT model with scaled uncertainties, presented in this work, and the data. The models can be divided into the following three categories:

\begin{enumerate}
\item {\bf Effective Models}
\begin{itemize}

\item \genie~hA is a simple, data-driven, effective model~\cite{genie, intranuke}. Instead of a full cascade model, the total cross section for each scattering process within the nucleus is calculated---effectively reducing the cascade to a single step. The model is normalized to pion/nucleon--iron scattering data and cross sections for targets other than iron are obtained by scaling by A$^{2/3}$~\cite{intranuke}. The hA model is the default in \genie~and has been used by most MINERvA and NOvA analyses published to date~\cite{Aliaga:2013uqz, nova-simulation}.

\item The \genie~hA 2014 model is a development version of the hA model and is included here for completeness. It includes a wider range of $\pi^{\pm}$--A data to reduce the need for A$^{2/3}$ extrapolation~\cite{genie-2015}.
\end{itemize}

\item {\bf Cascade models} all follow the same general principles as the NEUT cascade model described in Section~\ref{sec:neut-summary}. The implementation, input data, and theoretical approximations often differ to reflect the motivations and priorities of the simulation packages.
  
\begin{itemize}
\item \genie~hN is an alternative full cascade model for FSI available in \genie~\cite{genie, intranuke}. Only data on free nucleons is used as an input. The development version of this model, \genie~hN2015, is used for these comparisons. Work is ongoing~\cite{hN_trento} to incorporate the Oset \textit{et al.} model~\cite{salcedo:pionfsi} as is used in NEUT and NuWro.

\item The PEANUT (Pre-Equilibrium Approach to NUclear Thermalization) model is an intra-nuclear cascade model implemented in \fluka~\cite{fluka1,fluka2}. Similar to NEUT, it uses the Oset \textit{et al.} model~\cite{salcedo:pionfsi} to describe the absorptive width of the optical potential for pion momenta below 300 MeV$/c$. $\pi^\pm$--free nucleon cross sections are used to describe elastic, quasi-elastic, and charge exchange interactions.

\item The Bertini cascade model of Geant 4.9.4~\cite{bertini} is part of the QGSP\_BERT physics list. It is valid for pion momenta below 9.9 GeV$/c$. It also handles all other long-live hadrons. A detailed treatment of pre-equilibrium and evaporation physics is included, relevant at momenta below 200 MeV where the cascade model approach begins to fail, as the de Broglie wavelength of the probe is roughly the same as the distance between nucleons in the target nucleus. The CERN-HERA compilation of hadron-nucleus elementary cross section data is used as input~\cite{cern-hera}.

\item The \nuwro~event generator uses cascade model~\cite{NuWro} based on the Oset~\textit{et al.} model~\cite{salcedo:pionfsi}, as in NEUT. It introduces a phenomenological treatment of the formation zone (or time) effect, which can interplay with the pion absorption probability in a non-trivial momentum dependent way.

\end{itemize}

\item {\bf Transport models}

\begin{itemize}
\item The Giessen Boltzmann-Uehling-Uhlenbeck (GiBUU) model is an implementation of transport model for nuclear reactions~\cite{Buss20121}. It describes the dynamical evolution of the interacting nuclear system through a coupled set of semi-classical kinetic equations, while taking into account the hadronic potentials and the equation of state of nuclear matter within the Boltzmann-Uehling-Uhlenbeck (BUU) theory.
\end{itemize}
\end{enumerate}

\begin{figure*}[htbp]
  \centering
  \subfloat[Reactive]            {\includegraphics[width=0.33\linewidth]{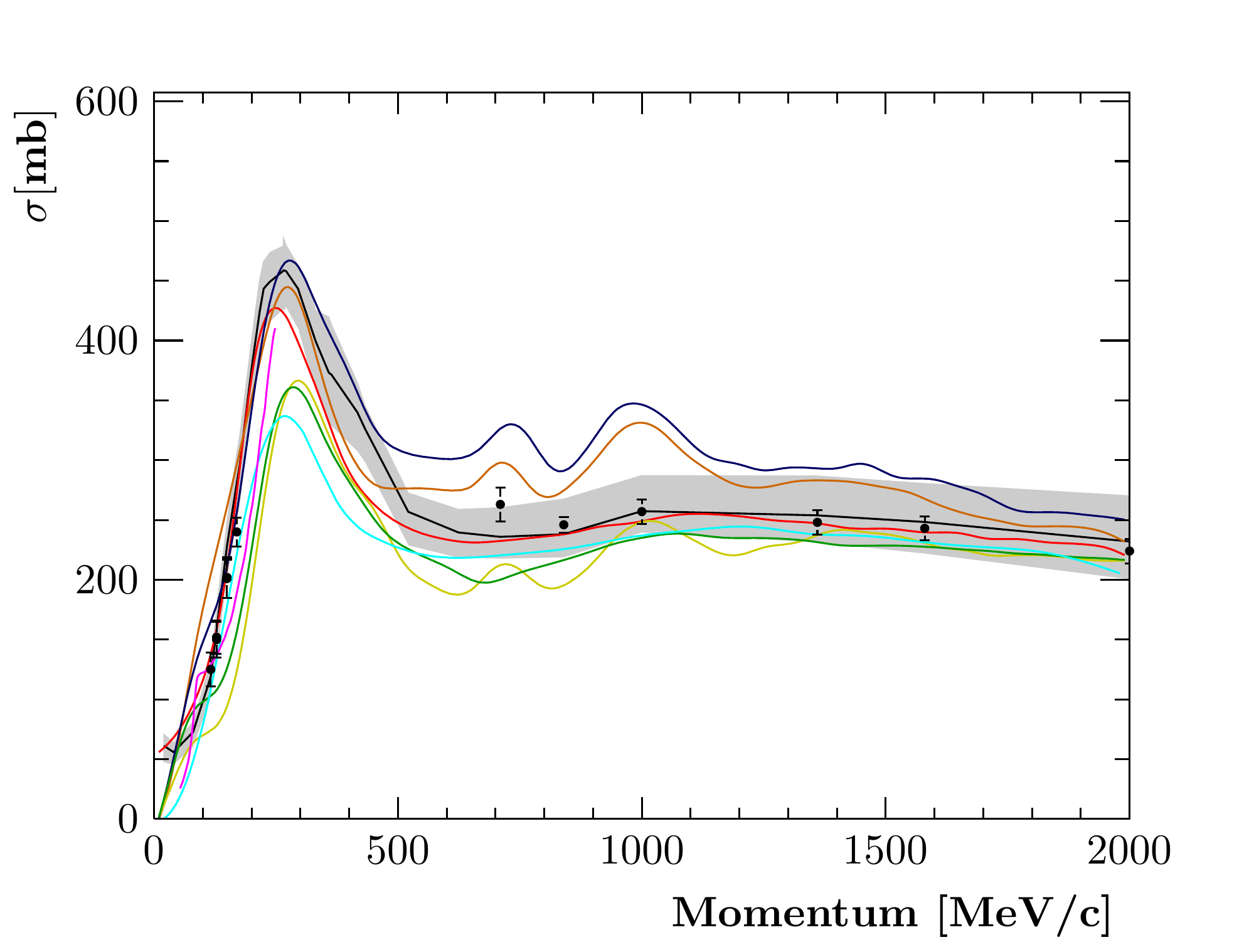}}
  \subfloat[Quasi-elastic]       {\includegraphics[width=0.33\linewidth]{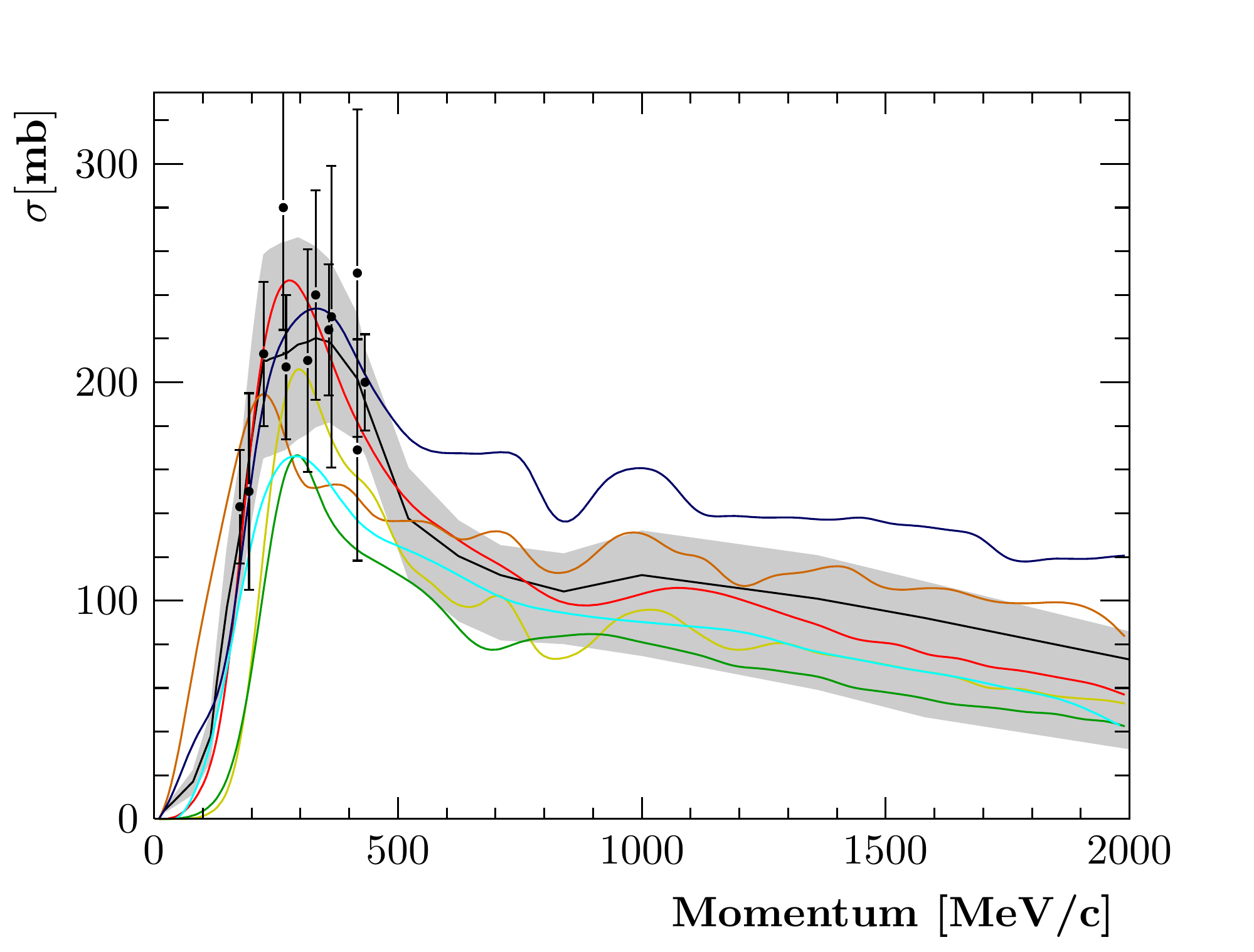}}
  \subfloat[Absorption (ABS)]    {\includegraphics[width=0.33\linewidth]{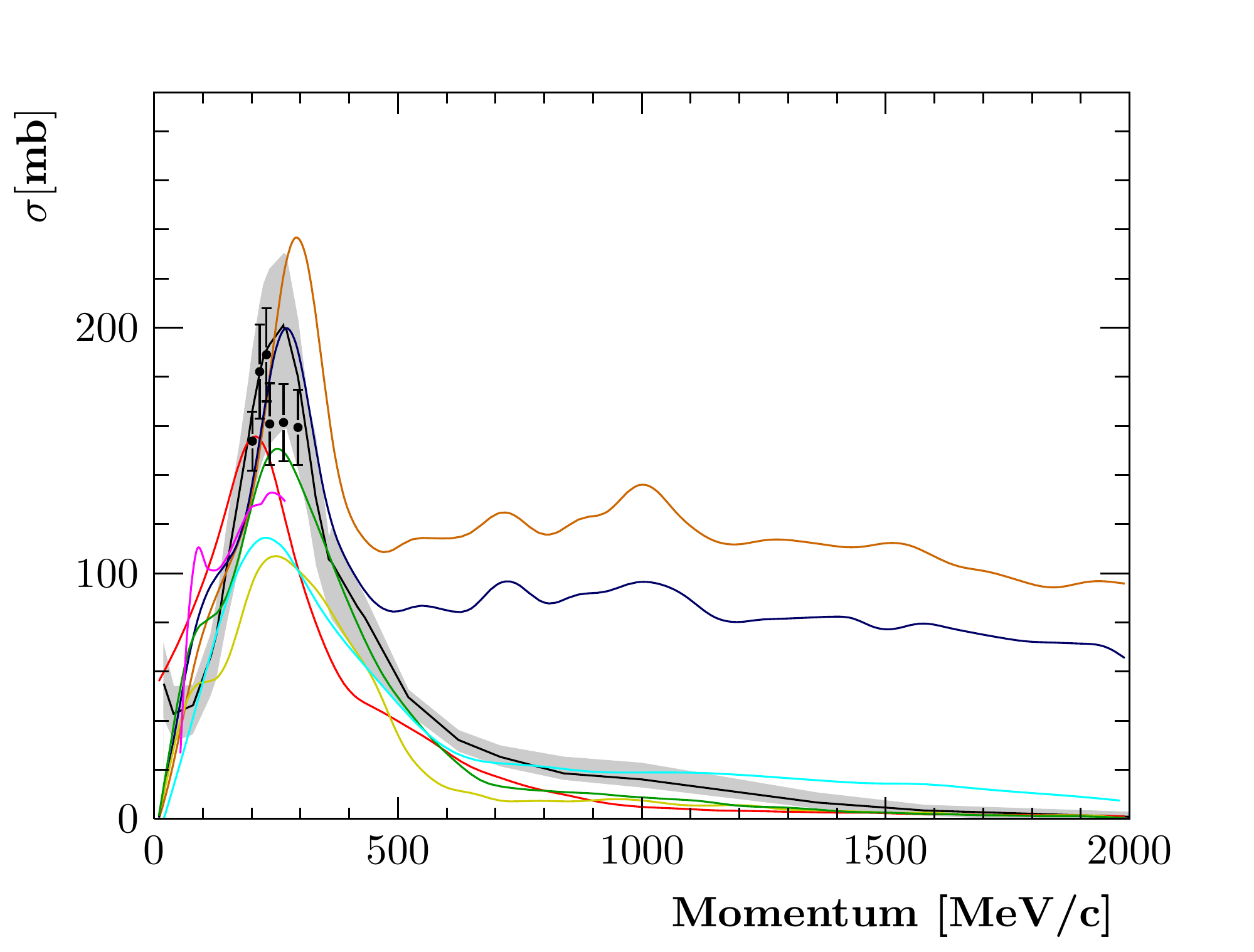}}\\\vspace{-12pt}
  \subfloat[Charge exchange (CX)]{\includegraphics[width=0.33\linewidth]{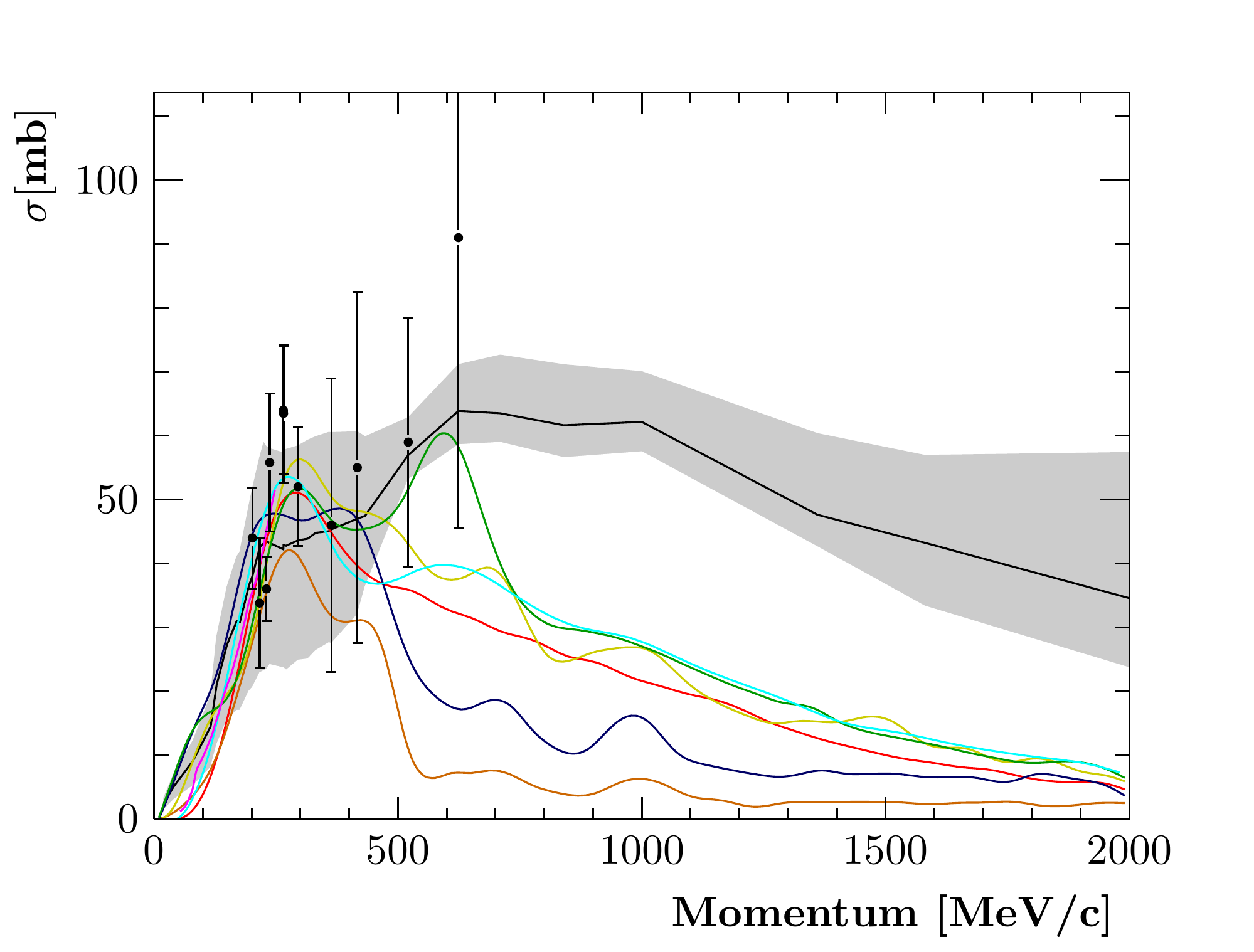}}
  \subfloat[ABS+CX]              {\includegraphics[width=0.33\linewidth]{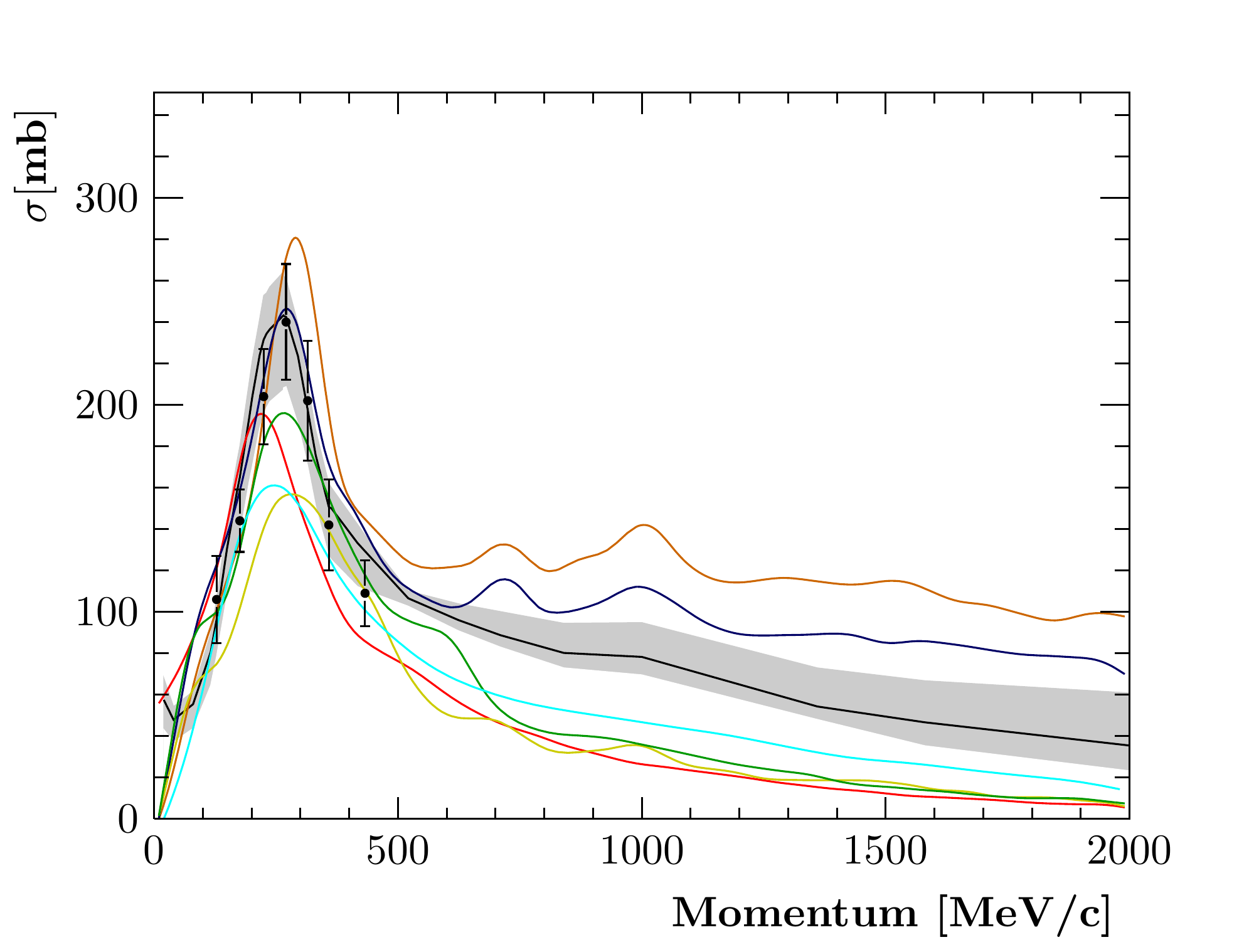}}
  \subfloat                      {\includegraphics[width=0.33\linewidth]{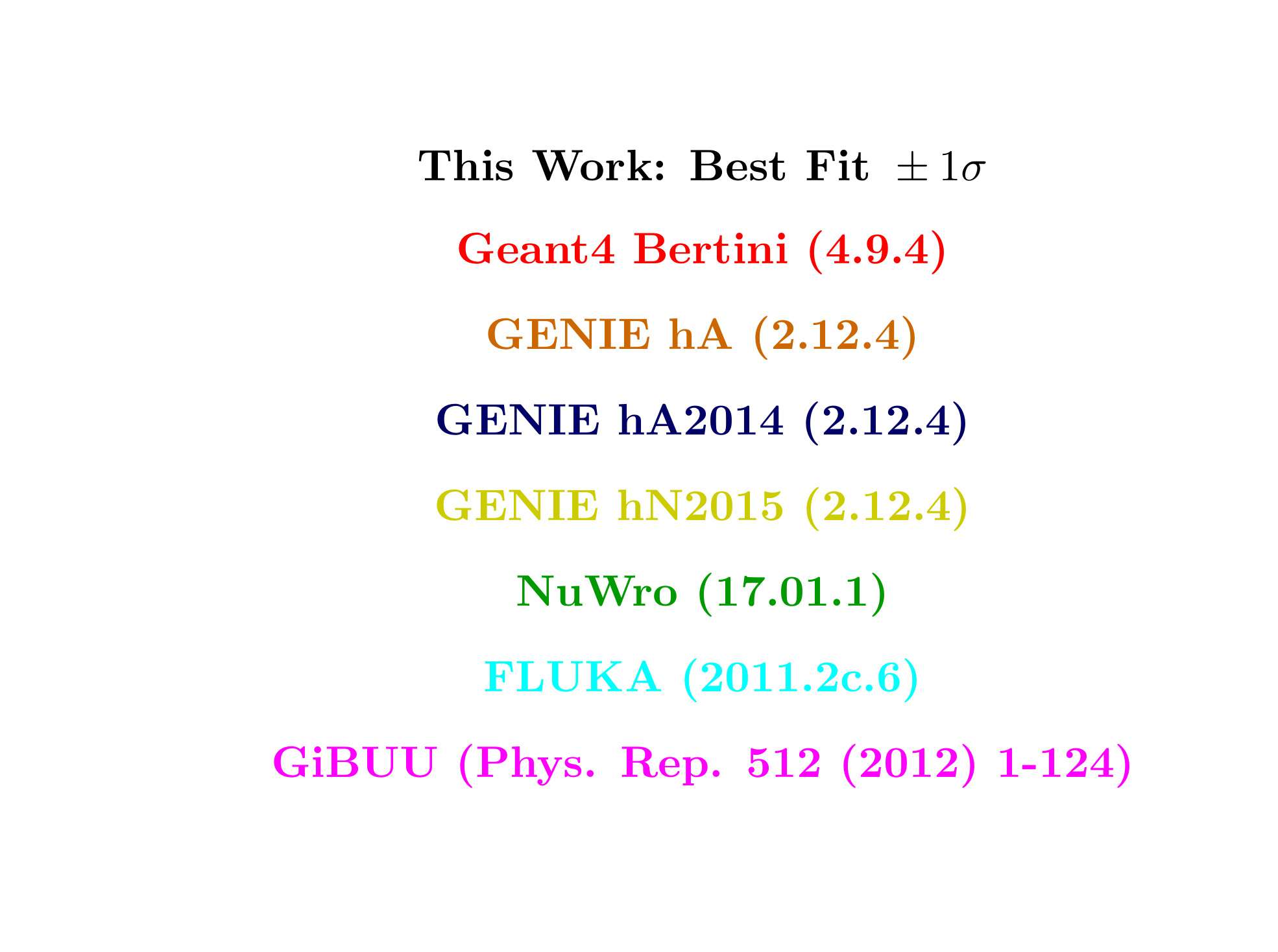}}
\caption{Comparison of the available $\pi^+$--$^{12}$C cross section external data with the NEUT best fit and its $1\sigma$ error band obtained in this work, and other models.}
\label{fig:models-c-pip}
\end{figure*}
\begin{figure*}[htbp]
  \centering
  \subfloat[Reactive]            {\includegraphics[width=0.33\linewidth]{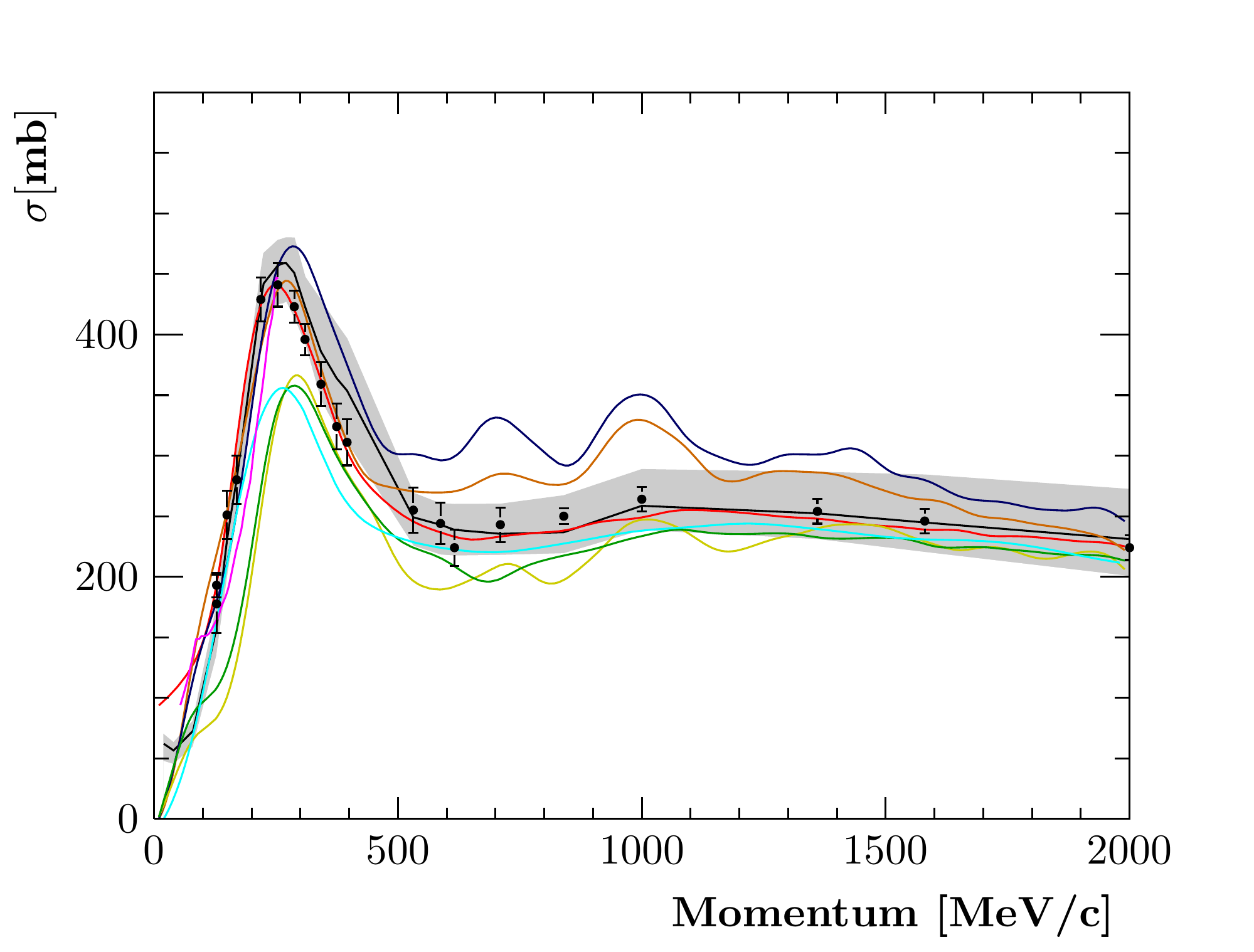}}
  \subfloat[Quasi-elastic]       {\includegraphics[width=0.33\linewidth]{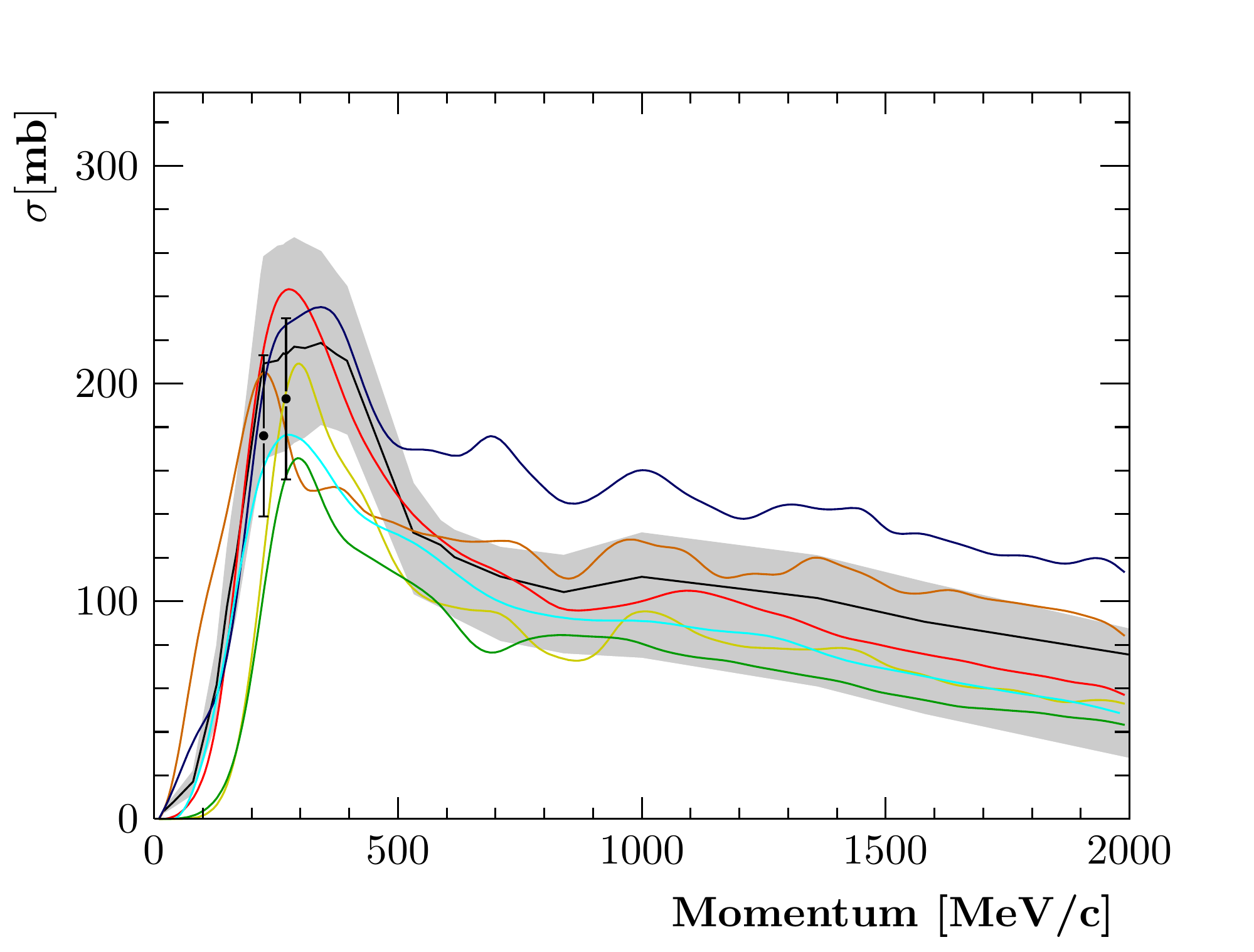}}
  \subfloat[Absorption (ABS)]    {\includegraphics[width=0.33\linewidth]{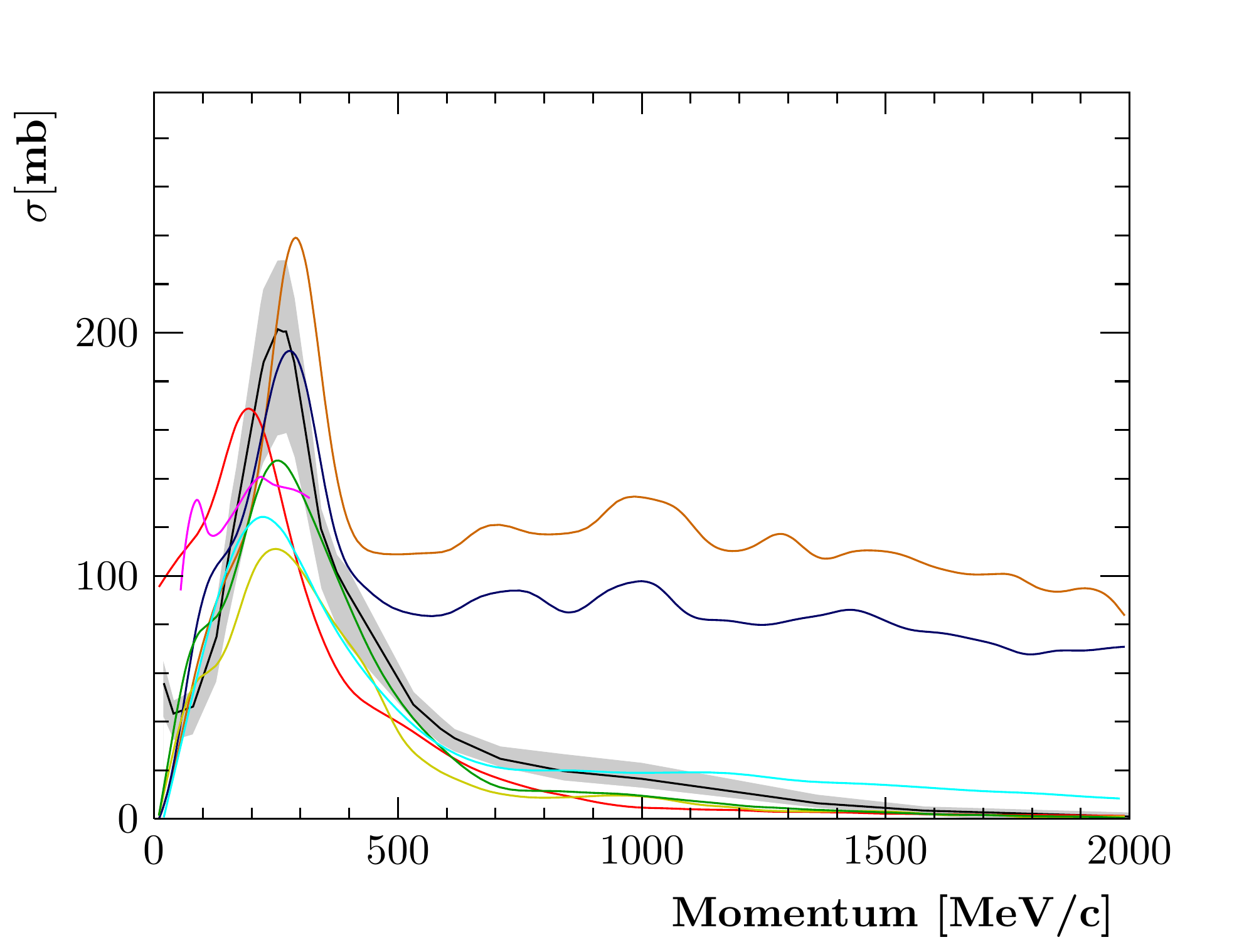}}\\\vspace{-12pt}
  \subfloat[Charge exchange (CX)]{\includegraphics[width=0.33\linewidth]{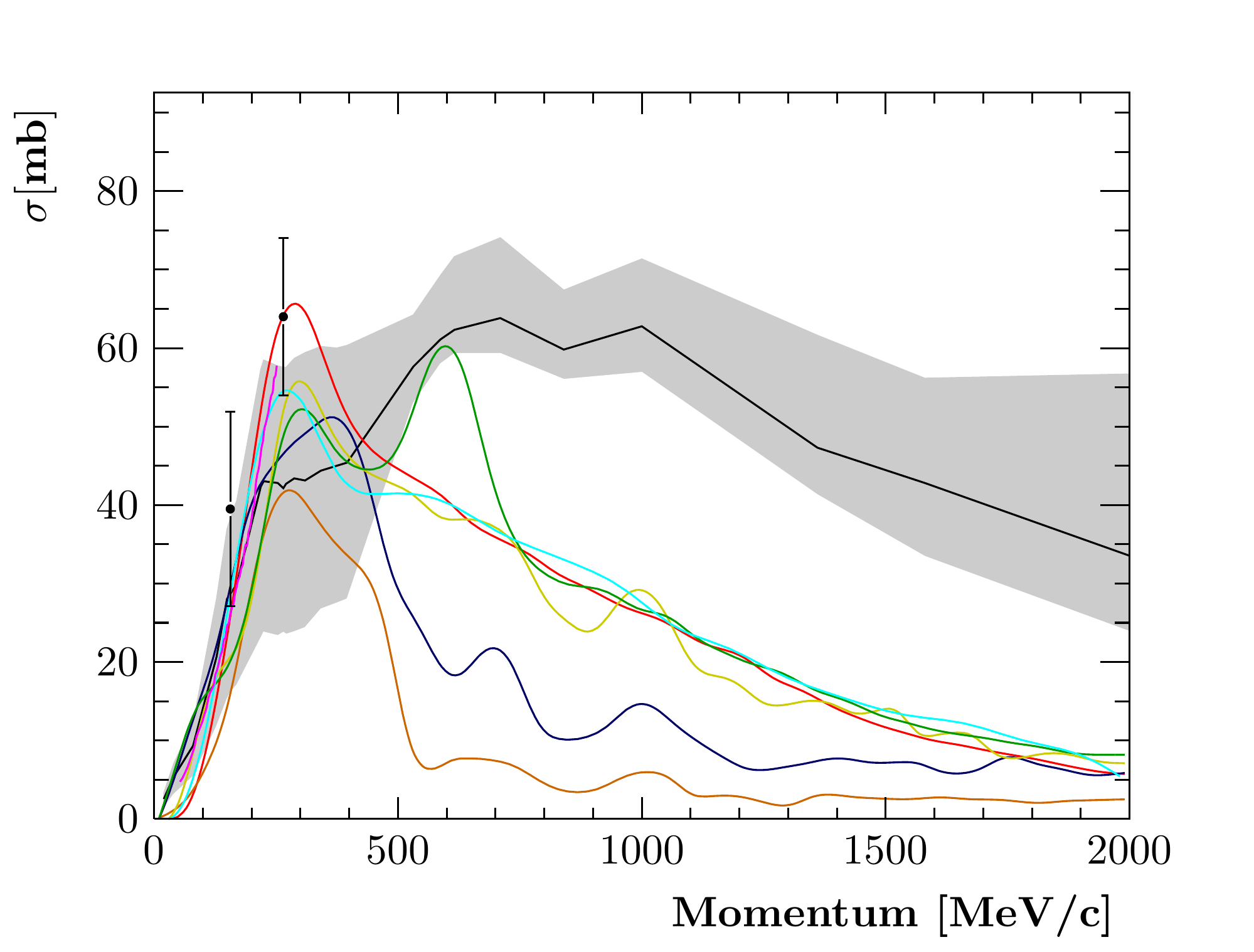}}
  \subfloat[ABS+CX]              {\includegraphics[width=0.33\linewidth]{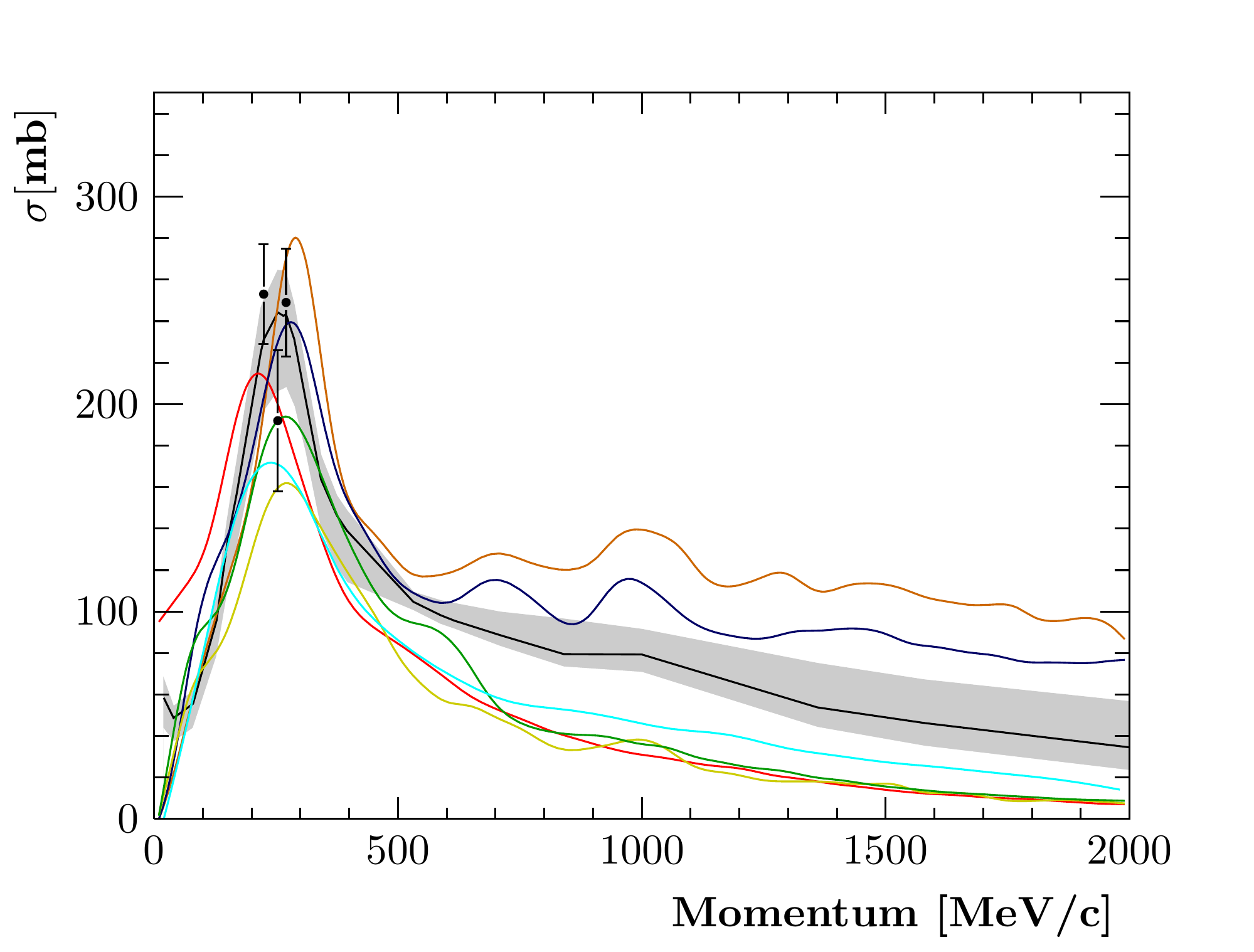}}
  \subfloat                      {\includegraphics[width=0.33\linewidth]{figures/pave.pdf}}
\caption{Comparison of the available $\pi^-$--$^{12}$C cross section external data with the NEUT best fit and its $1\sigma$ error band obtained in this work, and other models.}
\label{fig:models-c-pim}
\end{figure*}

Thin target particle gun simulations were used to produce microscopic cross sections for each of the models described above, and used to compare to data here, with the exception of GiBUU, where the the predictions for REAC and ABS of $\pi^{\pm}-^{12}$C and $^{63}$Cu were taken directly from Ref.~\cite{Buss20121}. To allow for a consistent comparison, the interactions channels were defined using only the final state particles, as described in Section~\ref{sec:neut-summary}.

Comparisons between the tuned NEUT results obtained in Section~\ref{sec:fit-results}, the data used in the tuning (summarized in Table~\ref{tab:piscatdat}), and the alternative models described above, are shown for carbon in Figures~\ref{fig:models-c-pip} and~\ref{fig:models-c-pim}. Similar plots for oxygen, aluminium, iron, copper and lead are found in Appendix~\ref{sec:more_model_comparisons}, including ratios of the models to data.

The NEUT prediction with scaled error bands covers $\sim$2/3 of the available data points as intended. In the energy regions where there is data, most of the models are in reasonable agreement. However, there are noticeable differences between the models outside the data range. Most notably the NEUT model predicts a much larger cross section for the single charge exchange channel at higher momenta. Given that the agreement is recovered in the reactive channel, this is likely to be caused by differences in the tagging of hadron production events (\pipm-A$\rightarrow\pi^+\pi^-X$ where the $\pi^+$ is absorbed). The effective GENIE models also predict a much larger $\pi^{\pm}$ absorption cross section at higher momenta. Additionally, most of the models fail to properly reproduce the data for the absorption cross section at low momenta for heavy nuclei (see Appendix~\ref{sec:more_model_comparisons}). One possible explanation is the lack of a model for the Coulomb attraction felt by the negatively charge pion, which is included for the GiBUU model, the only model which obtains good agreement with the data in this region.

\section{Summary and Outlook}
\label{sec:outlook}
The cascade model used to simulate pion interactions in NEUT has been tuned to a variety of available data on several nuclear targets, and uncertainties have been produced. These have recently been used in analyses by the T2K and Super-Kamiokande collaborations, for whom NEUT is the primary neutrino simulation software. The uncertainties produced in this work are reduced with respect to previous attempts to constrain the pion interaction parameters~\cite{long_prd, deperio-thesis} by $\sim$50\%, which should correspond to a similar reduction in the FSI and SI uncertainties, currently a large uncertainty in the T2K oscillation analysis~\cite{long_prd, Abe:2017vif}. Additionally, the methods developed can be applied for tuning other similar cascade simulation packages.

One of the additions to this fit over previous attempts is the new DUET data~\cite{duet-asbcx:piscat, duet:piscat}. DUET made measurements specifically targeted at T2K's needs and provided more information than available from most historical measurements from the 1950's--1990's. To reduce the uncertainties further for future experiments more modern data is needed. Fortunately, there are dedicated experiments~\cite{Cavanna:2014iqa, Abi:2017aow} in charged particle test beams aiming to make such measurements on argon to fulfill this need for the future DUNE oscillation program~\cite{Acciarri:2015uup}.

One limitation of this work is neglecting the measured pion kinematic distributions, for which only very limited data is available~\cite{levenson:piscat, fujii:piscat}. NEUT comparisons to these distributions can be found in Ref.~\cite{deperio-thesis}. Although this would require significant development of the NEUT cascade model, having such data from new $\pi^{\pm}$--A scattering experiments would provide a stringent constraint on models in the future.

\begin{acknowledgments}
  The authors would like to thank the members of the
  T2K collaboration and the authors of the NEUT generator
  for their help and support.
  We acknowledge the support of MEXT, Japan;
  NSERC (Grant No. SAPPJ-2014-00031), NRC and CFI, Canada;
  CEA and CNRS/IN2P3, France;
  SNSF and SERI, Switzerland;
  STFC, UK; and
  DOE, USA.
  In addition, participation of individual researchers and institutions has been further
  supported by funds from the Ontario Graduate Scholarship, the Queen Elizabeth II Graduate Scholarship in Science \& Technology, the Japan Society for the Promotion of Science (JSPS) Postdoctoral Fellowship for Research in Japan, the H2020 Grant
  No.  RISE-GA644294-JENNIFER, the Alfred P. Sloan Foundation and the DOE Early Career program, USA.
  Computations were performed on the GPC supercomputer at the SciNet HPC Consortium \cite{scinet}.
  SciNet is funded by: the Canada Foundation for Innovation under the auspices of Compute Canada; 
  the Government of Ontario; Ontario Research Fund --- Research Excellence; and the University of Toronto.
\end{acknowledgments}

\bibliography{bibliography.bib}

\appendix
\section{Additional model comparisons}
\label{sec:more_model_comparisons}

\begin{figure*}[htbp]
  \centering
  \subfloat[Reactive]            {\includegraphics[width=0.33\linewidth]{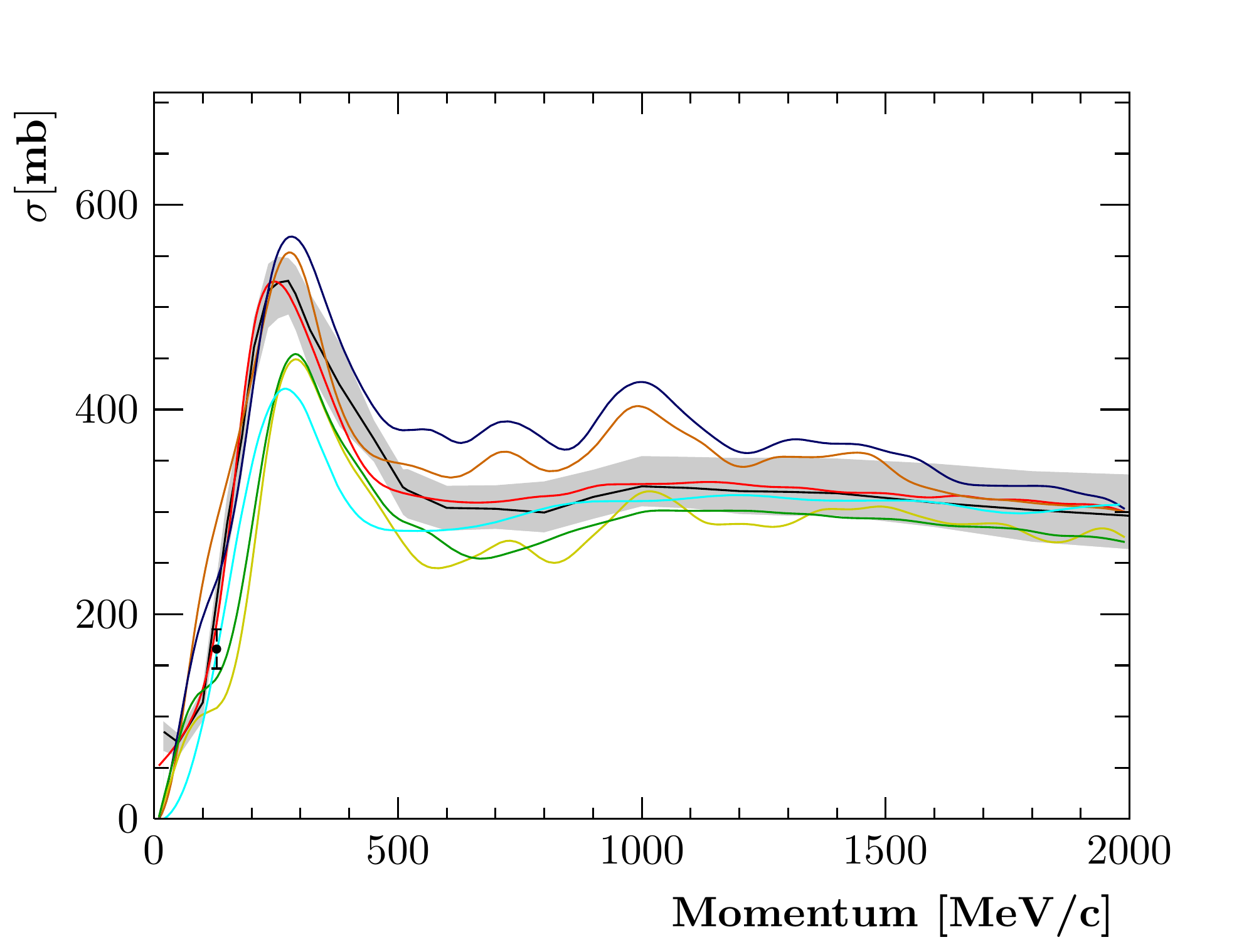}}
  \subfloat[Quasi-elastic]       {\includegraphics[width=0.33\linewidth]{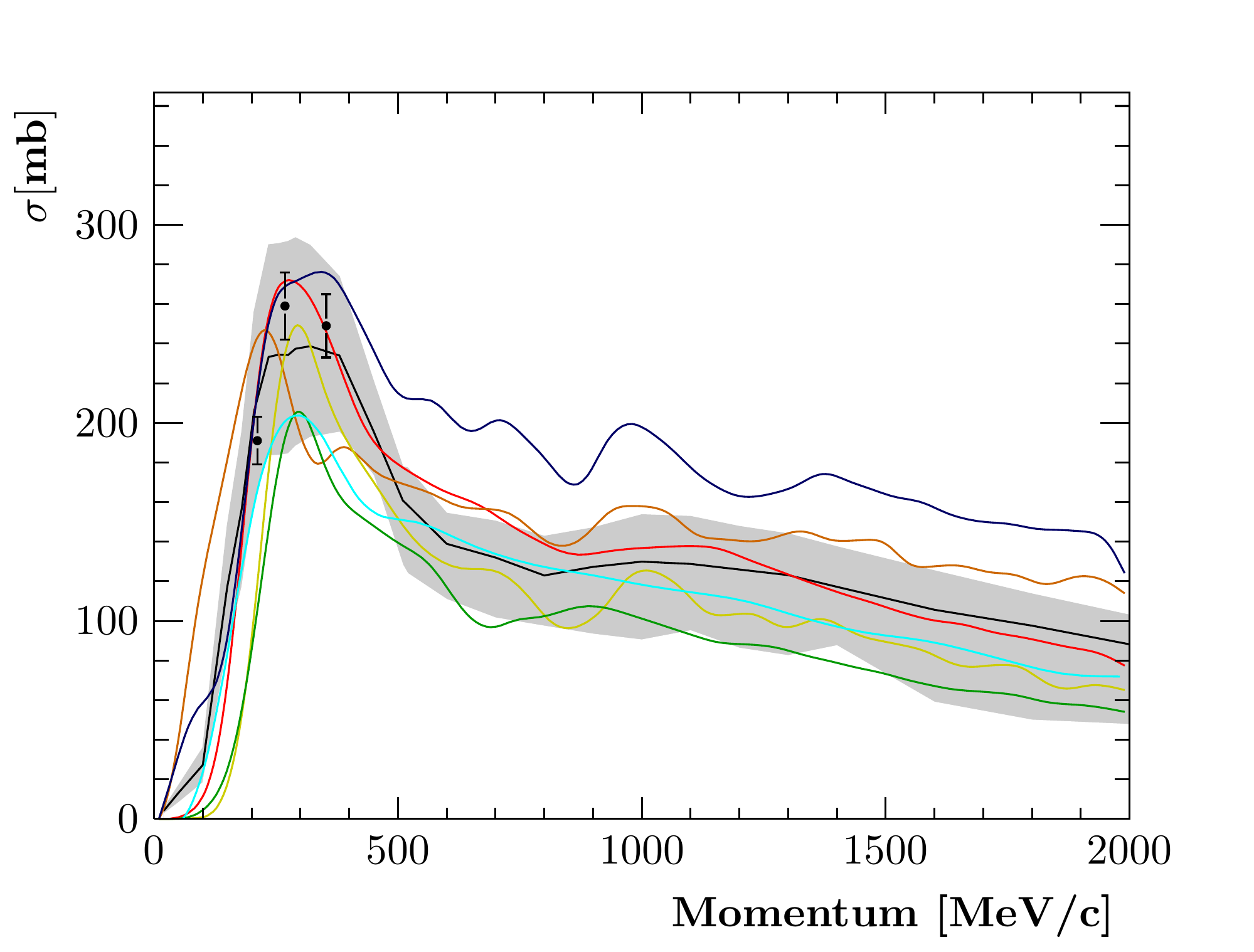}}
  \subfloat[Absorption (ABS)]    {\includegraphics[width=0.33\linewidth]{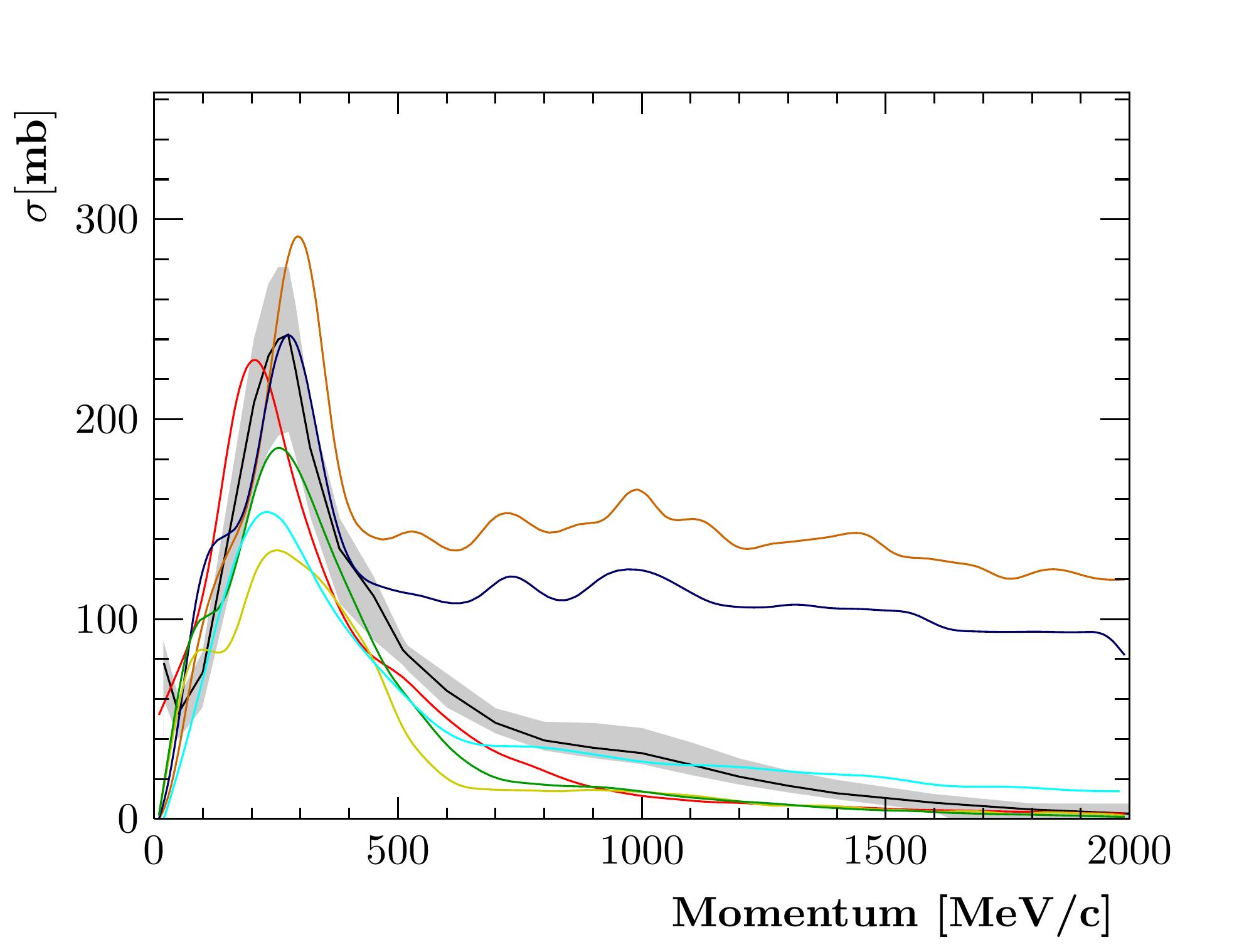}}\\\vspace{-12pt}
  \subfloat[Charge exchange (CX)]{\includegraphics[width=0.33\linewidth]{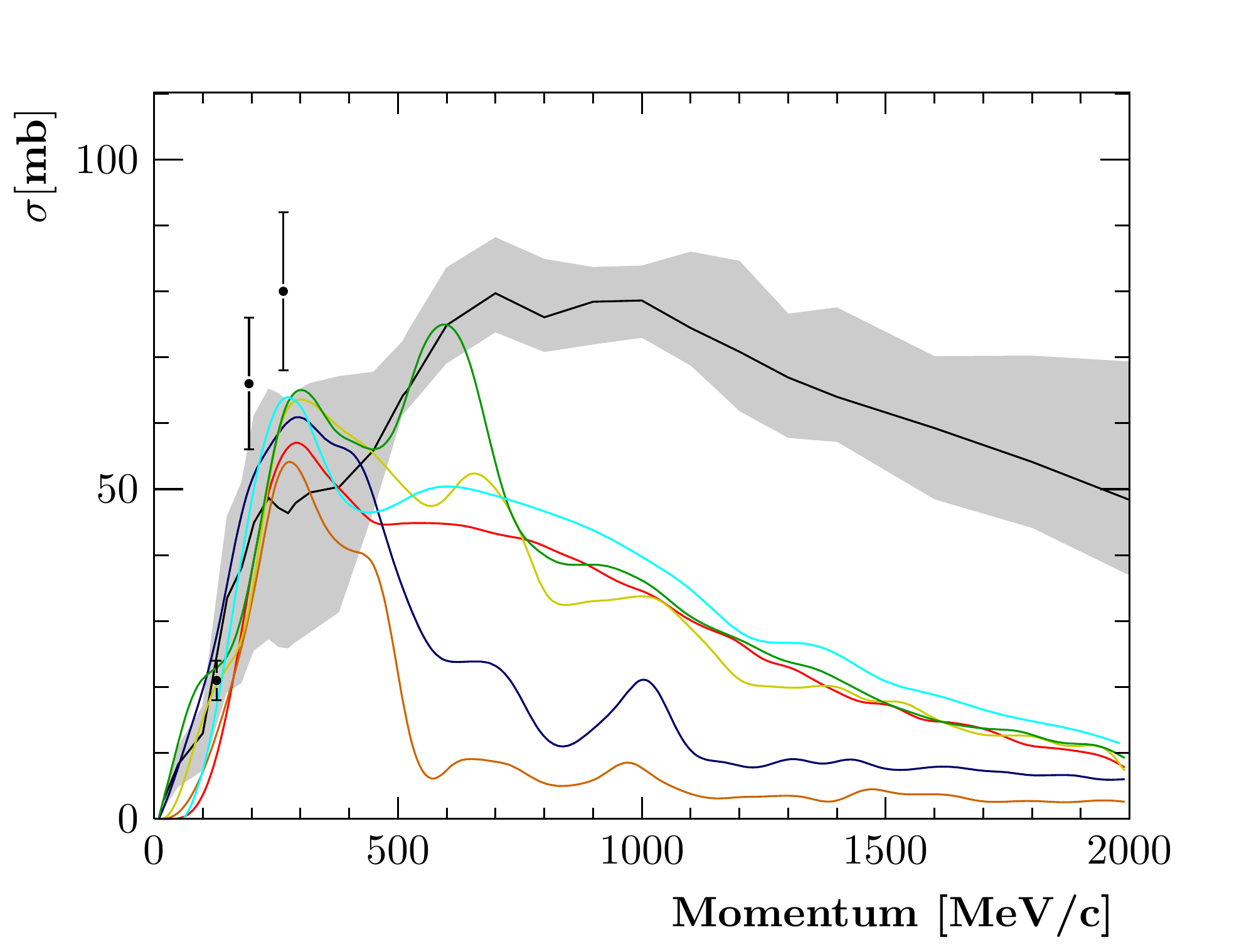}}
  \subfloat[ABS+CX]              {\includegraphics[width=0.33\linewidth]{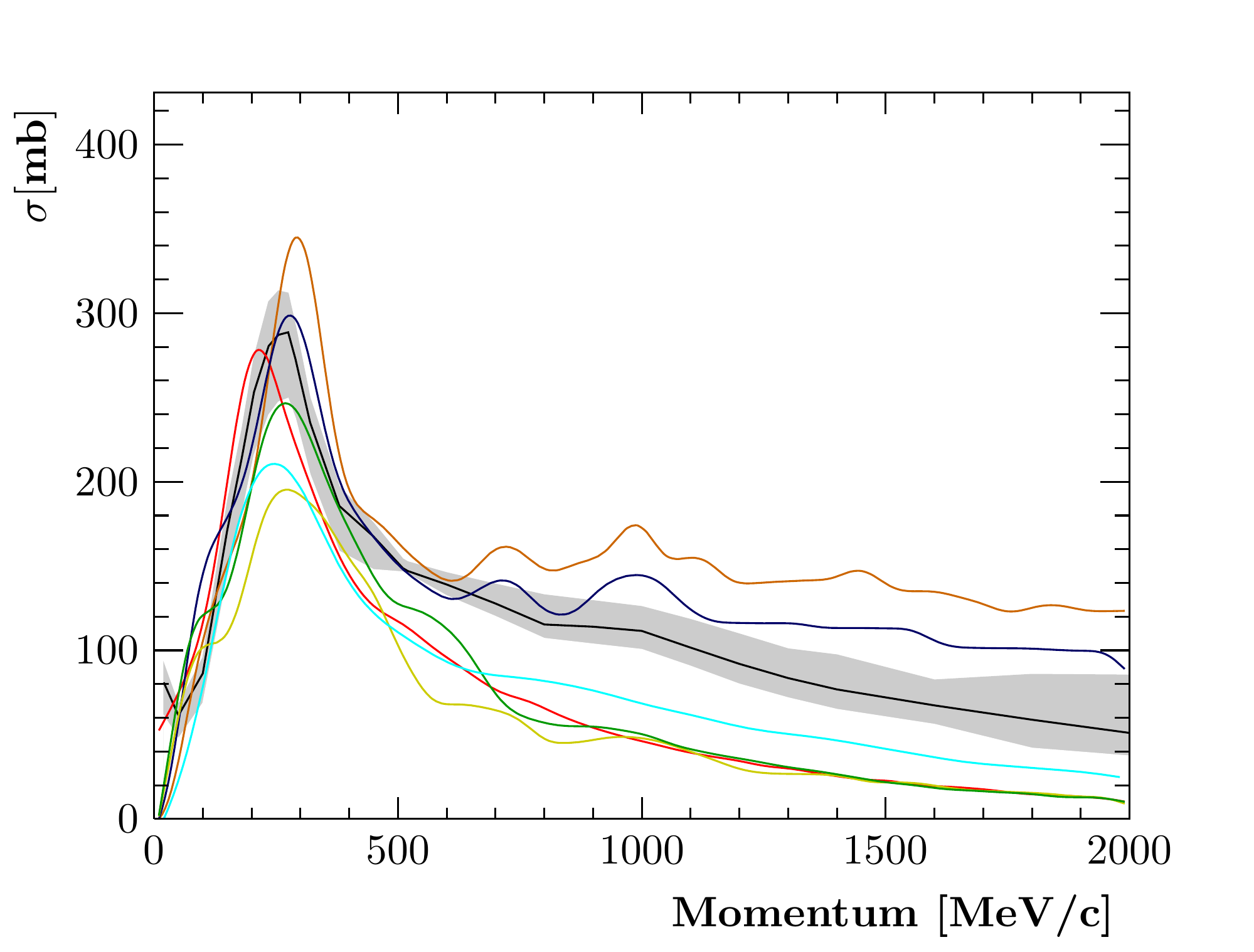}}
  \subfloat                      {\includegraphics[width=0.33\linewidth]{figures/pave.pdf}}
\caption{Comparison of the available $\pi^+$--$^{16}$O cross section external data with the NEUT best fit and its $1\sigma$ error band obtained in this work, and other models.}
\label{fig:models-o-pip}
\end{figure*}

\begin{figure*}[htbp]
  \centering
  \subfloat[Reactive]            {\includegraphics[width=0.33\linewidth]{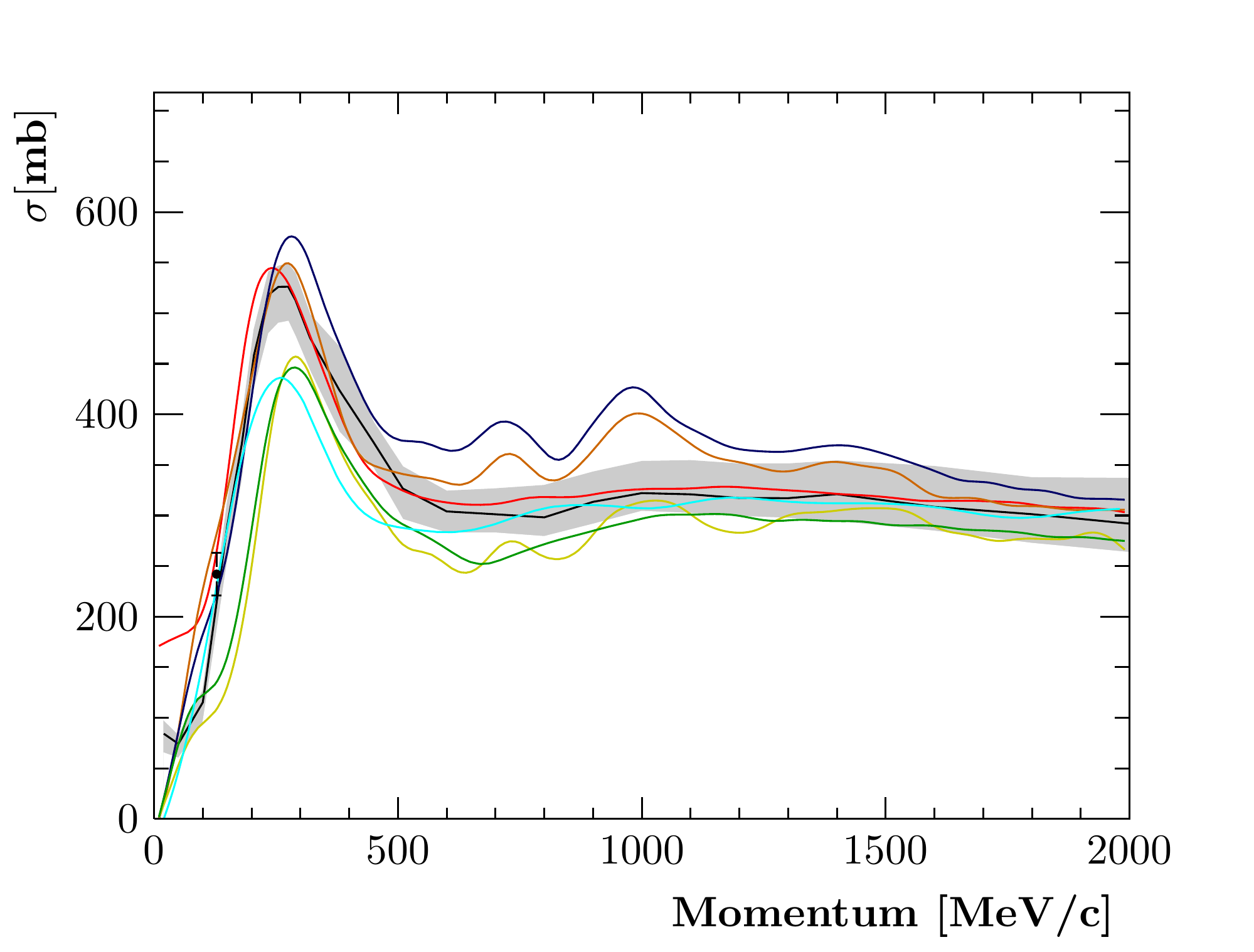}}
  \subfloat[Quasi-elastic]       {\includegraphics[width=0.33\linewidth]{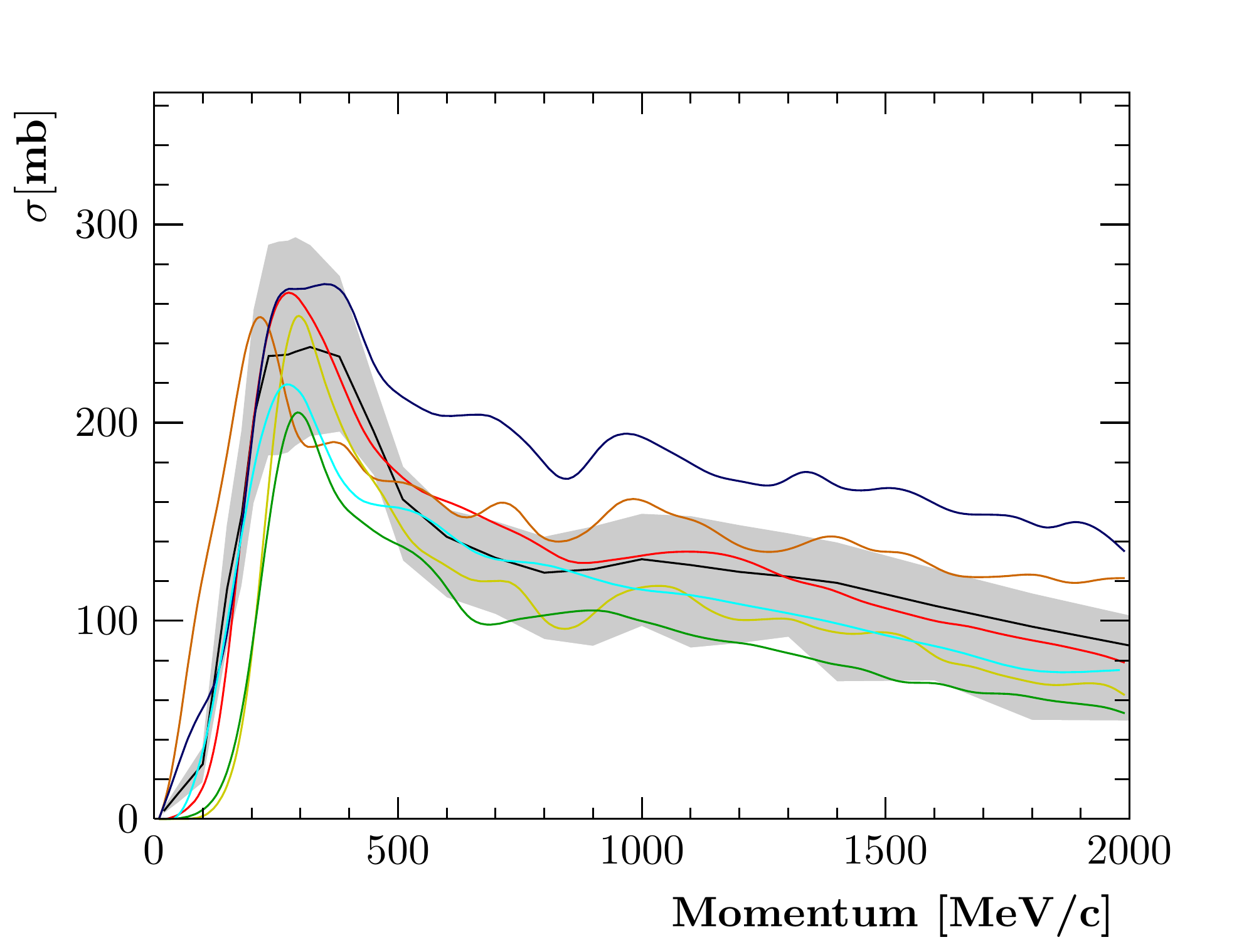}}
  \subfloat[Absorption (ABS)]    {\includegraphics[width=0.33\linewidth]{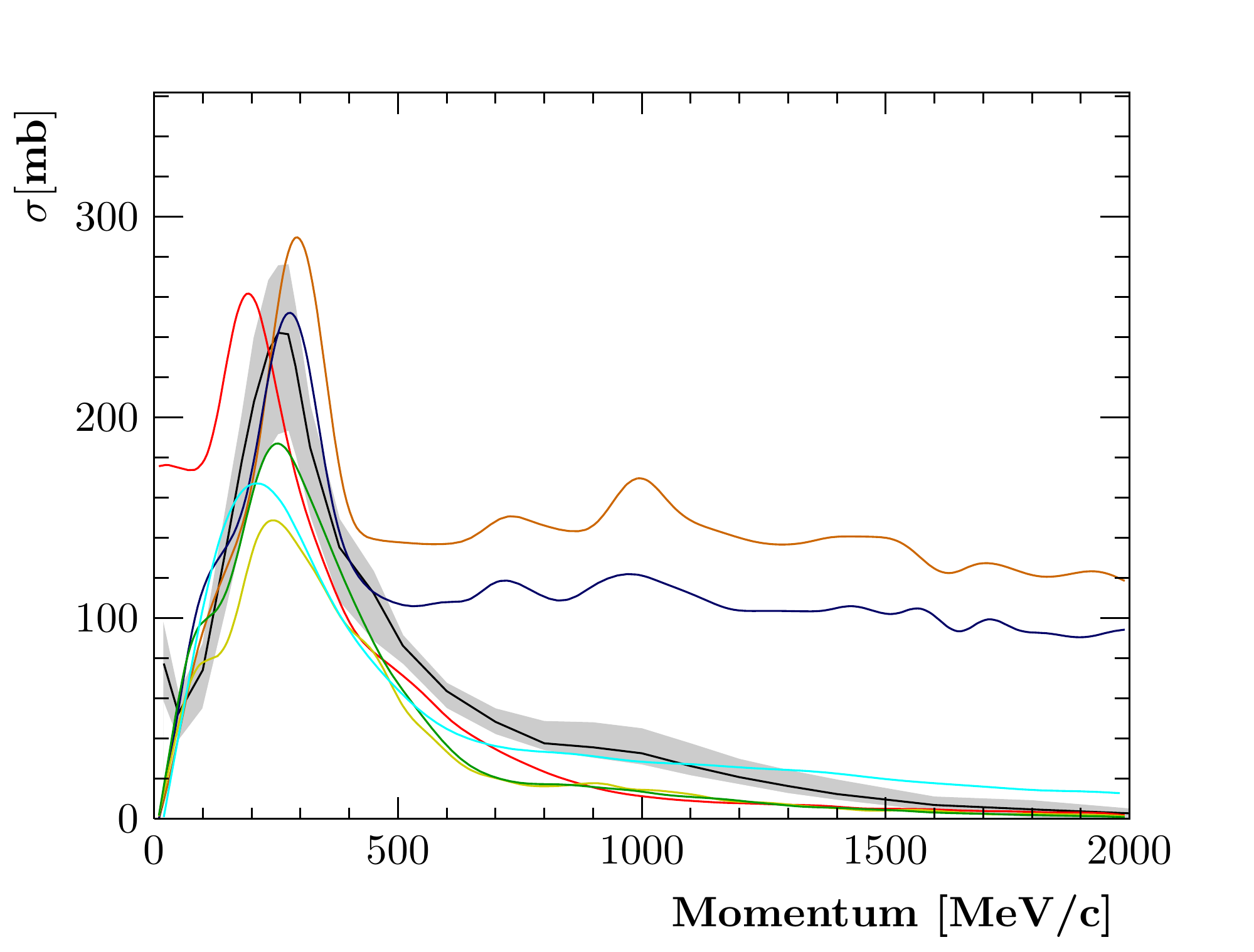}}\\\vspace{-12pt}
  \subfloat[Charge exchange (CX)]{\includegraphics[width=0.33\linewidth]{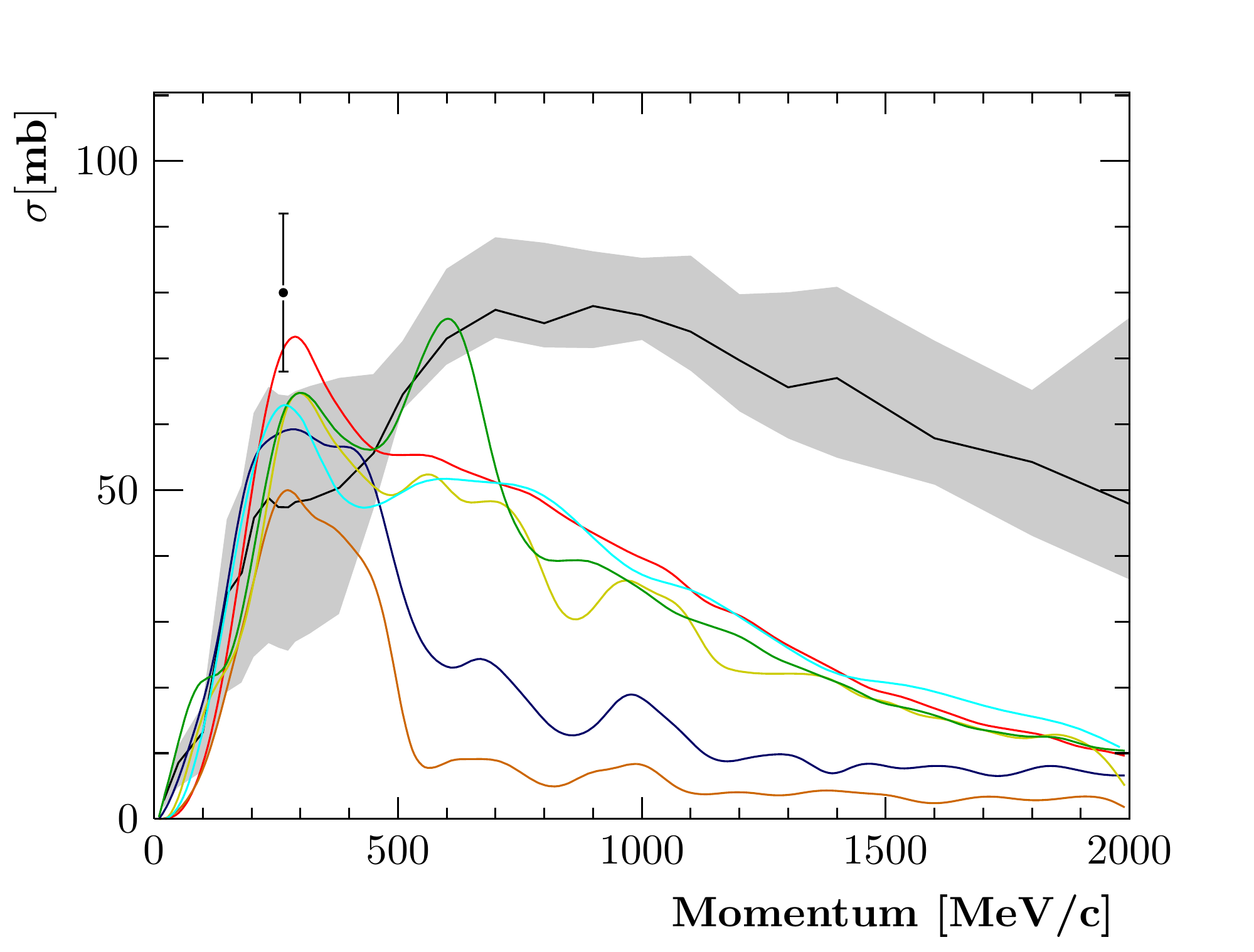}}
  \subfloat[ABS+CX]              {\includegraphics[width=0.33\linewidth]{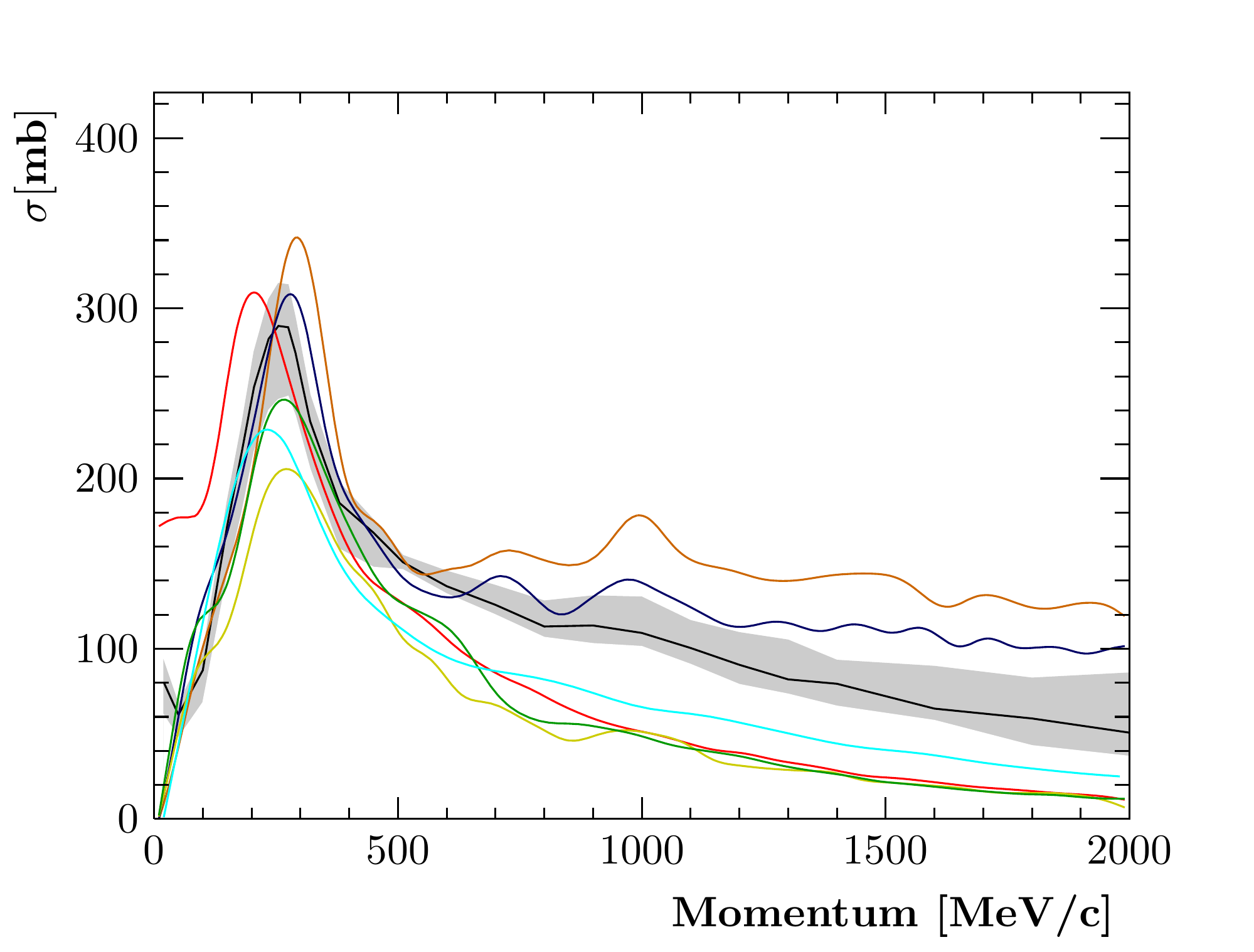}}
  \subfloat                      {\includegraphics[width=0.33\linewidth]{figures/pave.pdf}}
\caption{Comparison of the available $\pi^-$--$^{16}$O cross section external data with the NEUT best fit and its $1\sigma$ error band obtained in this work, and other models.}
\label{fig:models-o-pim}
\end{figure*}

\begin{figure*}[htbp]
  \centering
  \subfloat[Reactive]            {\includegraphics[width=0.33\linewidth]{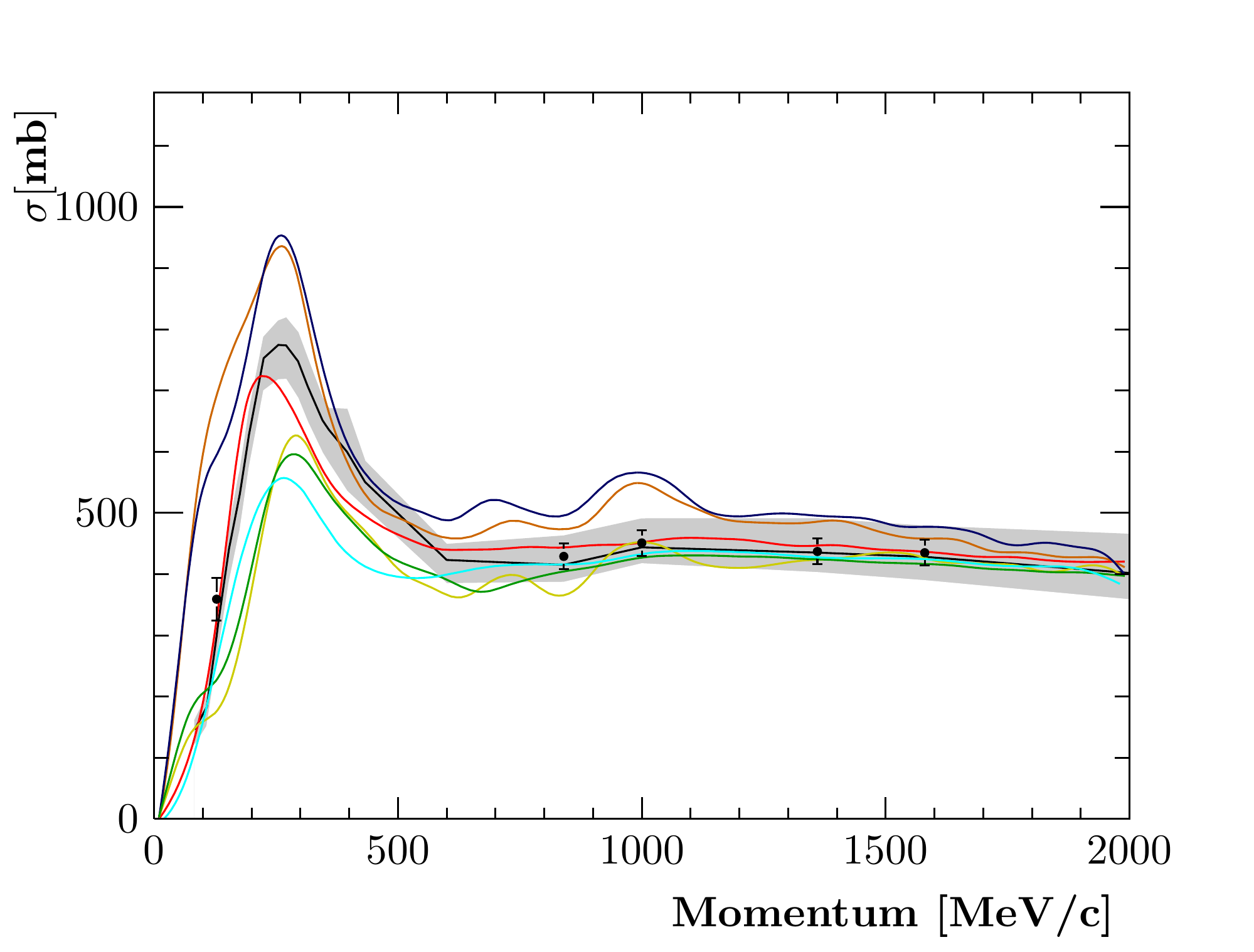}}
  \subfloat[Quasi-elastic]       {\includegraphics[width=0.33\linewidth]{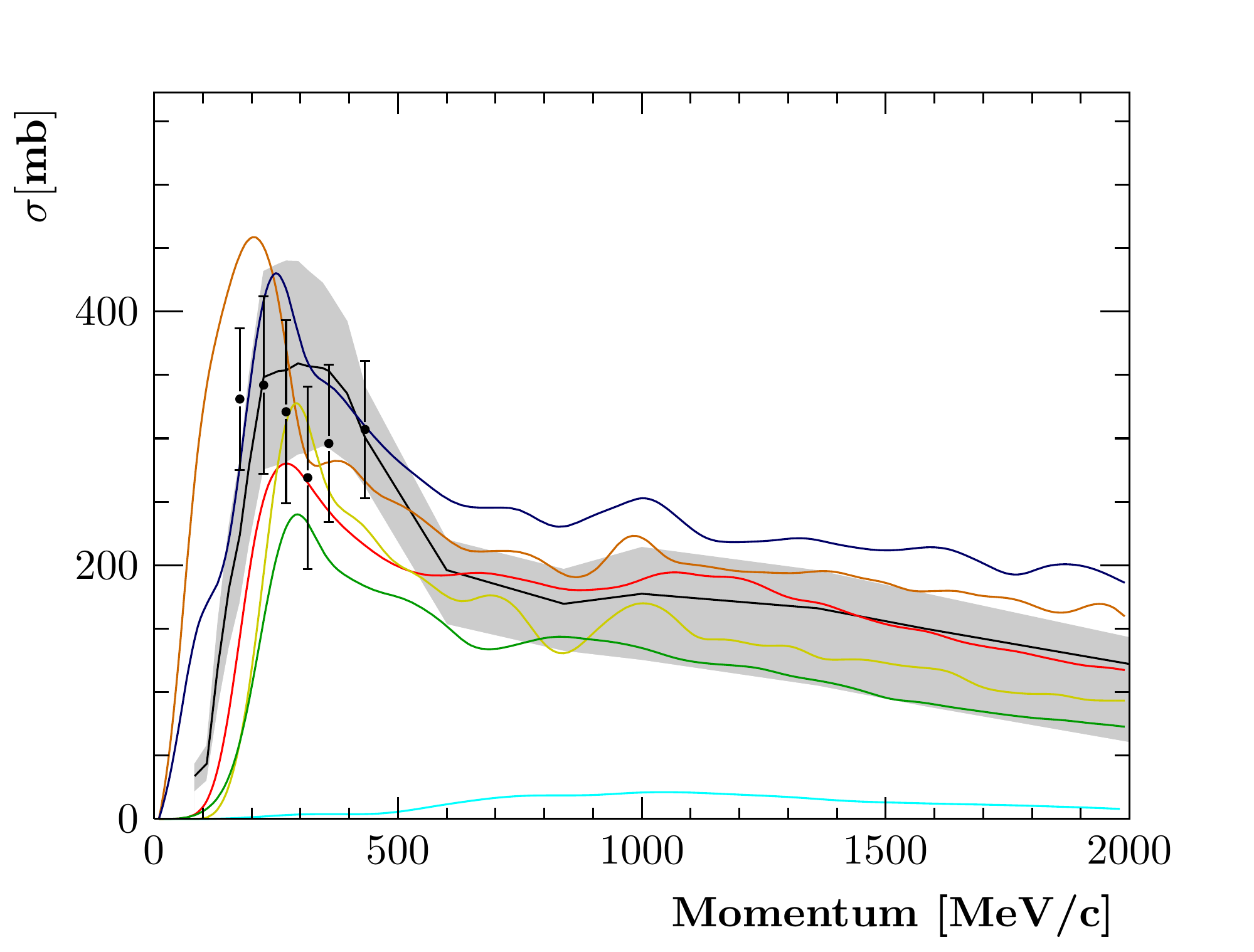}}
  \subfloat[Absorption (ABS)]    {\includegraphics[width=0.33\linewidth]{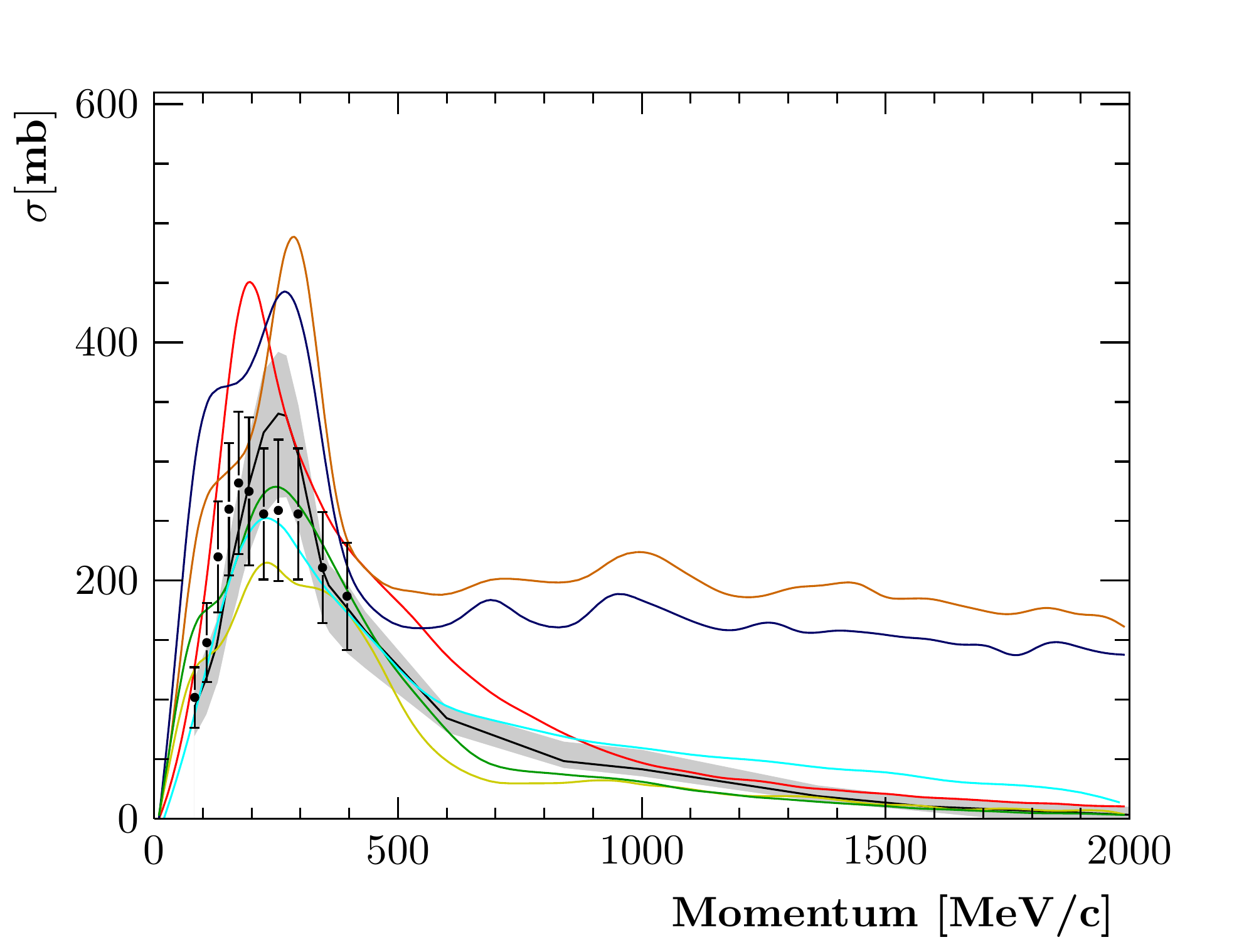}}\\\vspace{-12pt}
  \subfloat[Charge exchange (CX)]{\includegraphics[width=0.33\linewidth]{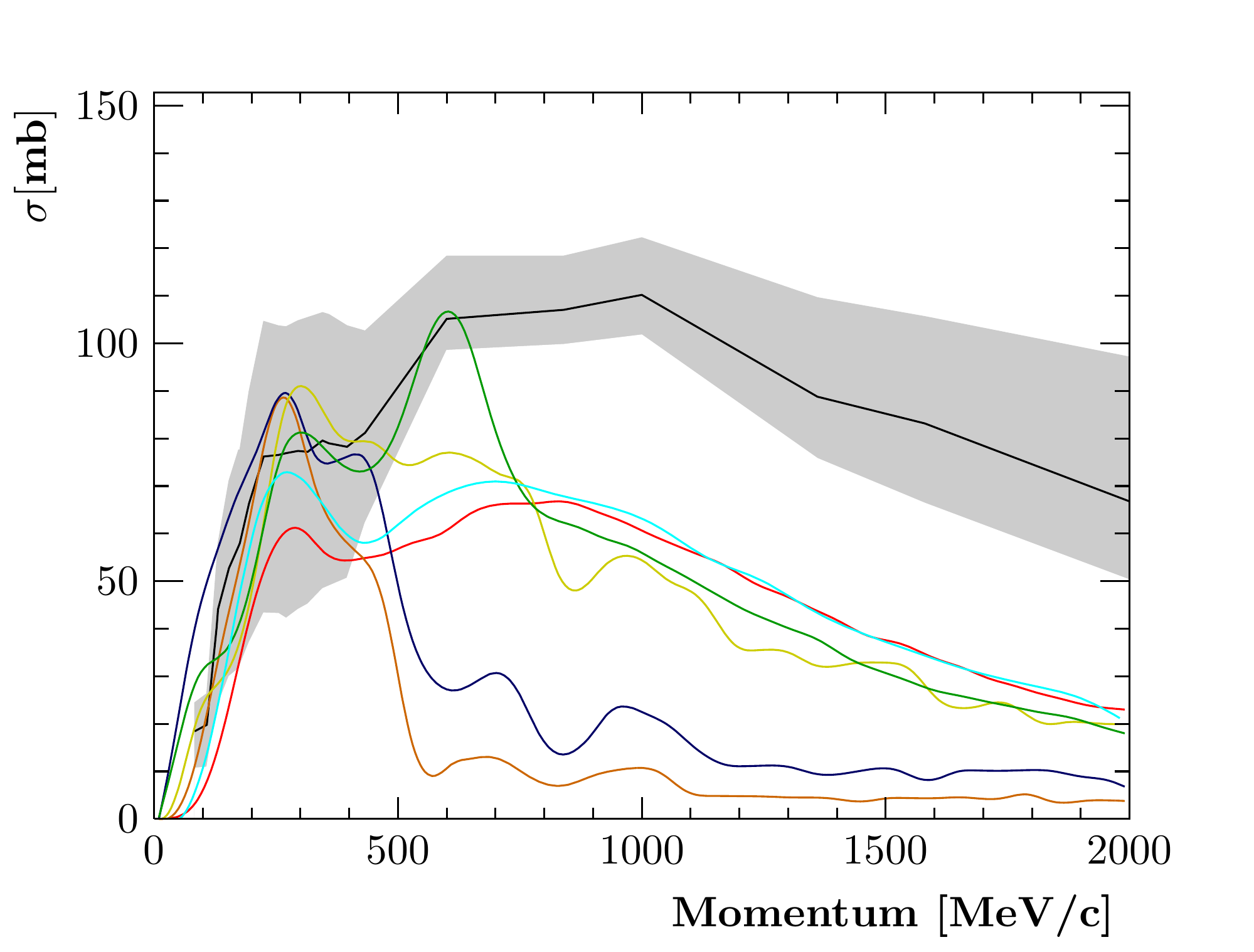}}
  \subfloat[ABS+CX]              {\includegraphics[width=0.33\linewidth]{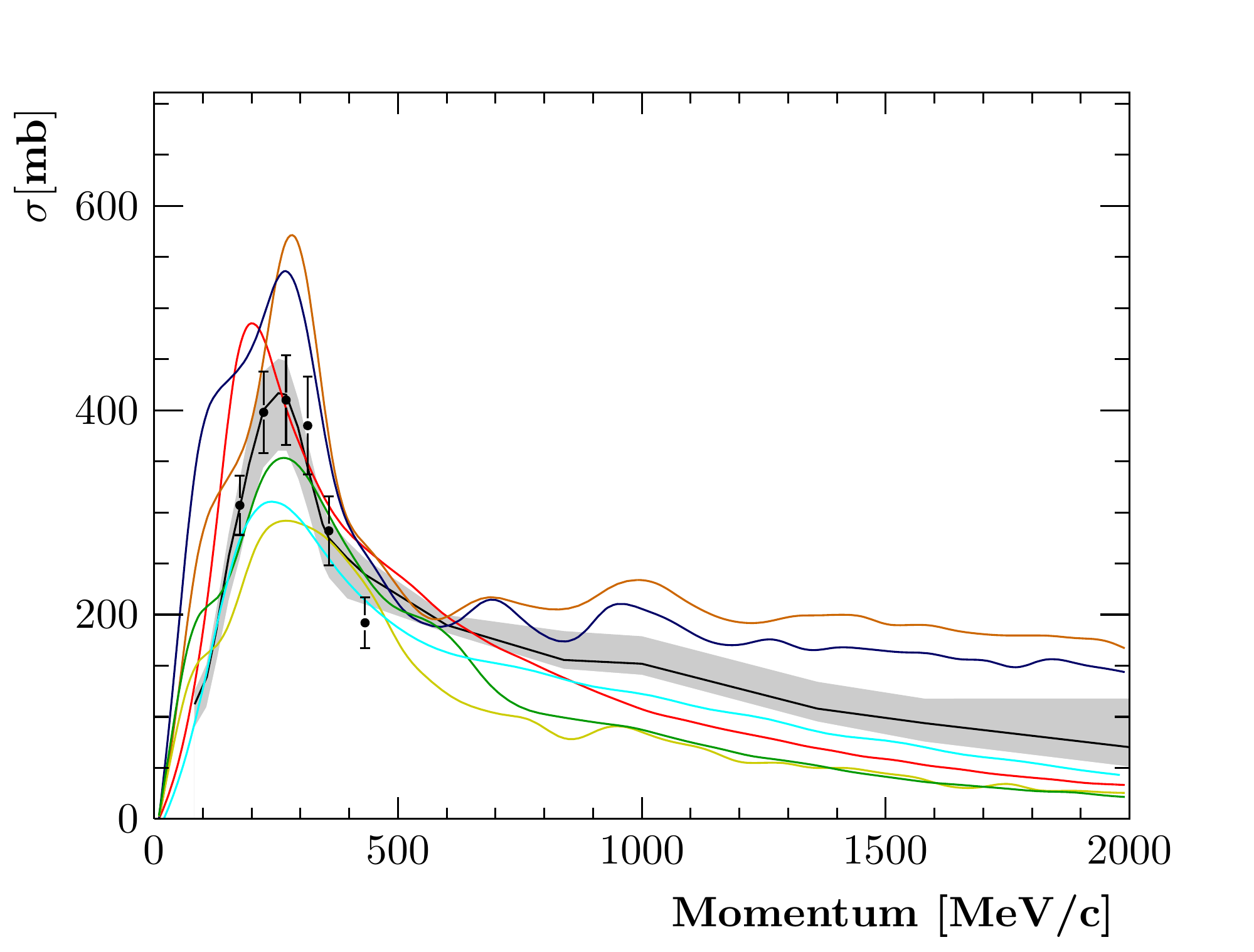}}
  \subfloat                      {\includegraphics[width=0.33\linewidth]{figures/pave.pdf}}
\caption{Comparison of the available $\pi^+$--$^{27}$Al cross section external data with the NEUT best fit and its $1\sigma$ error band obtained in this work, and other models.}
\label{fig:models-al-pip}
\end{figure*}

\begin{figure*}[htbp]
  \centering
  \subfloat[Reactive]            {\includegraphics[width=0.33\linewidth]{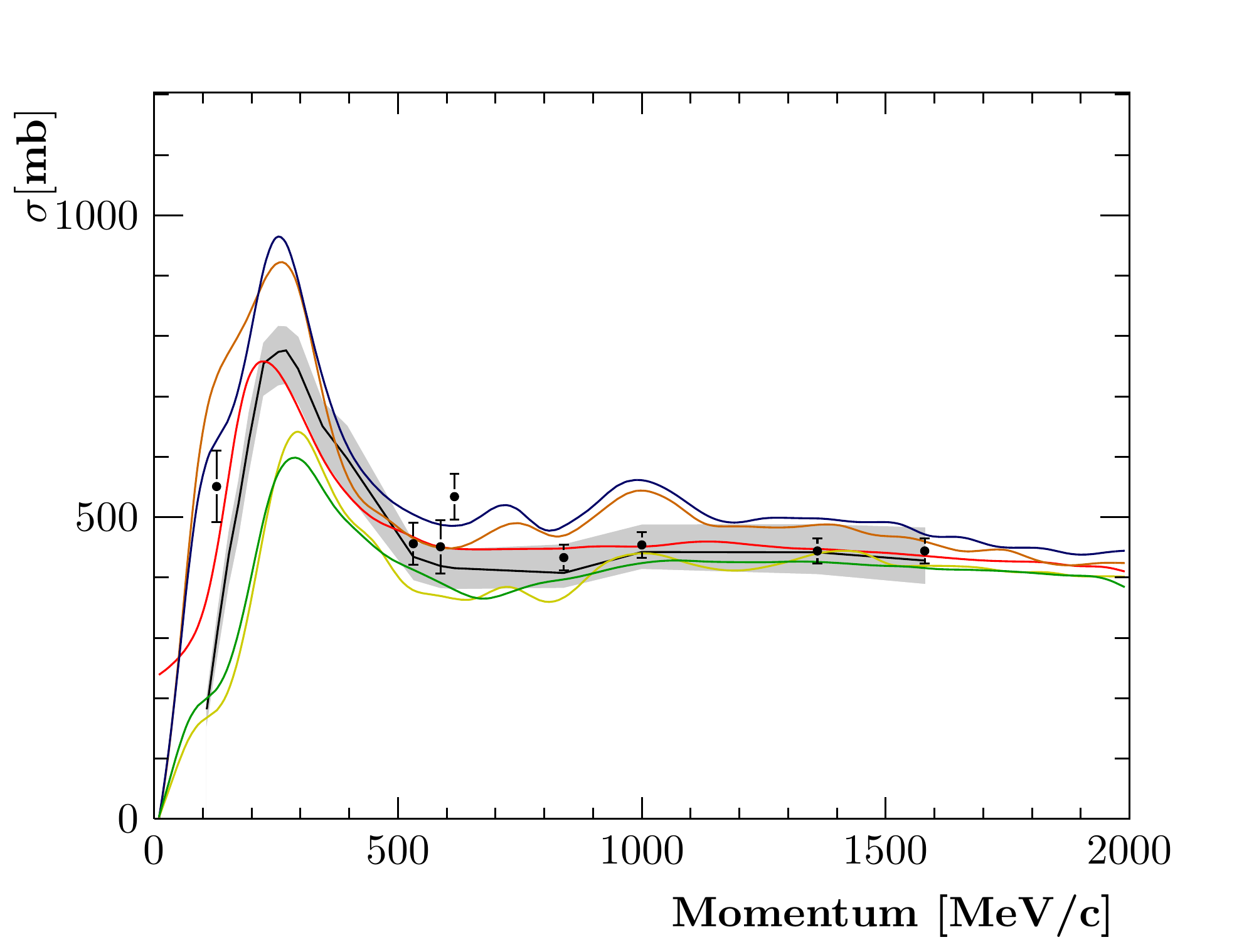}}
  \subfloat[Quasi-elastic]       {\includegraphics[width=0.33\linewidth]{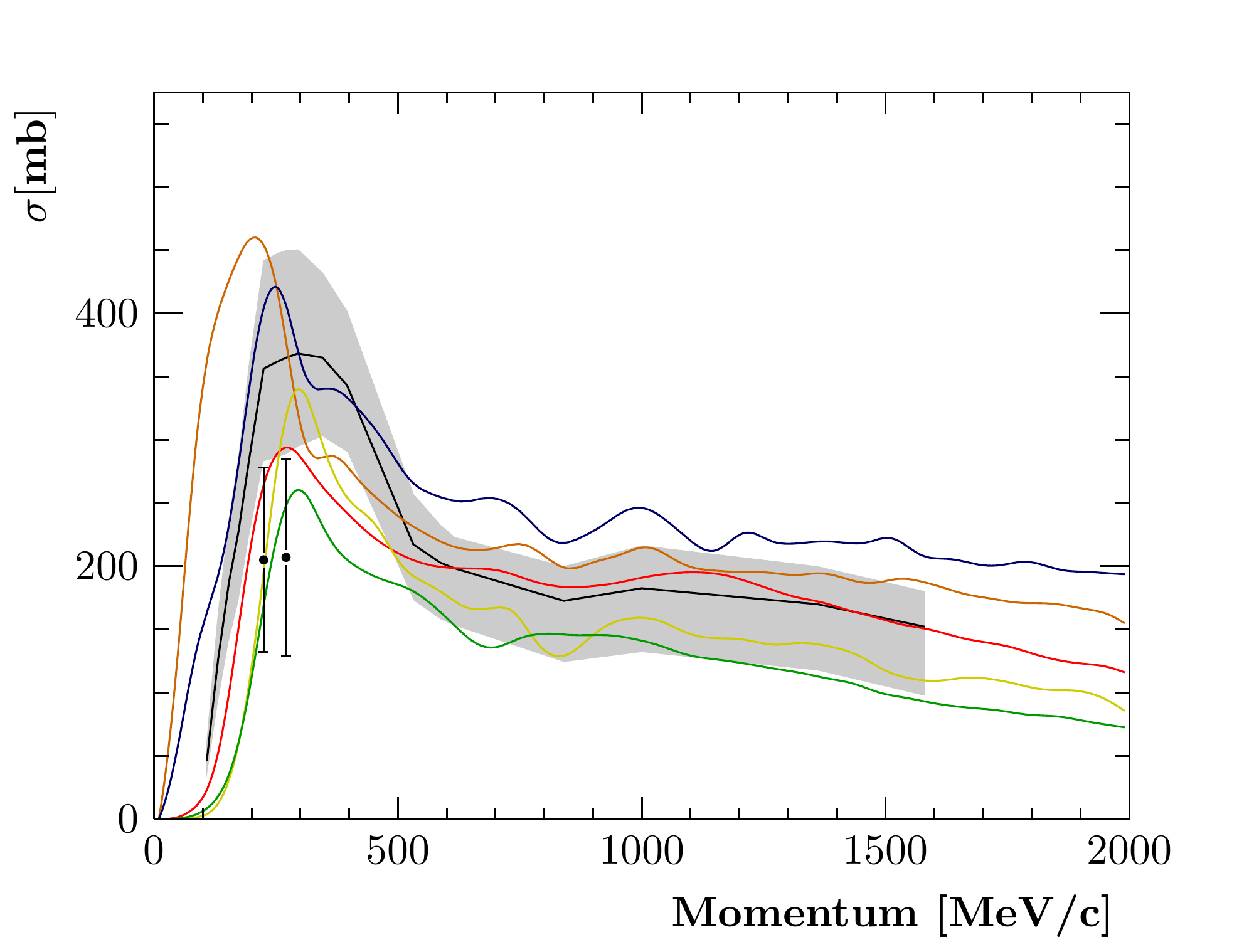}}
  \subfloat[Absorption (ABS)]    {\includegraphics[width=0.33\linewidth]{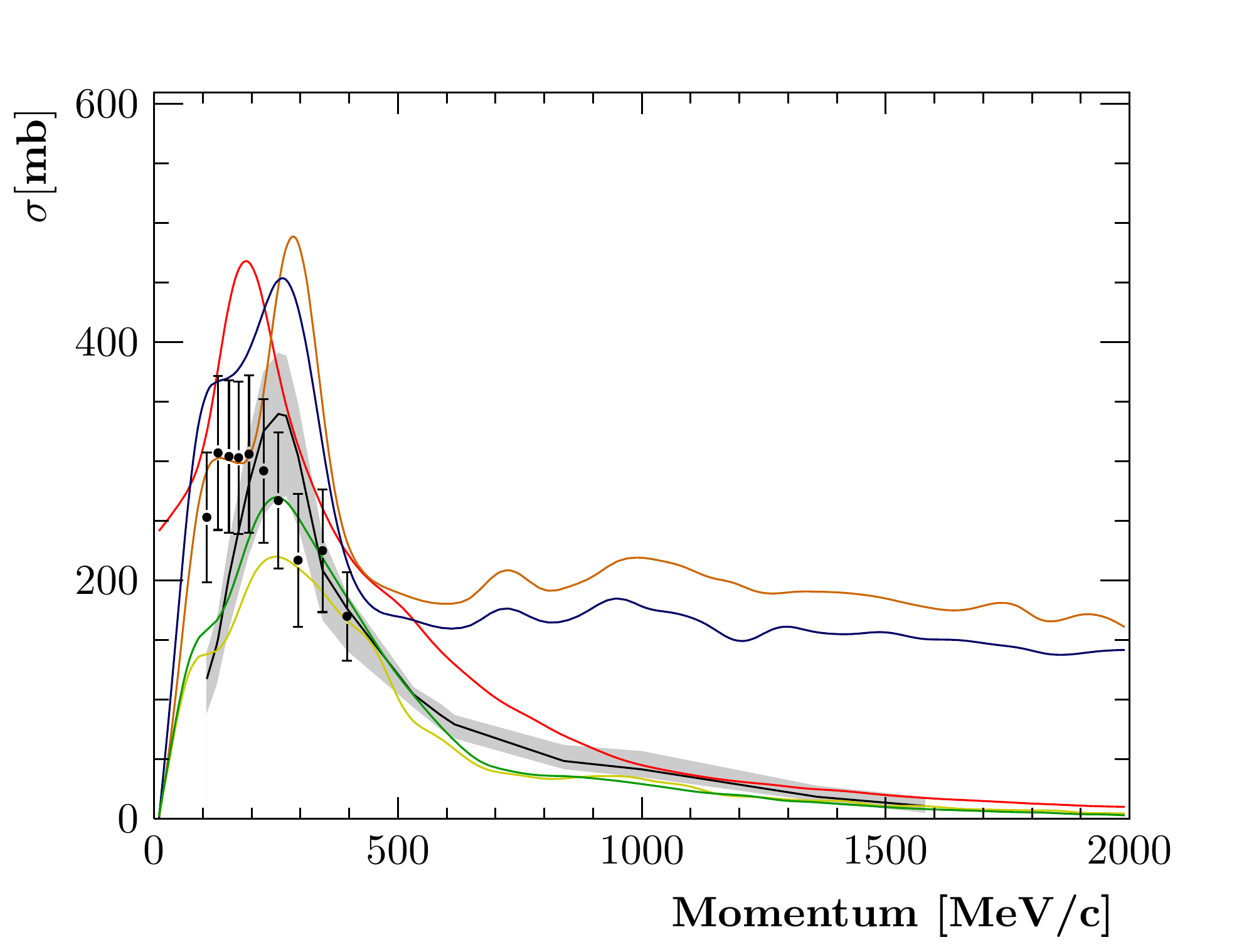}}\\\vspace{-12pt}
  \subfloat[Charge exchange (CX)]{\includegraphics[width=0.33\linewidth]{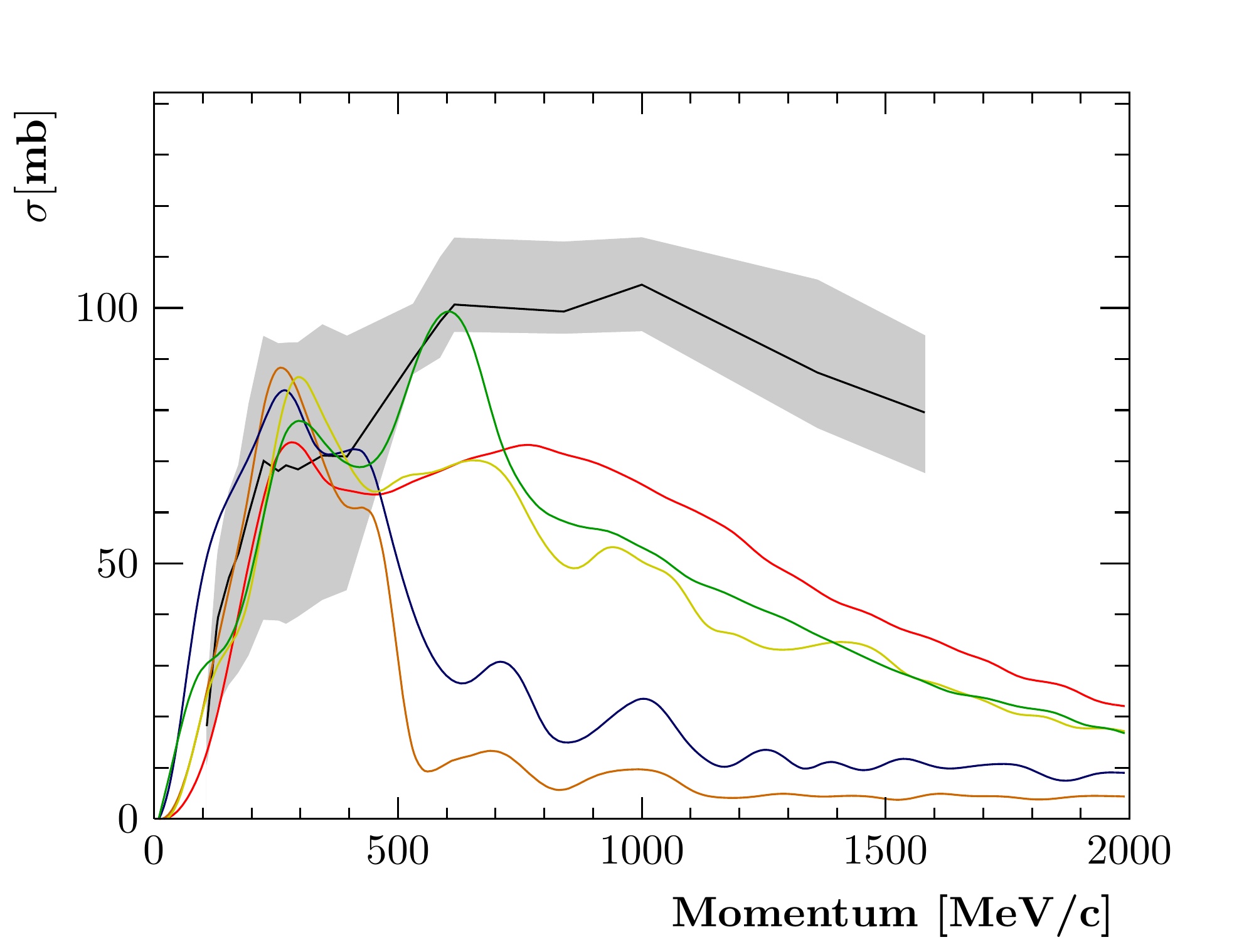}}
  \subfloat[ABS+CX]              {\includegraphics[width=0.33\linewidth]{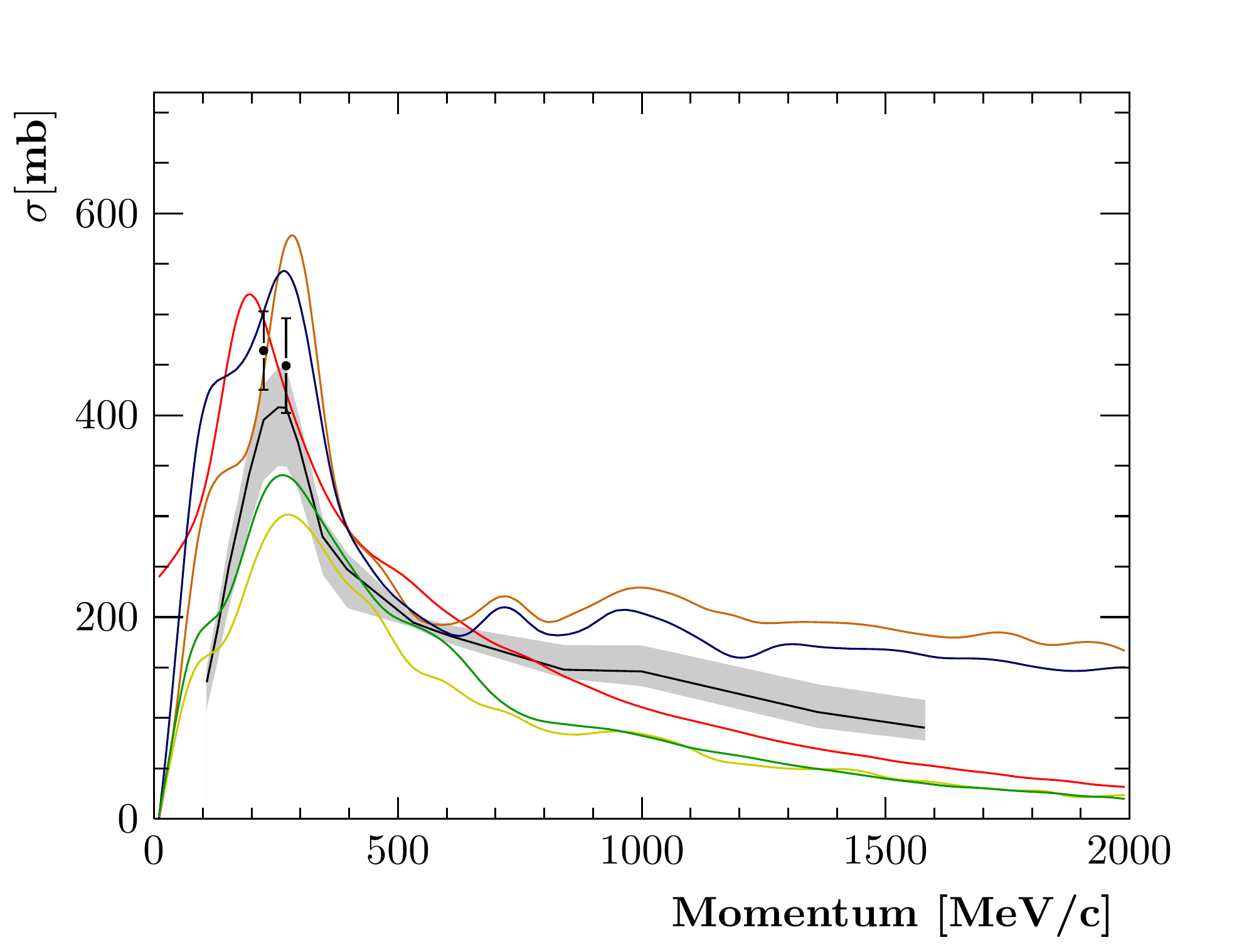}}
  \subfloat                      {\includegraphics[width=0.33\linewidth]{figures/pave.pdf}}
\caption{Comparison of the available $\pi^-$--$^{27}$Al cross section external data with the NEUT best fit and its $1\sigma$ error band obtained in this work, and other models.}
\label{fig:models-al-pim}
\end{figure*}

\begin{figure*}[htbp]
  \centering
  \subfloat[Reactive]            {\includegraphics[width=0.33\linewidth]{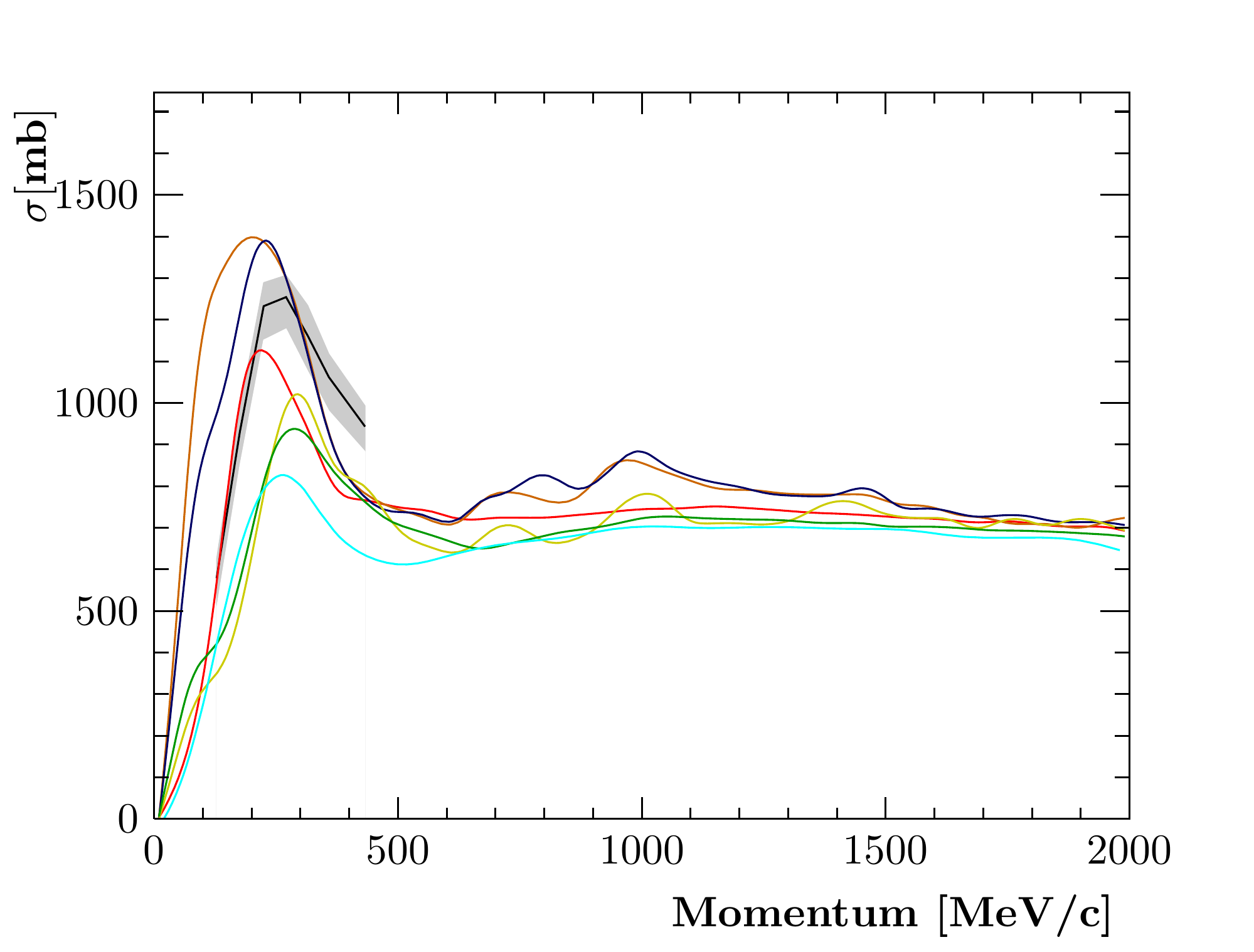}}
  \subfloat[Quasi-elastic]       {\includegraphics[width=0.33\linewidth]{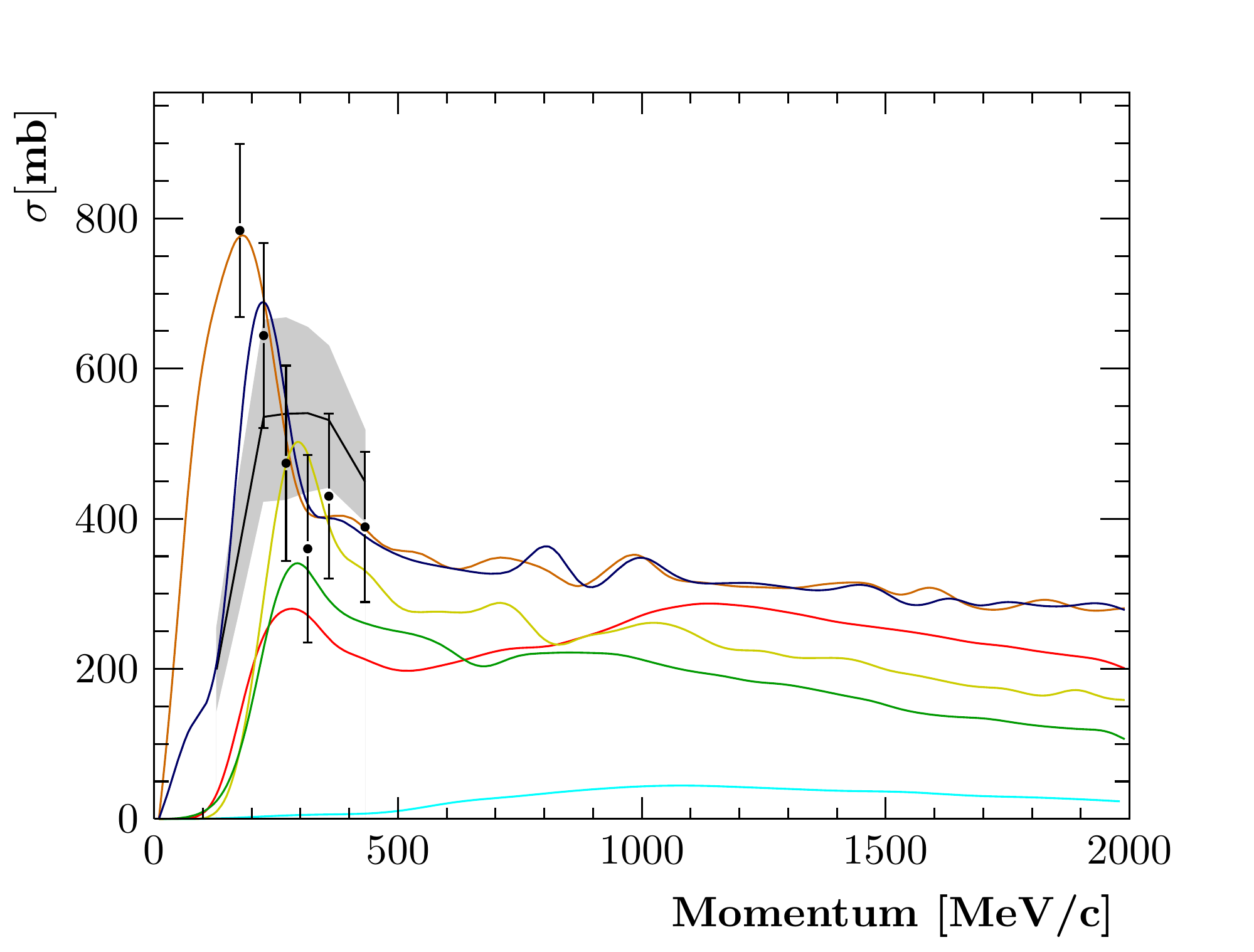}}
  \subfloat[Absorption (ABS)]    {\includegraphics[width=0.33\linewidth]{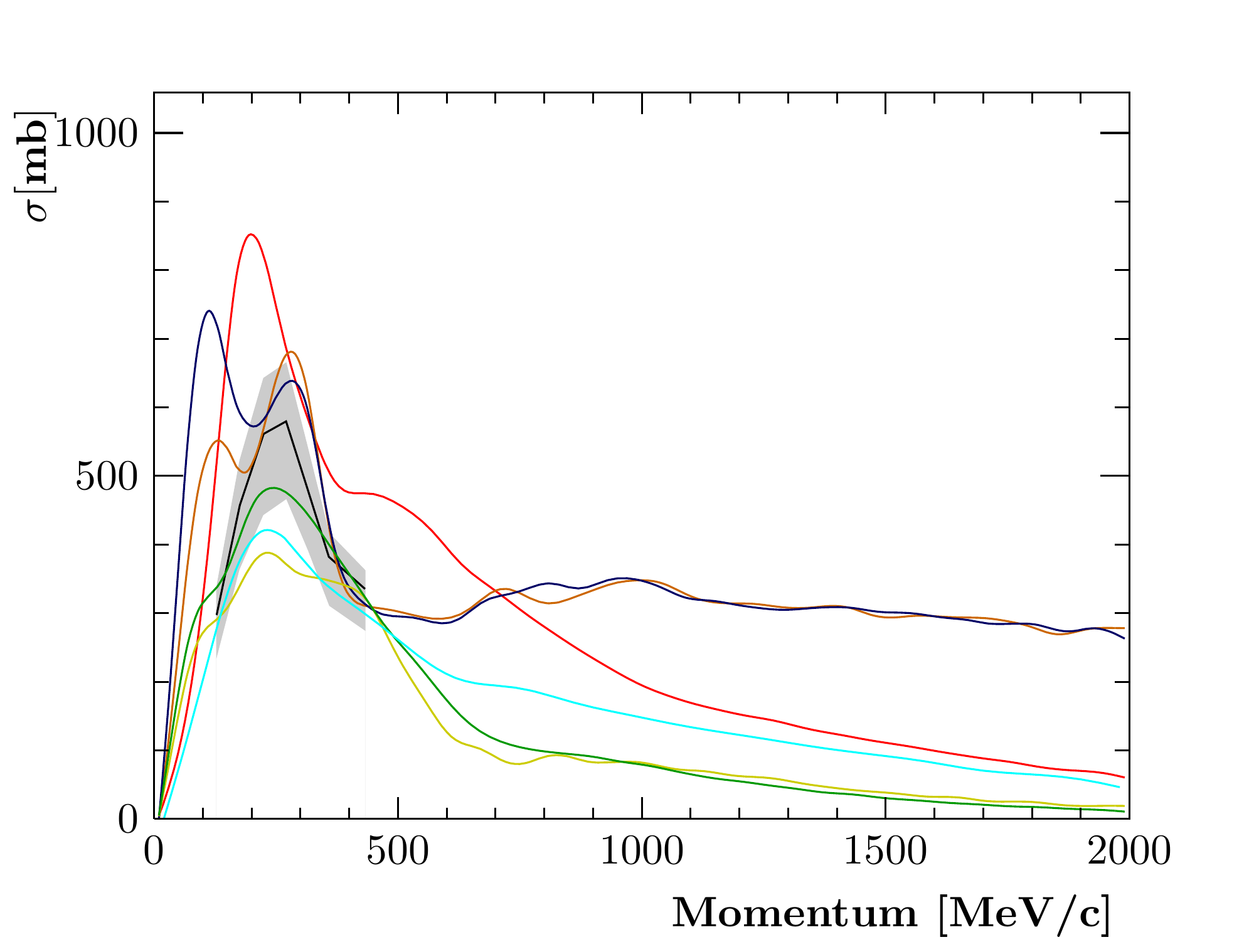}}\\\vspace{-12pt}
  \subfloat[Charge exchange (CX)]{\includegraphics[width=0.33\linewidth]{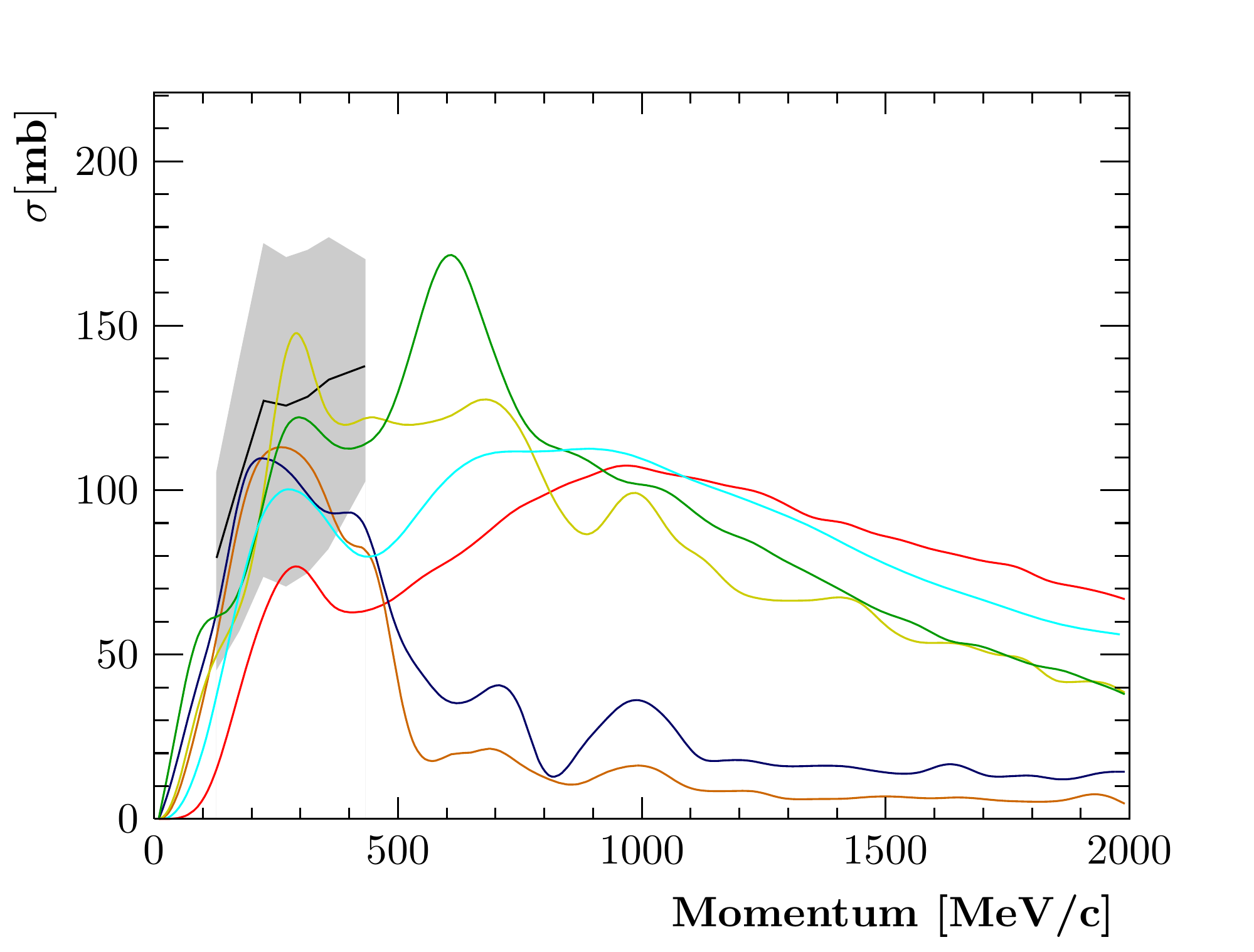}}
  \subfloat[ABS+CX]              {\includegraphics[width=0.33\linewidth]{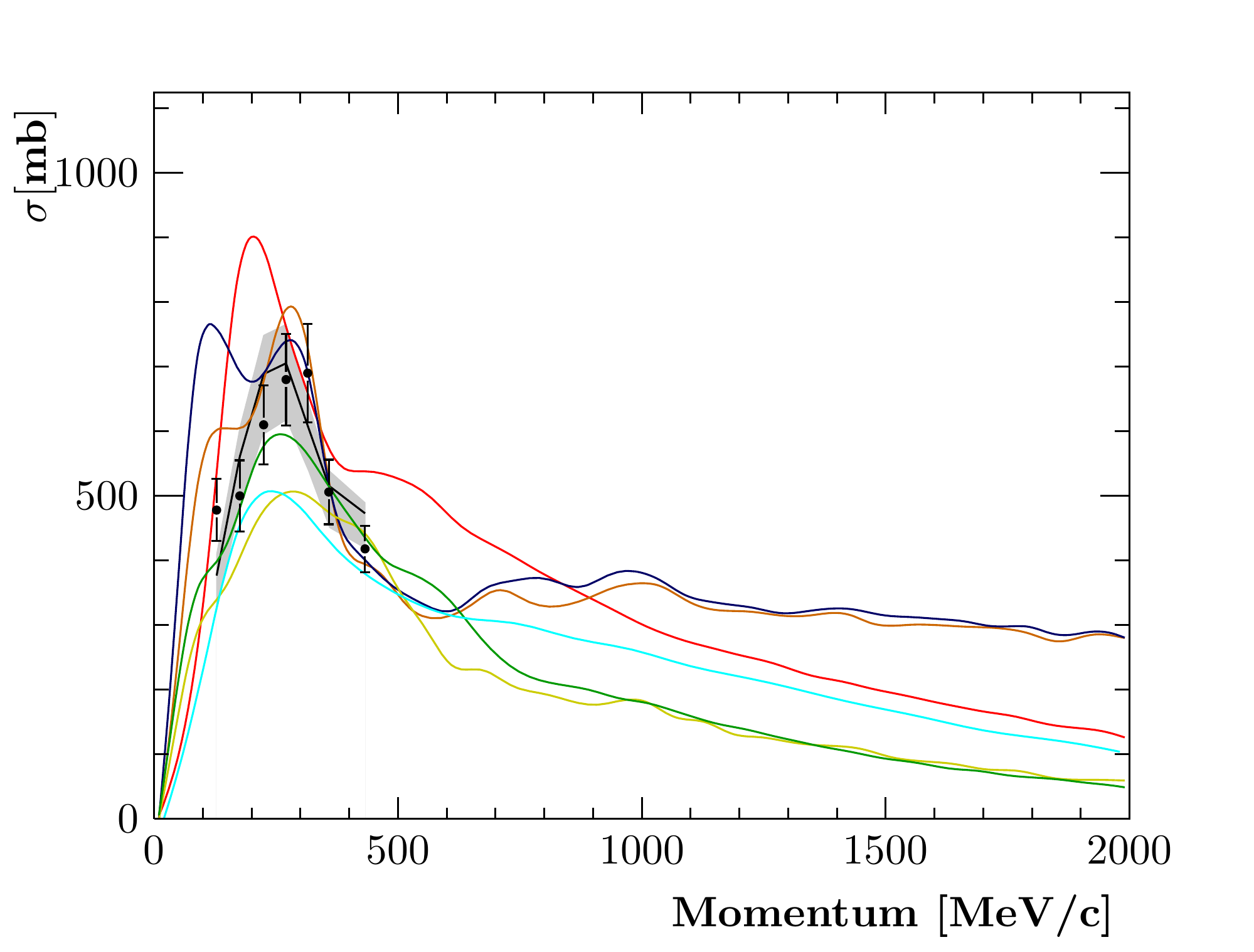}}
  \subfloat                      {\includegraphics[width=0.33\linewidth]{figures/pave.pdf}}
\caption{Comparison of the available $\pi^+$--$^{56}$Fe cross section external data with the NEUT best fit and its $1\sigma$ error band obtained in this work, and other models.}
\label{fig:models-fe-pip}
\end{figure*}

\begin{figure*}[htbp]
  \centering
  \subfloat[Reactive]            {\includegraphics[width=0.33\linewidth]{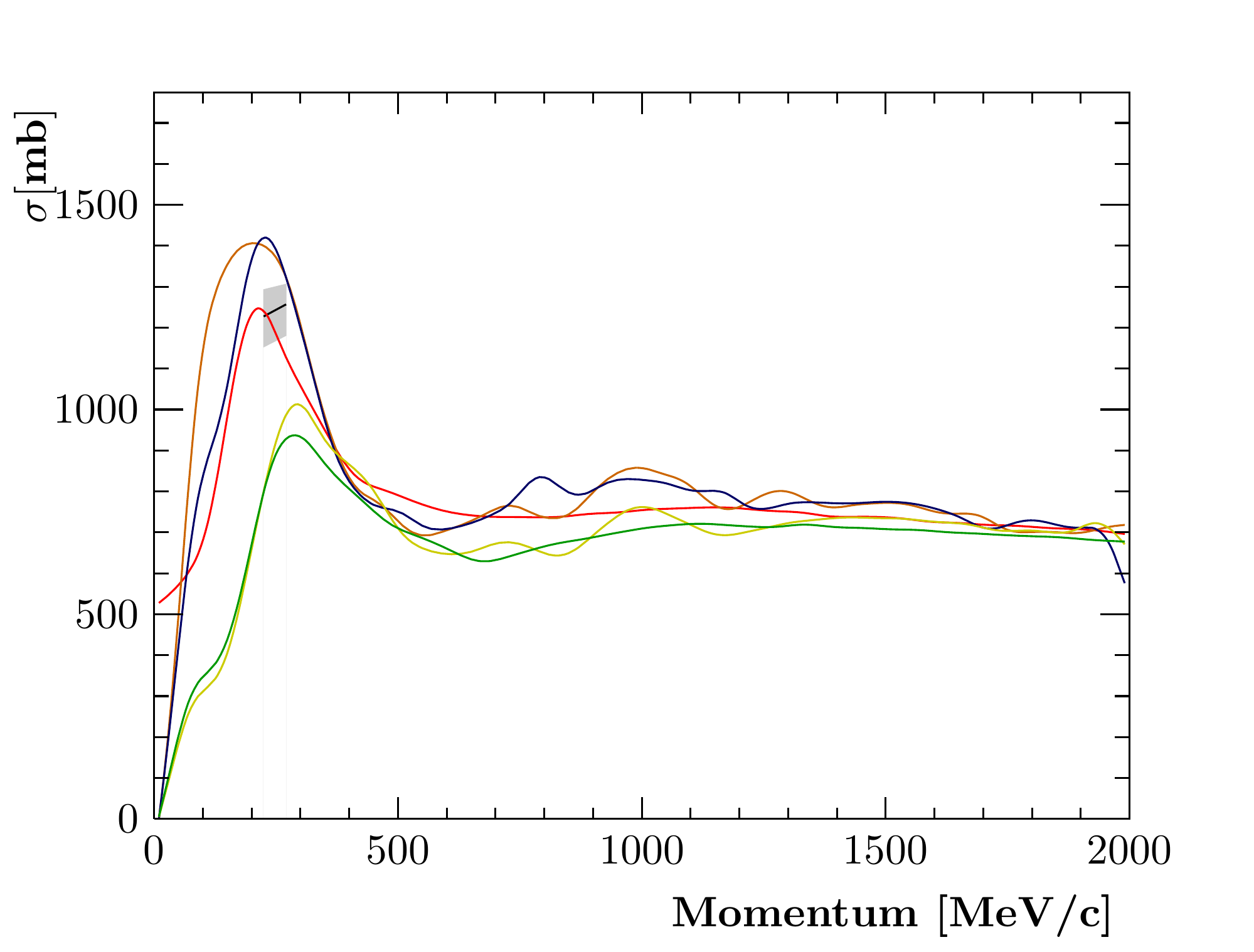}}
  \subfloat[Quasi-elastic]       {\includegraphics[width=0.33\linewidth]{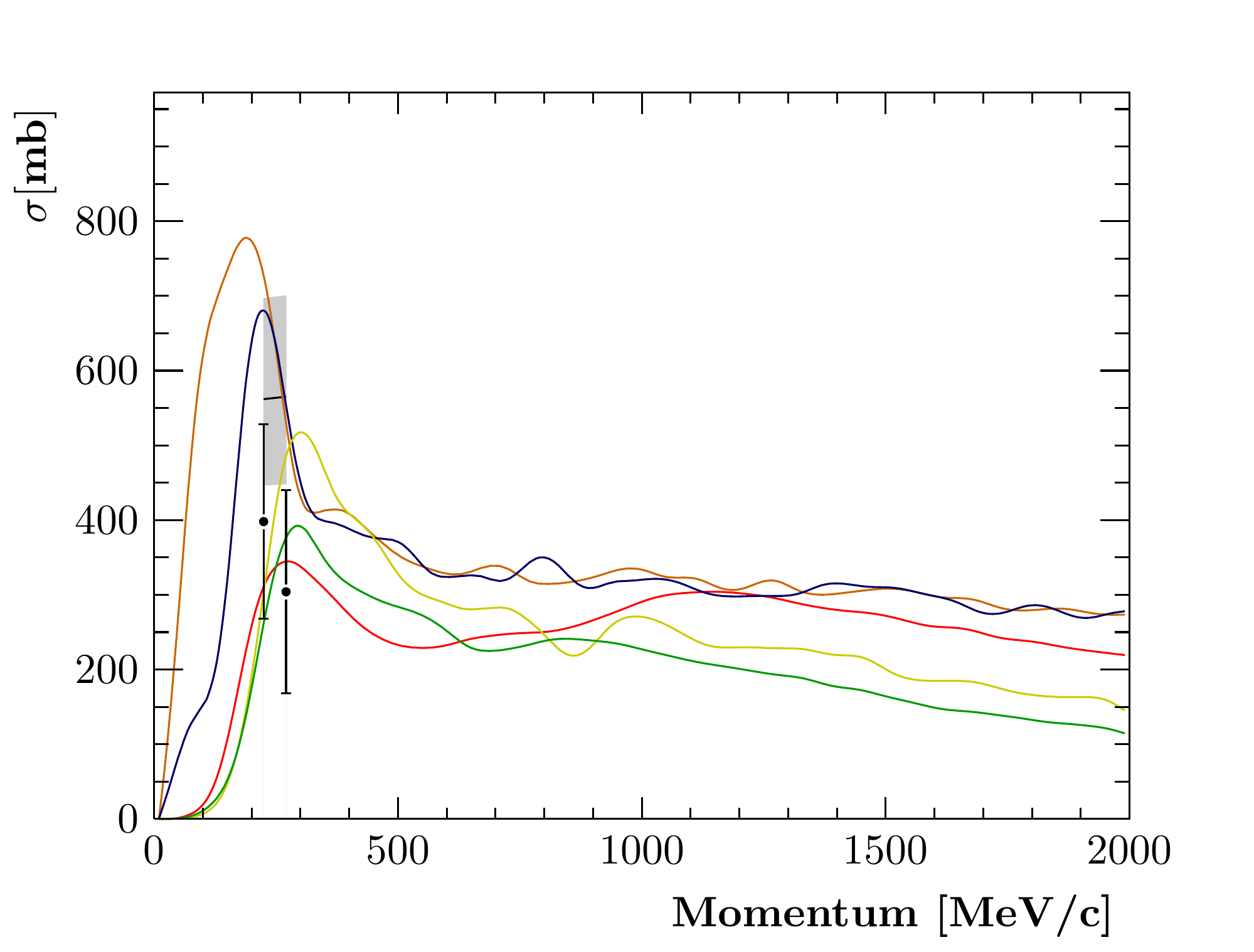}}
  \subfloat[Absorption (ABS)]    {\includegraphics[width=0.33\linewidth]{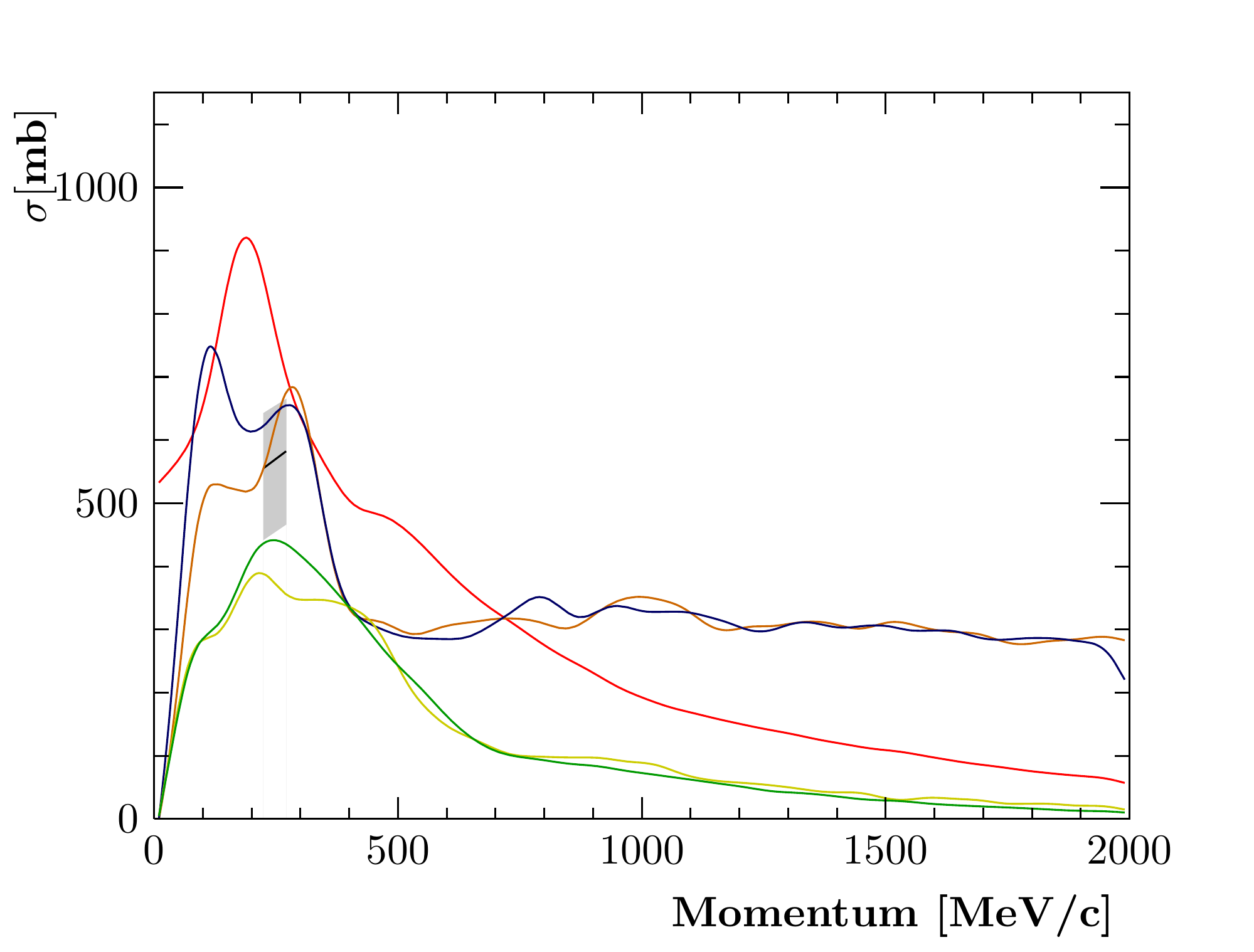}}\\\vspace{-12pt}
  \subfloat[Charge exchange (CX)]{\includegraphics[width=0.33\linewidth]{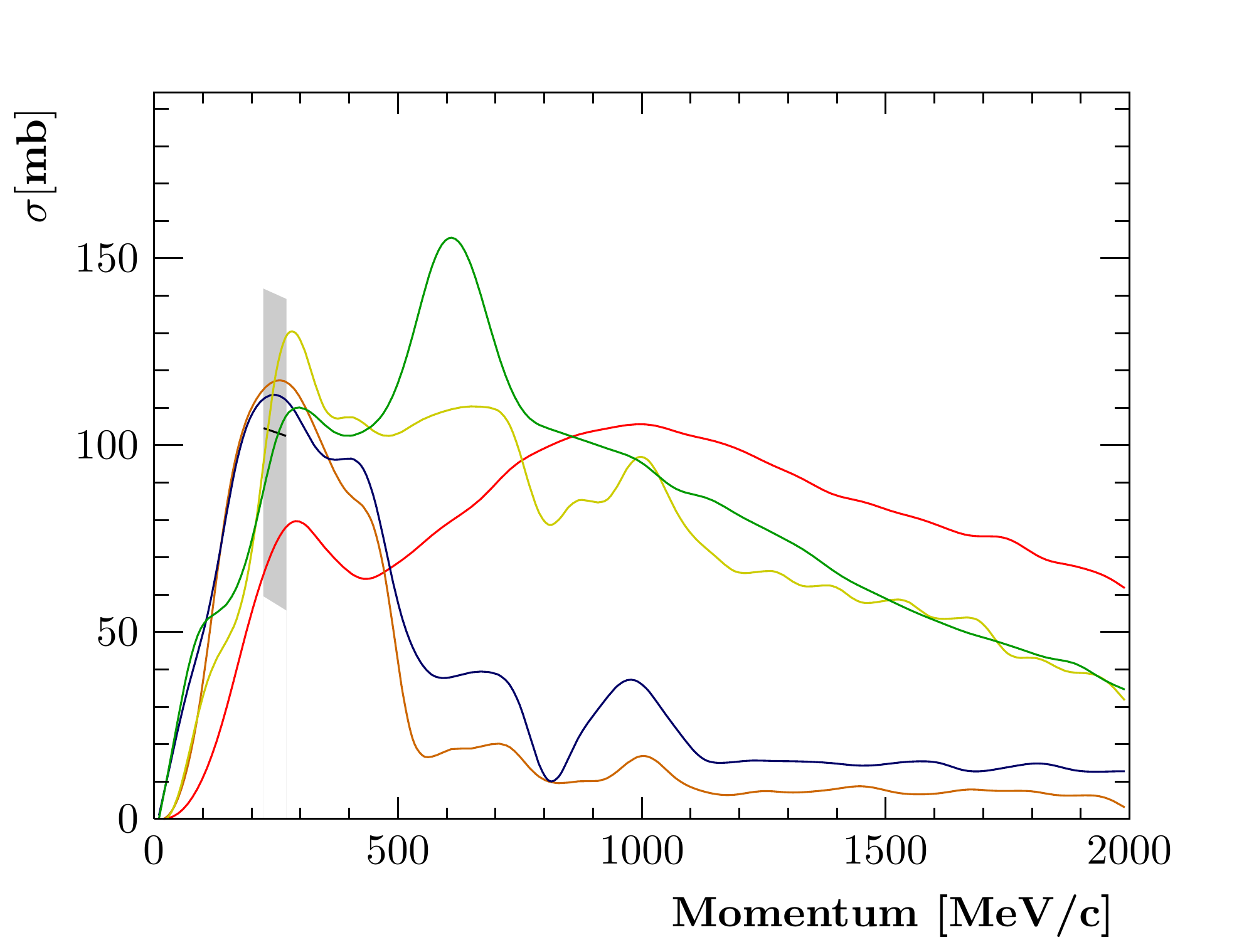}}
  \subfloat[ABS+CX]              {\includegraphics[width=0.33\linewidth]{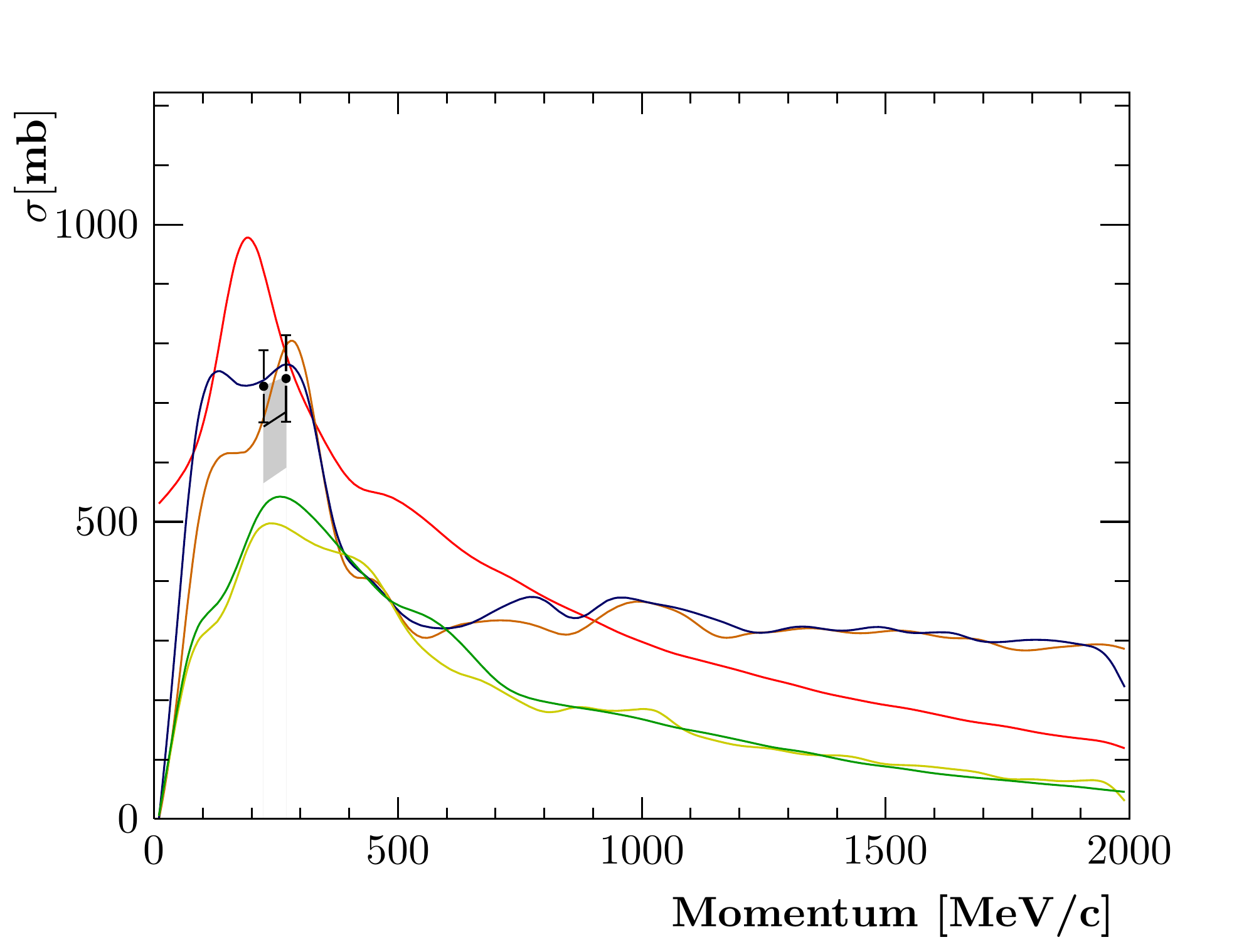}}
  \subfloat                      {\includegraphics[width=0.33\linewidth]{figures/pave.pdf}}
\caption{Comparison of the available $\pi^-$--$^{56}$Fe cross section external data with the NEUT best fit and its $1\sigma$ error band obtained in this work, and other models.}
\label{fig:models-fe-pim}
\end{figure*}

\begin{figure*}[htbp]
  \centering
  \subfloat[Reactive]            {\includegraphics[width=0.33\linewidth]{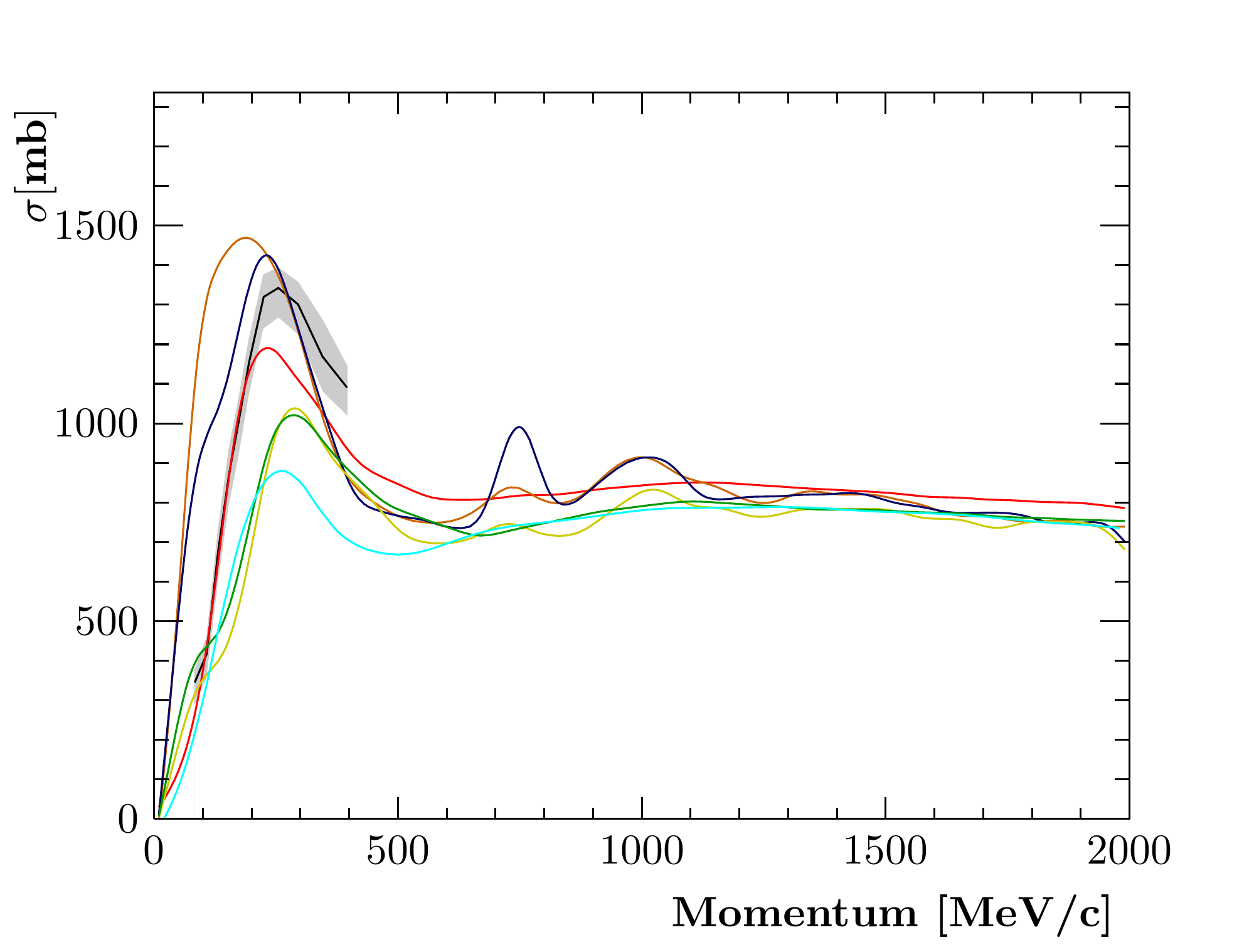}}
  \subfloat[Quasi-elastic]       {\includegraphics[width=0.33\linewidth]{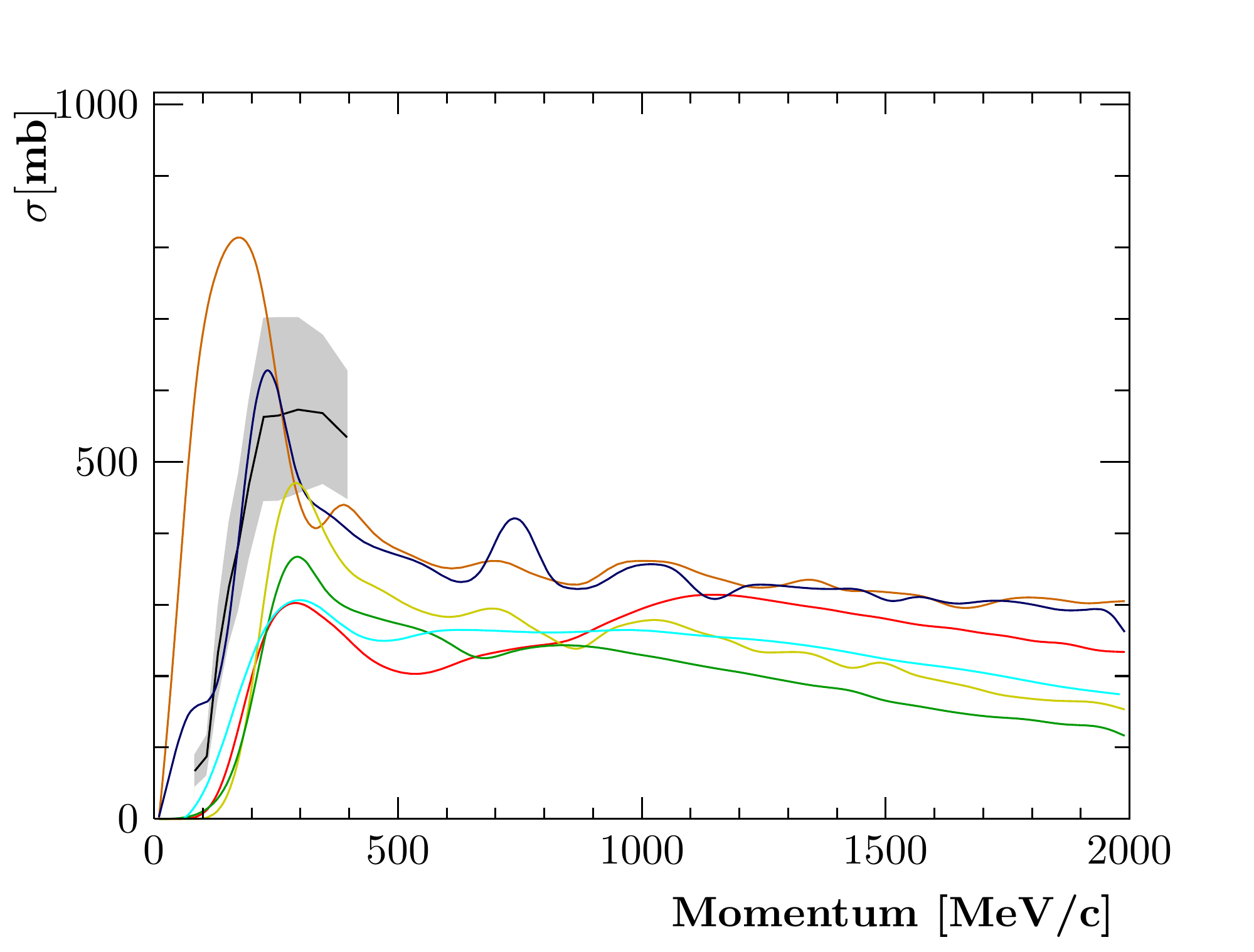}}
  \subfloat[Absorption (ABS)]    {\includegraphics[width=0.33\linewidth]{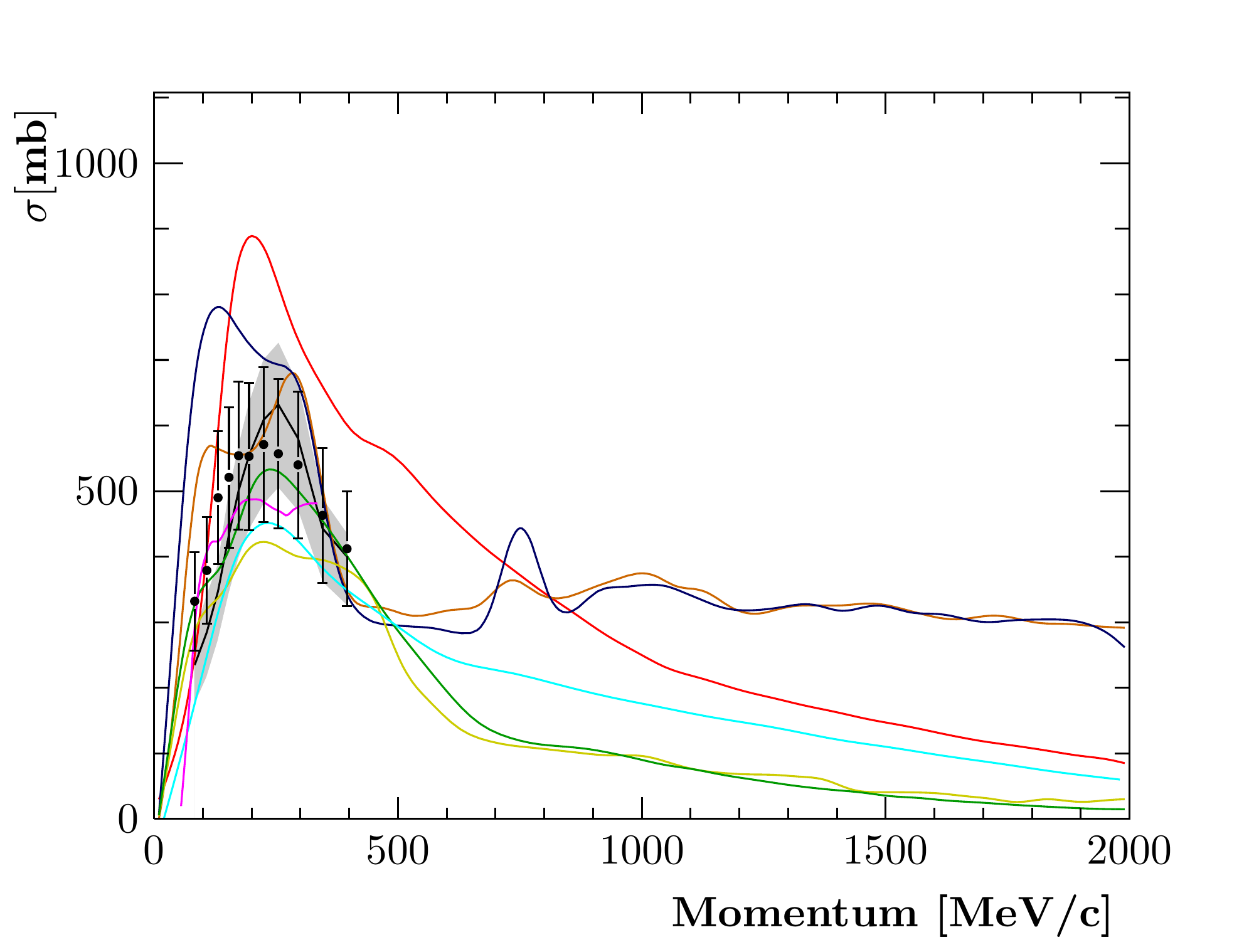}}\\\vspace{-12pt}
  \subfloat[Charge exchange (CX)]{\includegraphics[width=0.33\linewidth]{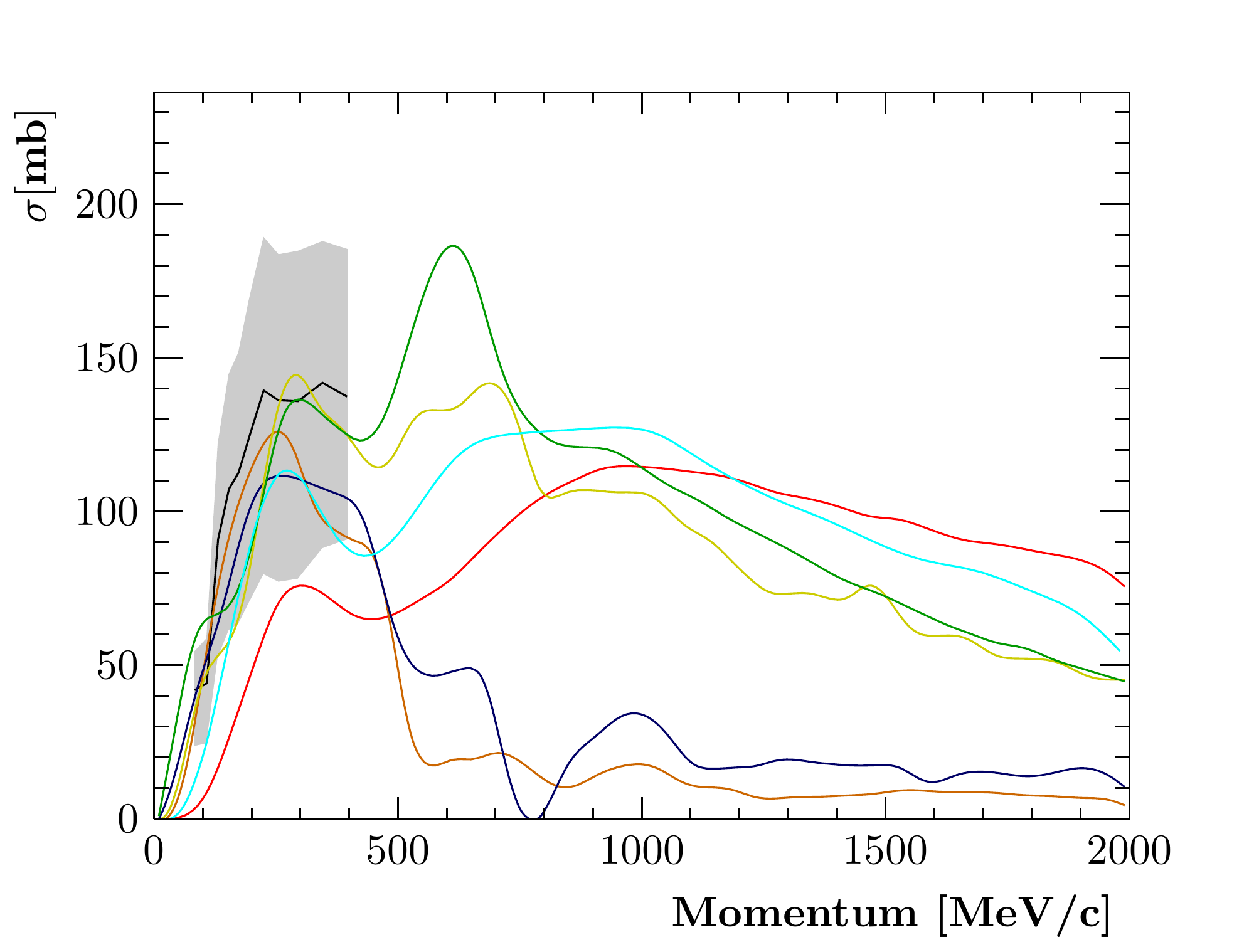}}
  \subfloat[ABS+CX]              {\includegraphics[width=0.33\linewidth]{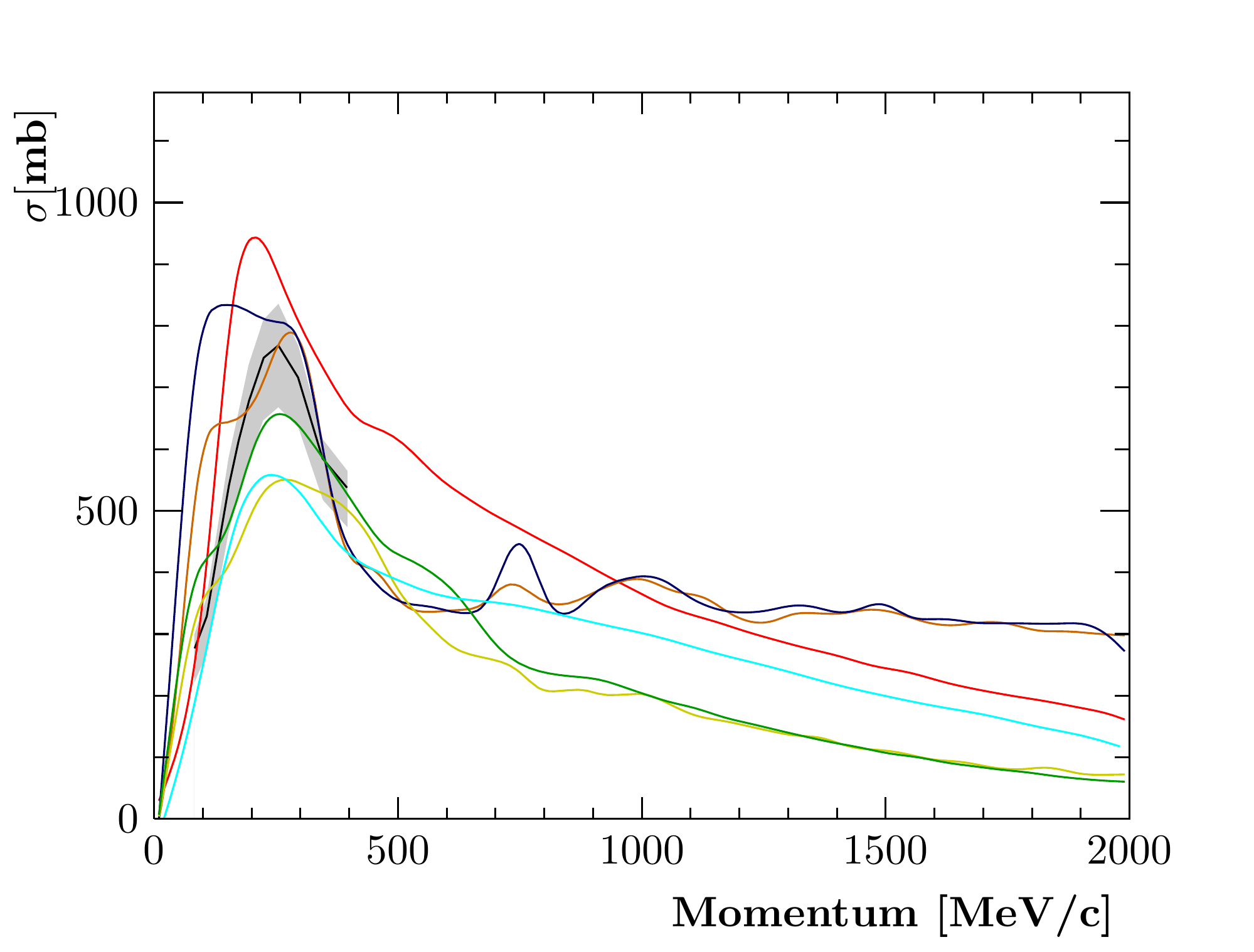}}
  \subfloat                      {\includegraphics[width=0.33\linewidth]{figures/pave.pdf}}
\caption{Comparison of the available $\pi^+$--$^{63}$Cu cross section external data with the NEUT best fit and its $1\sigma$ error band obtained in this work, and other models.}
\label{fig:models-cu-pip}
\end{figure*}

\begin{figure*}[htbp]
  \centering
  \subfloat[Reactive]            {\includegraphics[width=0.33\linewidth]{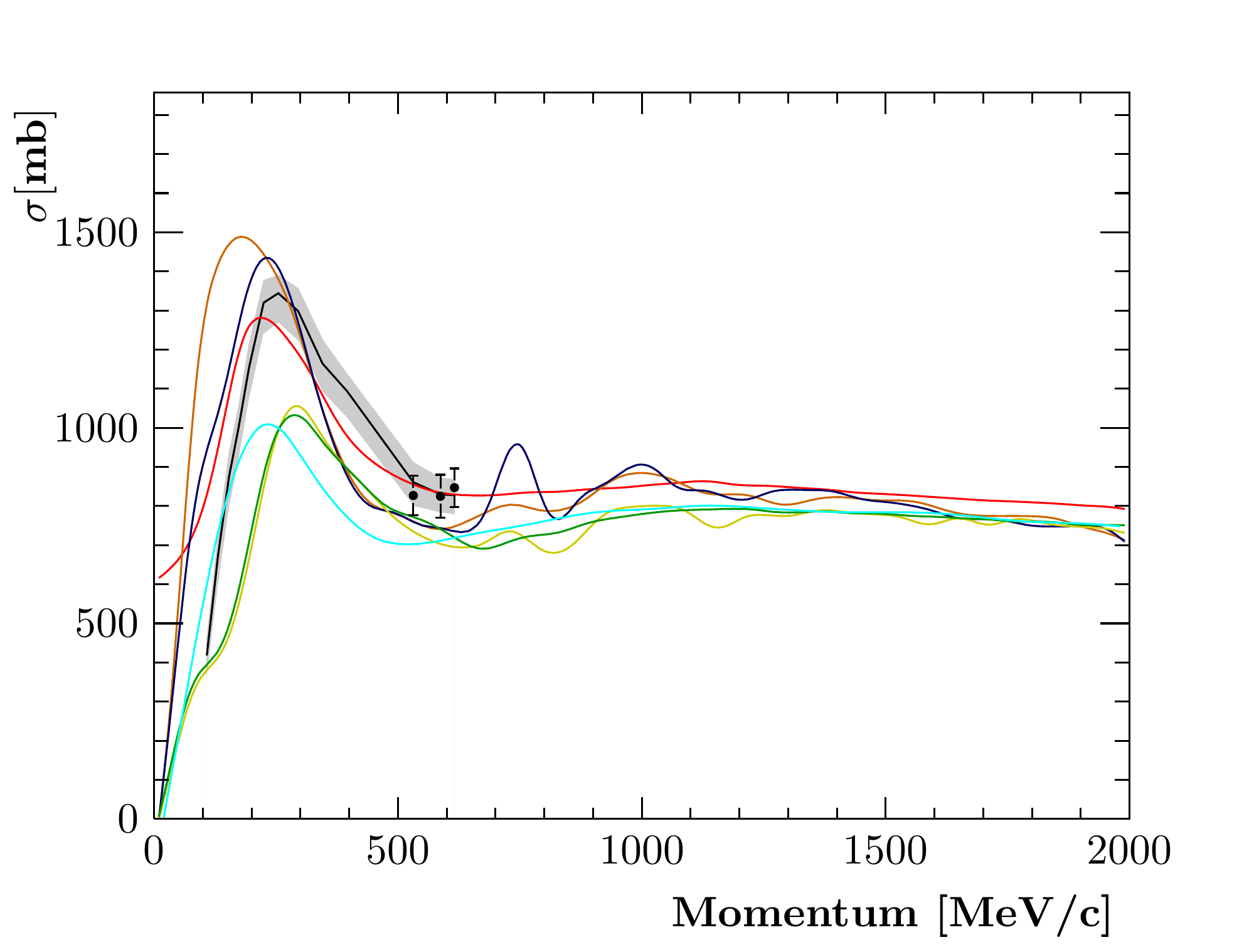}}
  \subfloat[Quasi-elastic]       {\includegraphics[width=0.33\linewidth]{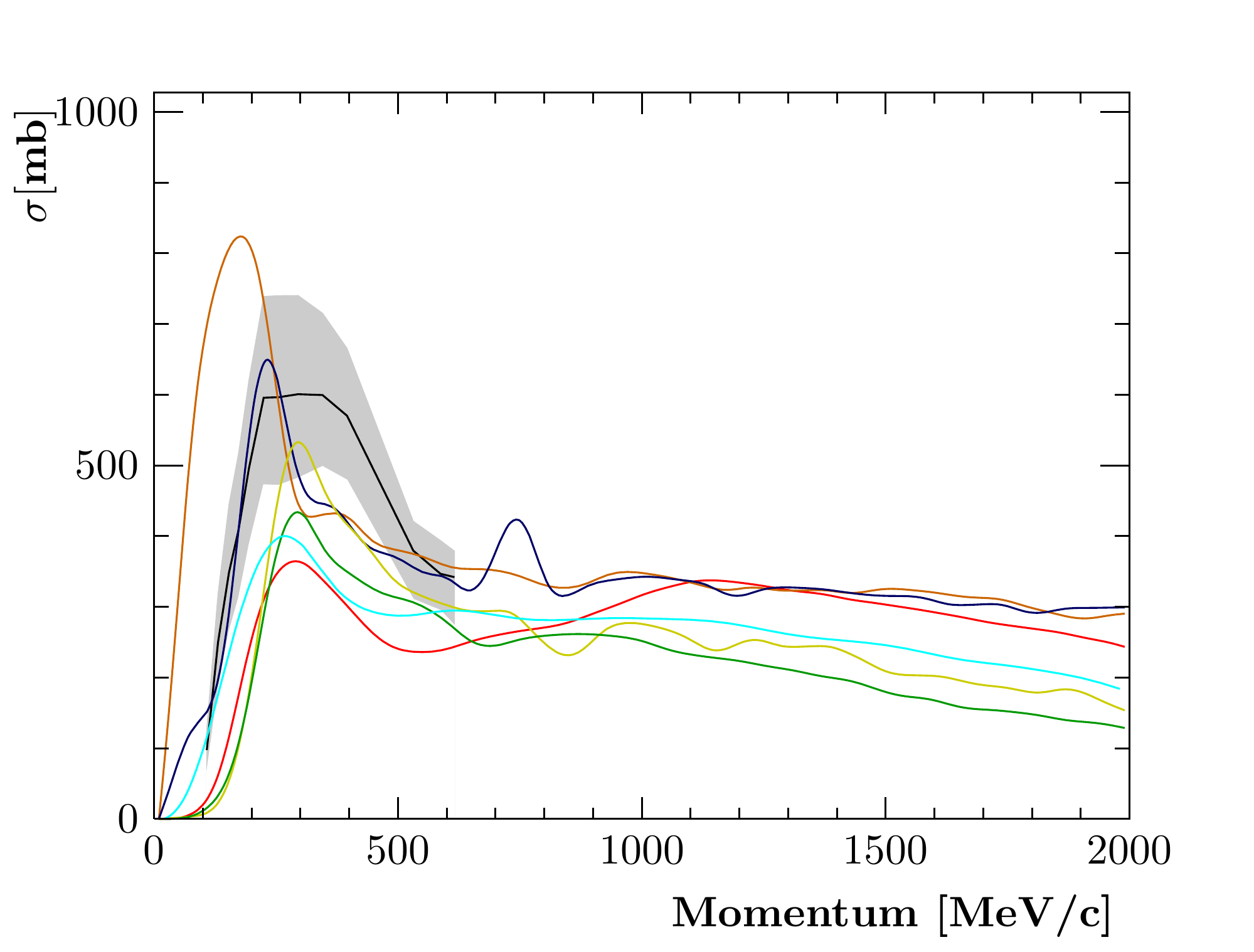}}
  \subfloat[Absorption (ABS)]    {\includegraphics[width=0.33\linewidth]{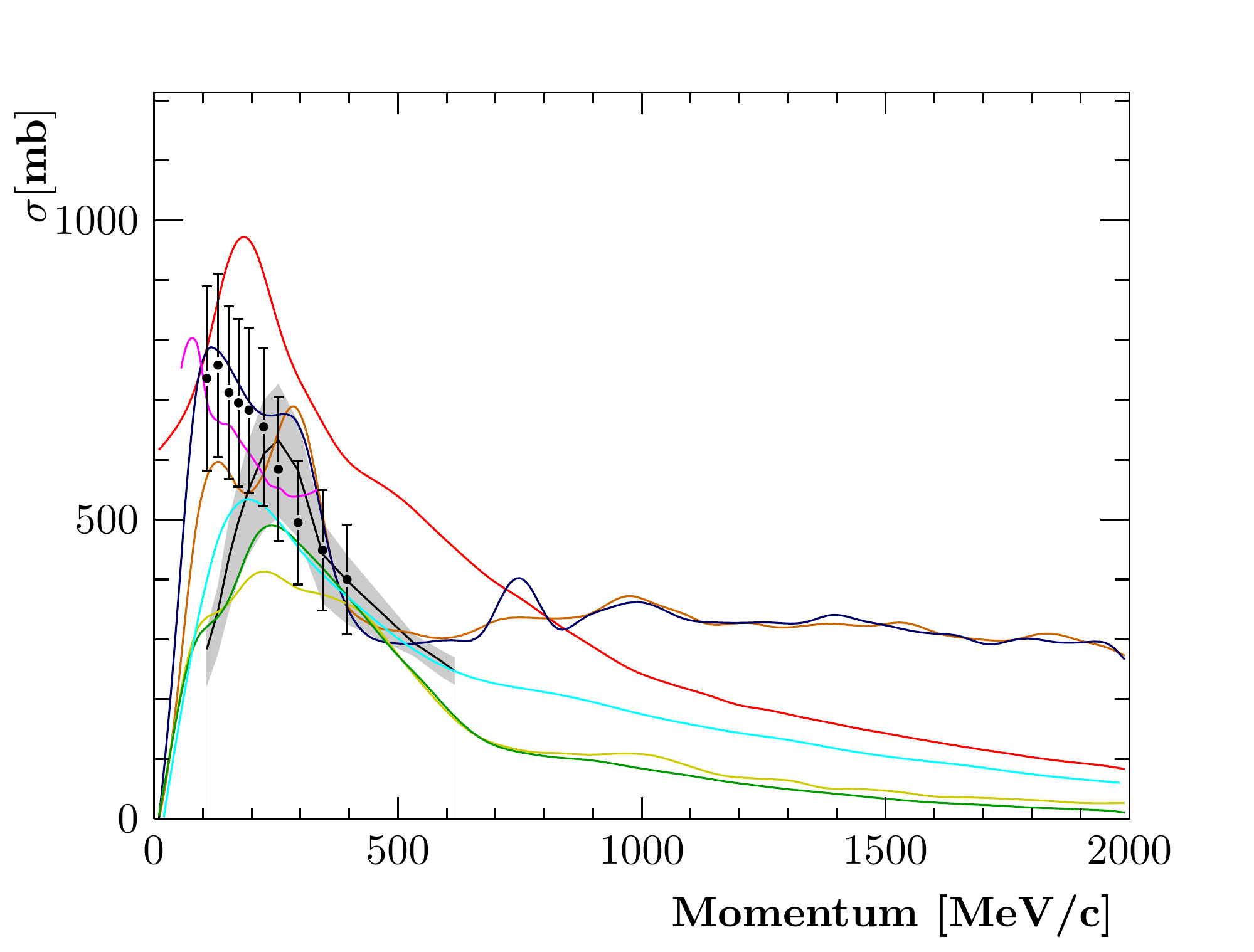}}\\\vspace{-12pt}
  \subfloat[Charge exchange (CX)]{\includegraphics[width=0.33\linewidth]{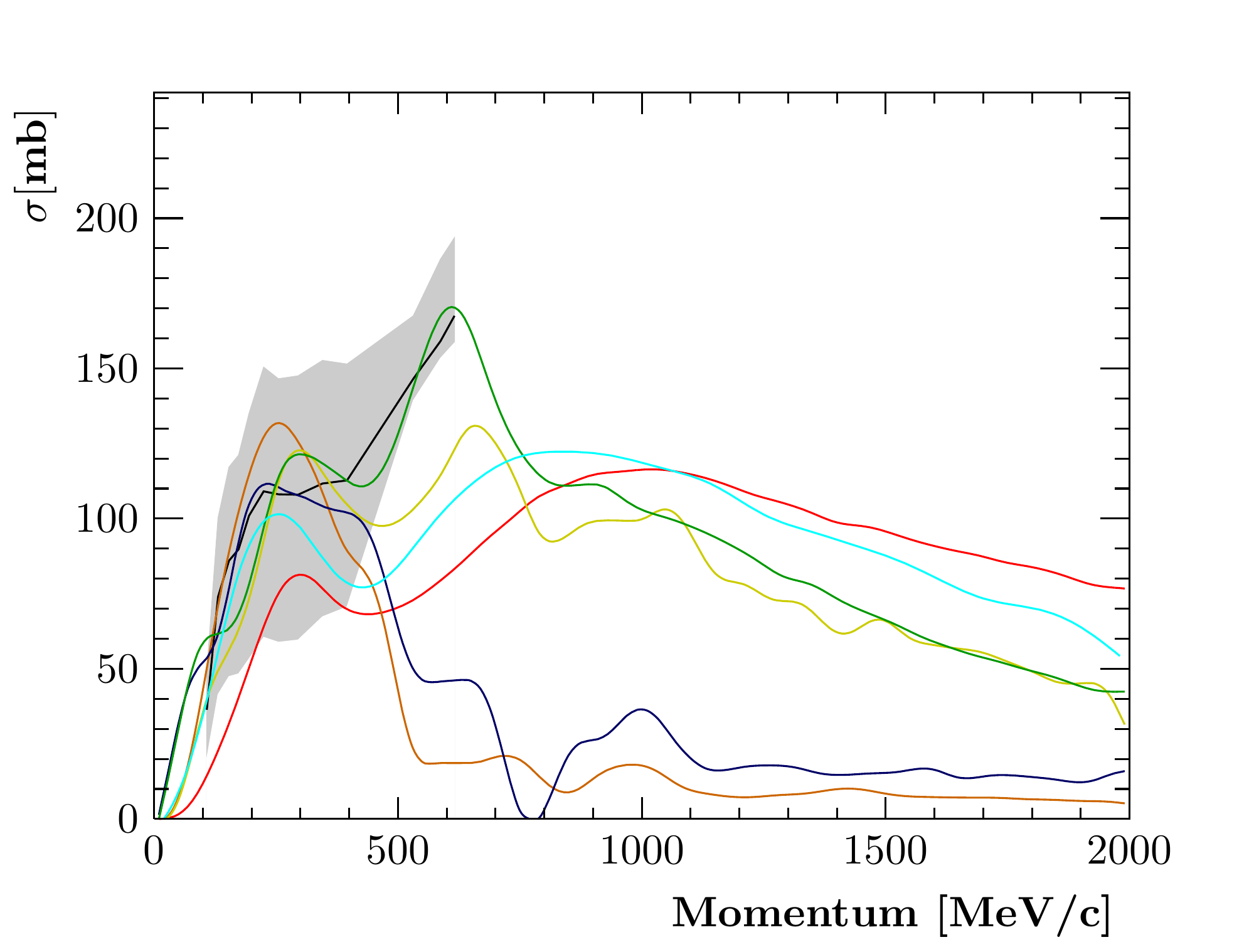}}
  \subfloat[ABS+CX]              {\includegraphics[width=0.33\linewidth]{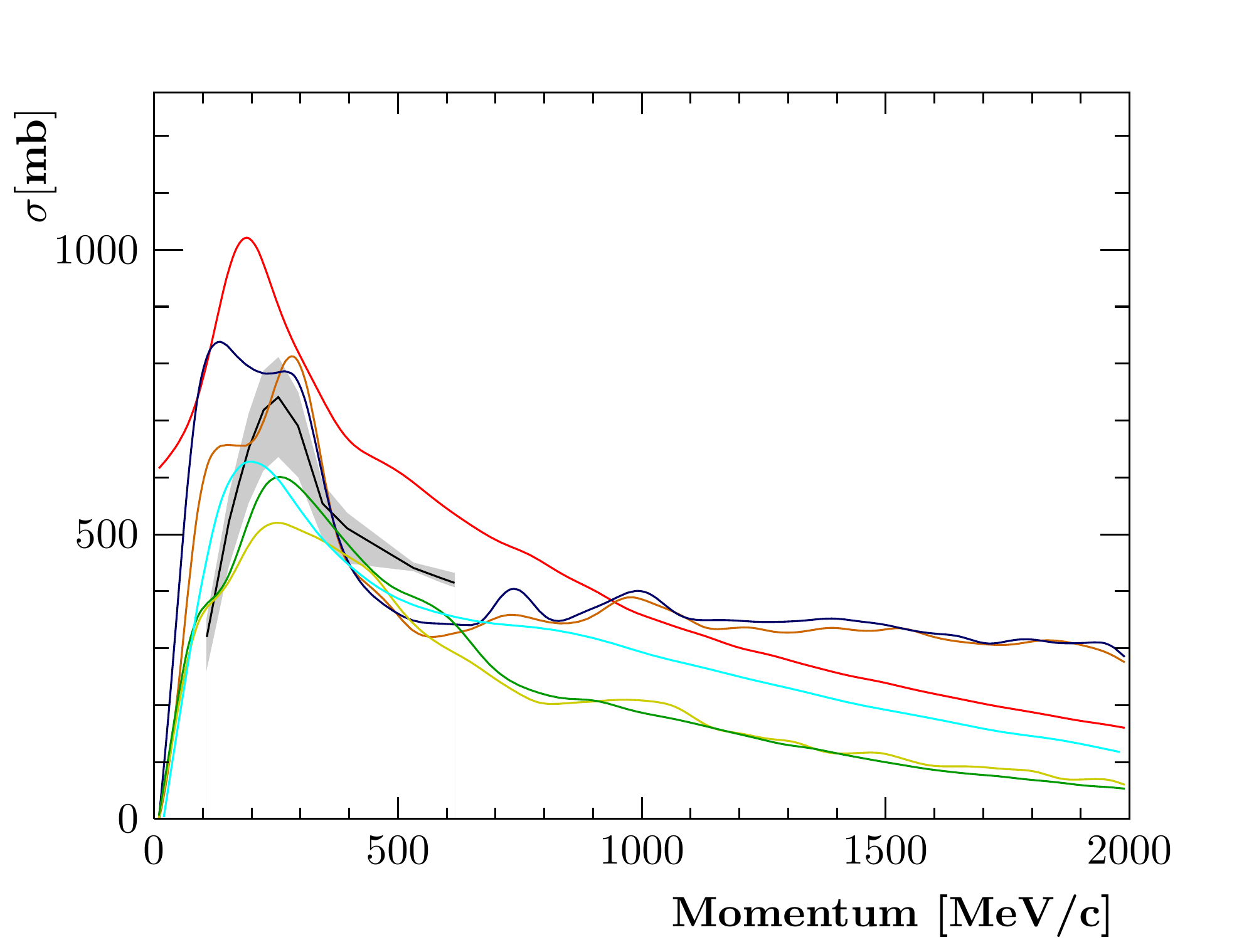}}
  \subfloat                      {\includegraphics[width=0.33\linewidth]{figures/pave.pdf}}
\caption{Comparison of the available $\pi^-$--$^{63}$Cu cross section external data with the NEUT best fit and its $1\sigma$ error band obtained in this work, and other models.}
\label{fig:models-cu-pim}
\end{figure*}

\begin{figure*}[htbp]
  \centering
  \subfloat[Reactive]            {\includegraphics[width=0.33\linewidth]{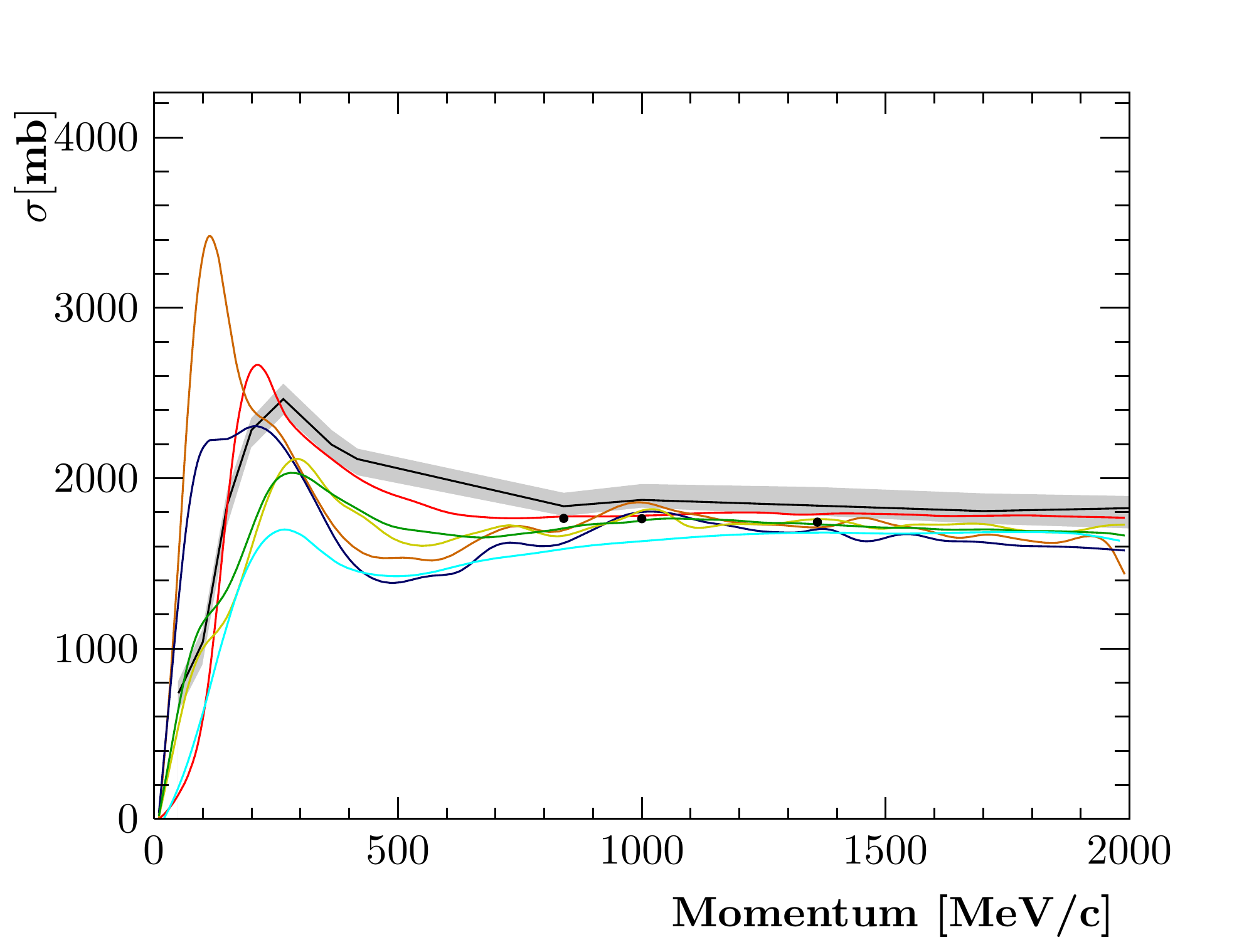}}
  \subfloat[Quasi-elastic]       {\includegraphics[width=0.33\linewidth]{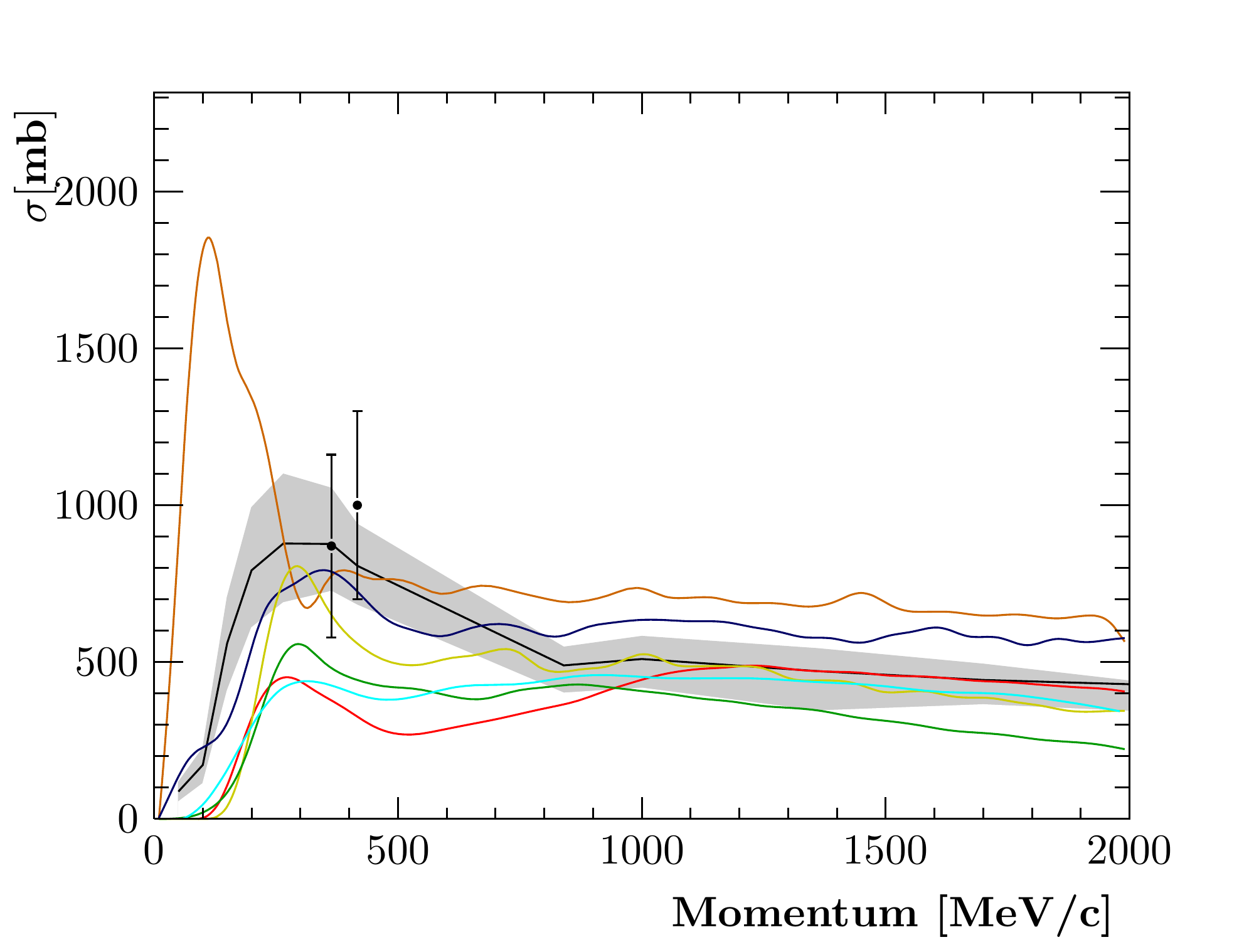}}
  \subfloat[Absorption (ABS)]    {\includegraphics[width=0.33\linewidth]{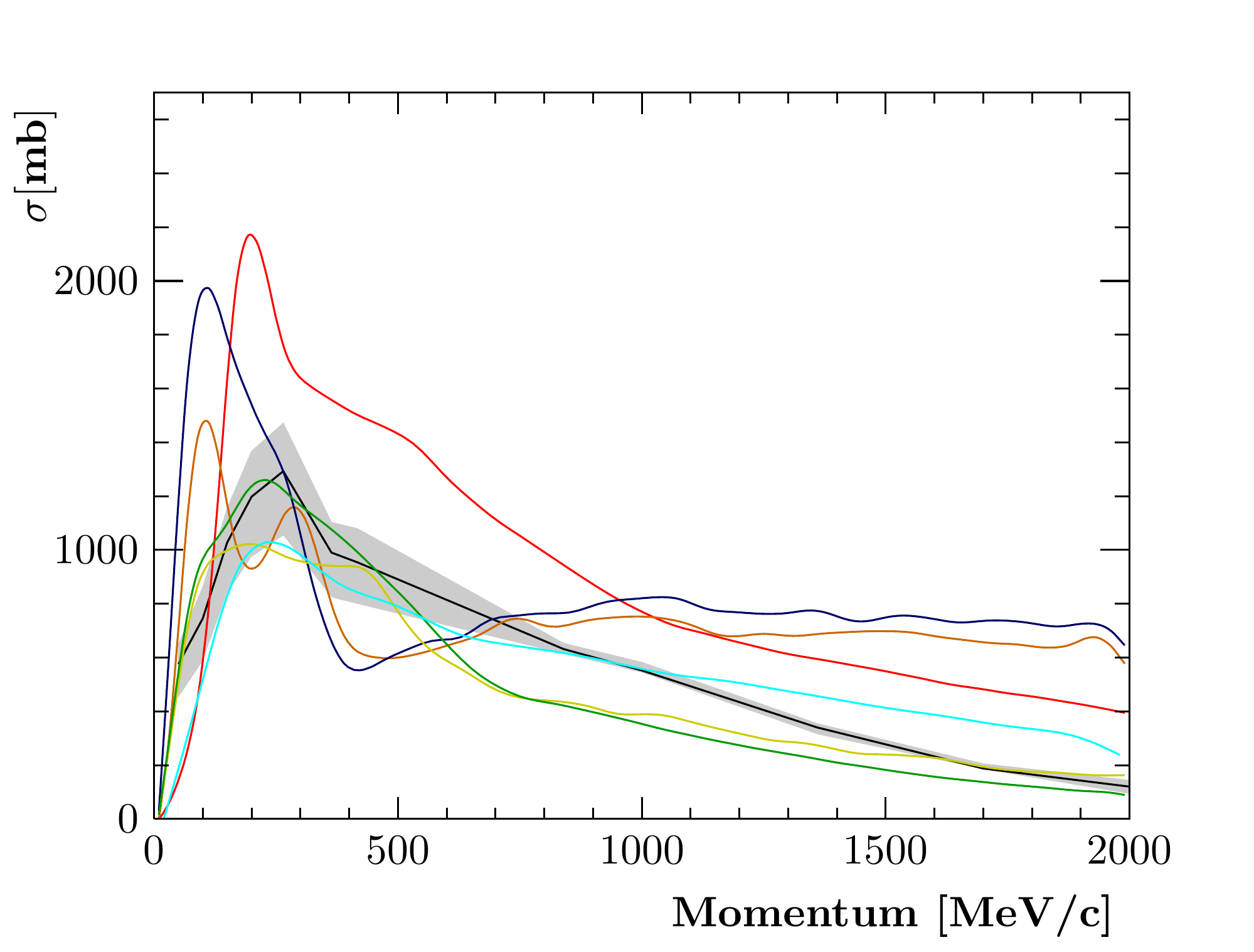}}\\\vspace{-12pt}
  \subfloat[Charge exchange (CX)]{\includegraphics[width=0.33\linewidth]{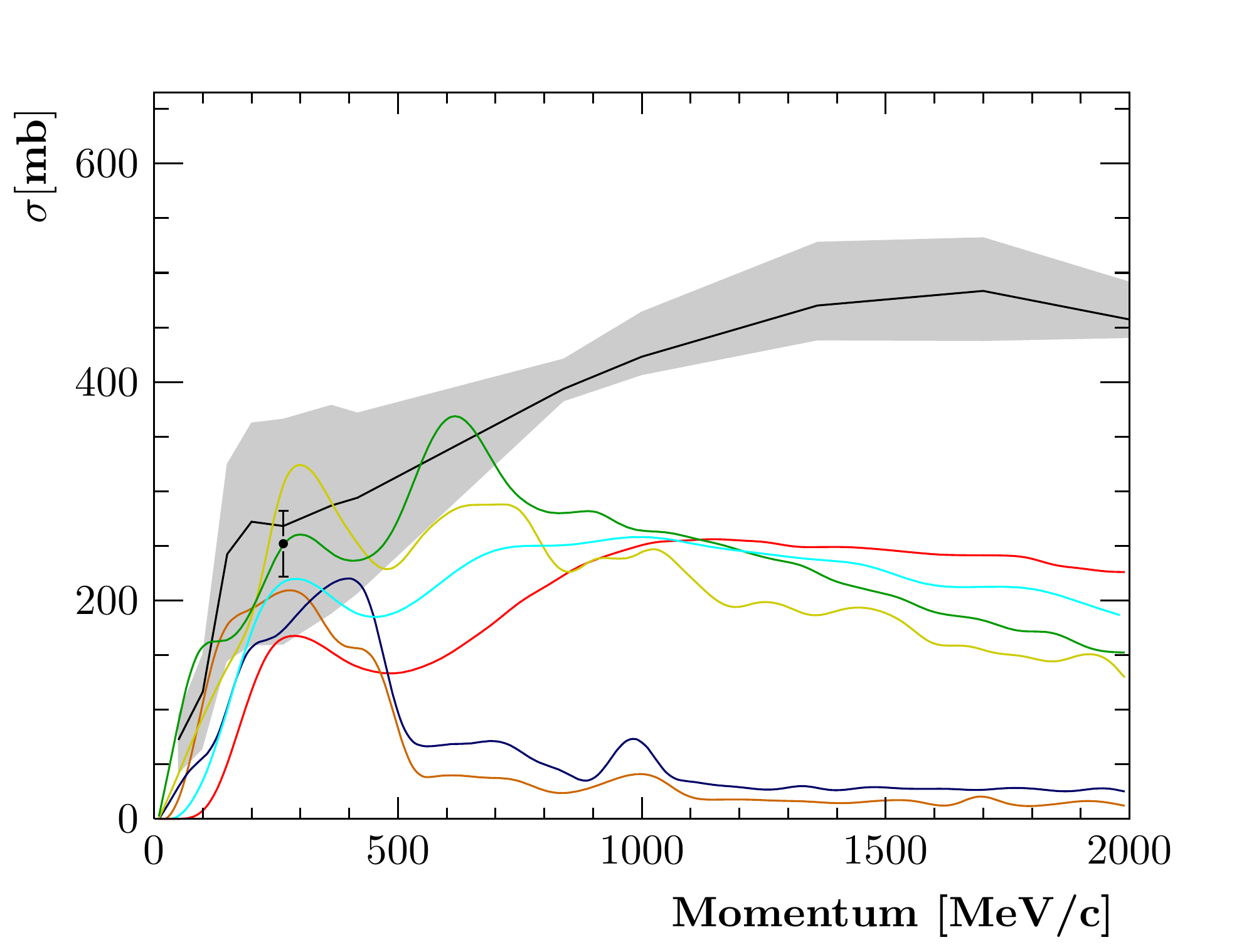}}
  \subfloat[ABS+CX]              {\includegraphics[width=0.33\linewidth]{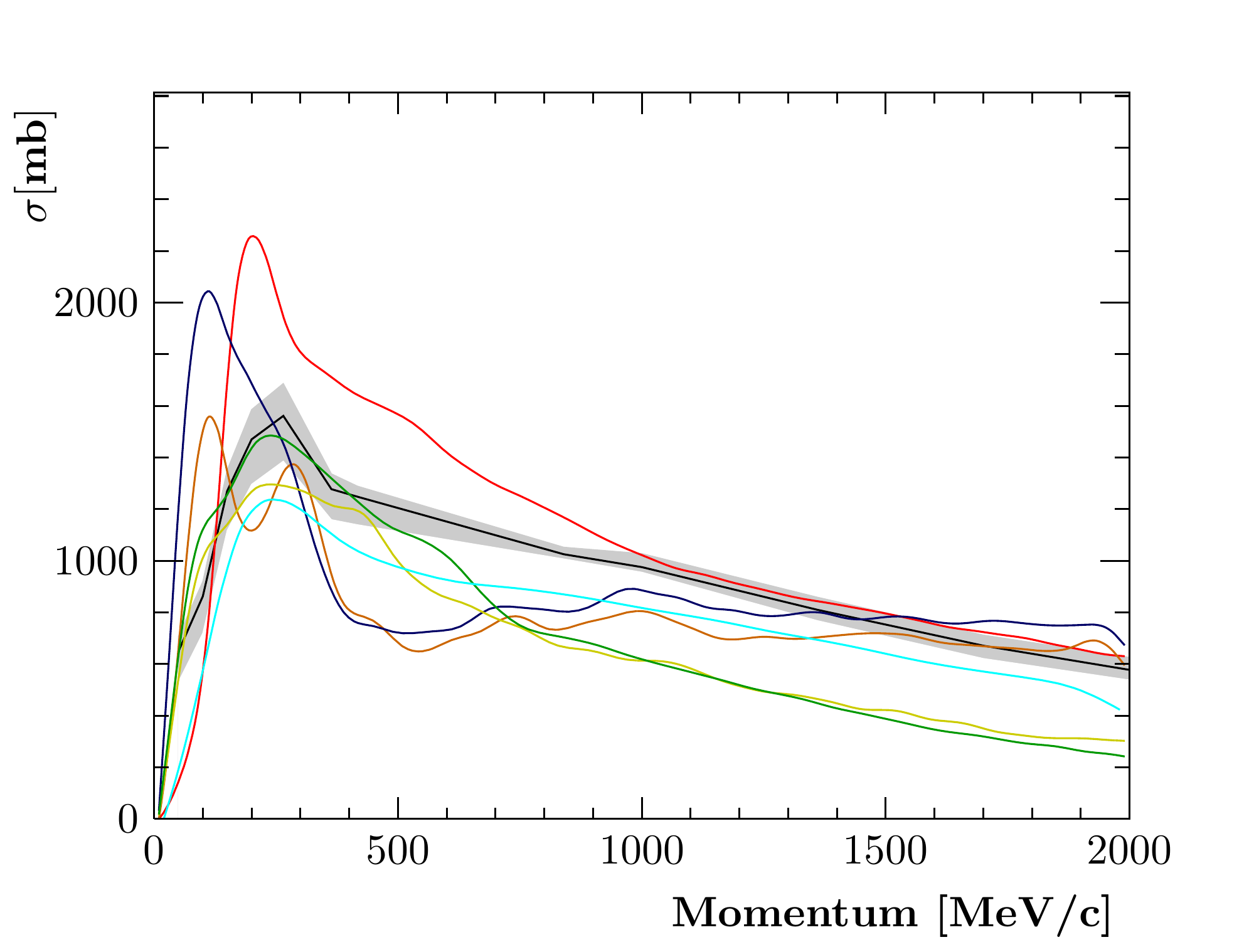}}
  \subfloat                      {\includegraphics[width=0.33\linewidth]{figures/pave.pdf}}
\caption{Comparison of the available $\pi^+$--$^{207}$Pb cross section external data with the NEUT best fit and its $1\sigma$ error band obtained in this work, and other models.}
\label{fig:models-pb-pip}
\end{figure*}

\begin{figure*}[htbp]
  \subfloat[Reactive]            {\includegraphics[width=0.33\linewidth]{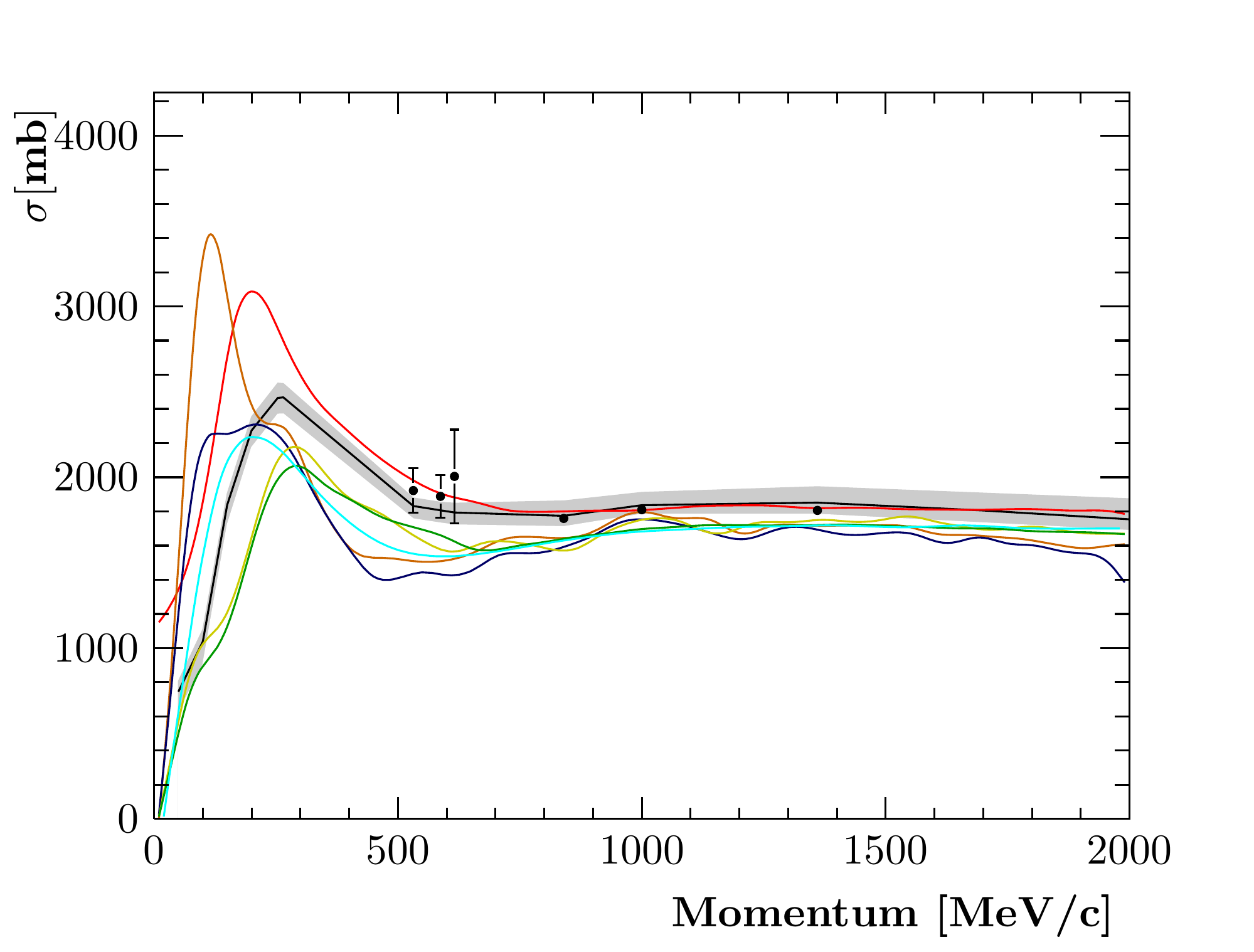}}
  \subfloat[Quasi-elastic]       {\includegraphics[width=0.33\linewidth]{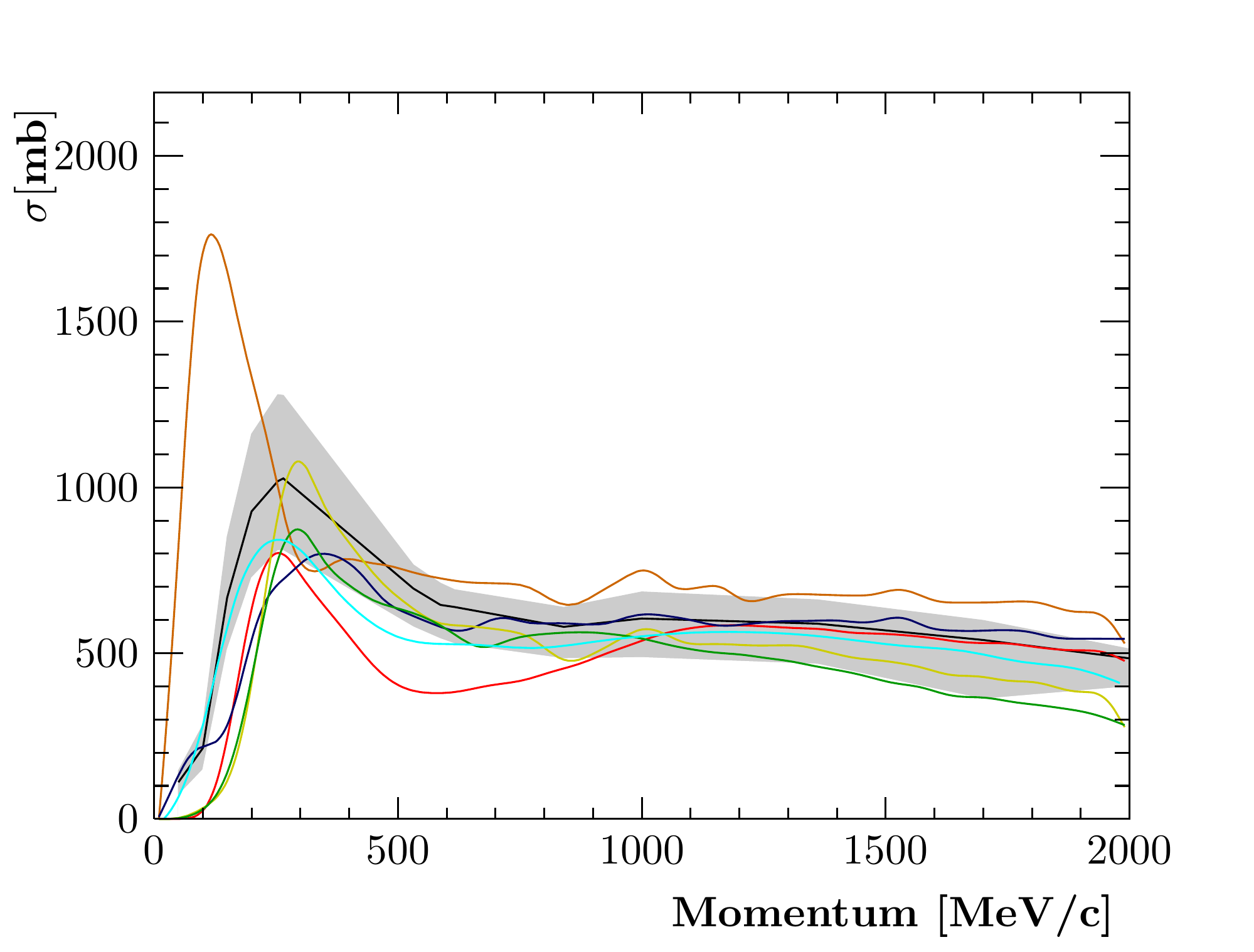}}
  \subfloat[Absorption (ABS)]    {\includegraphics[width=0.33\linewidth]{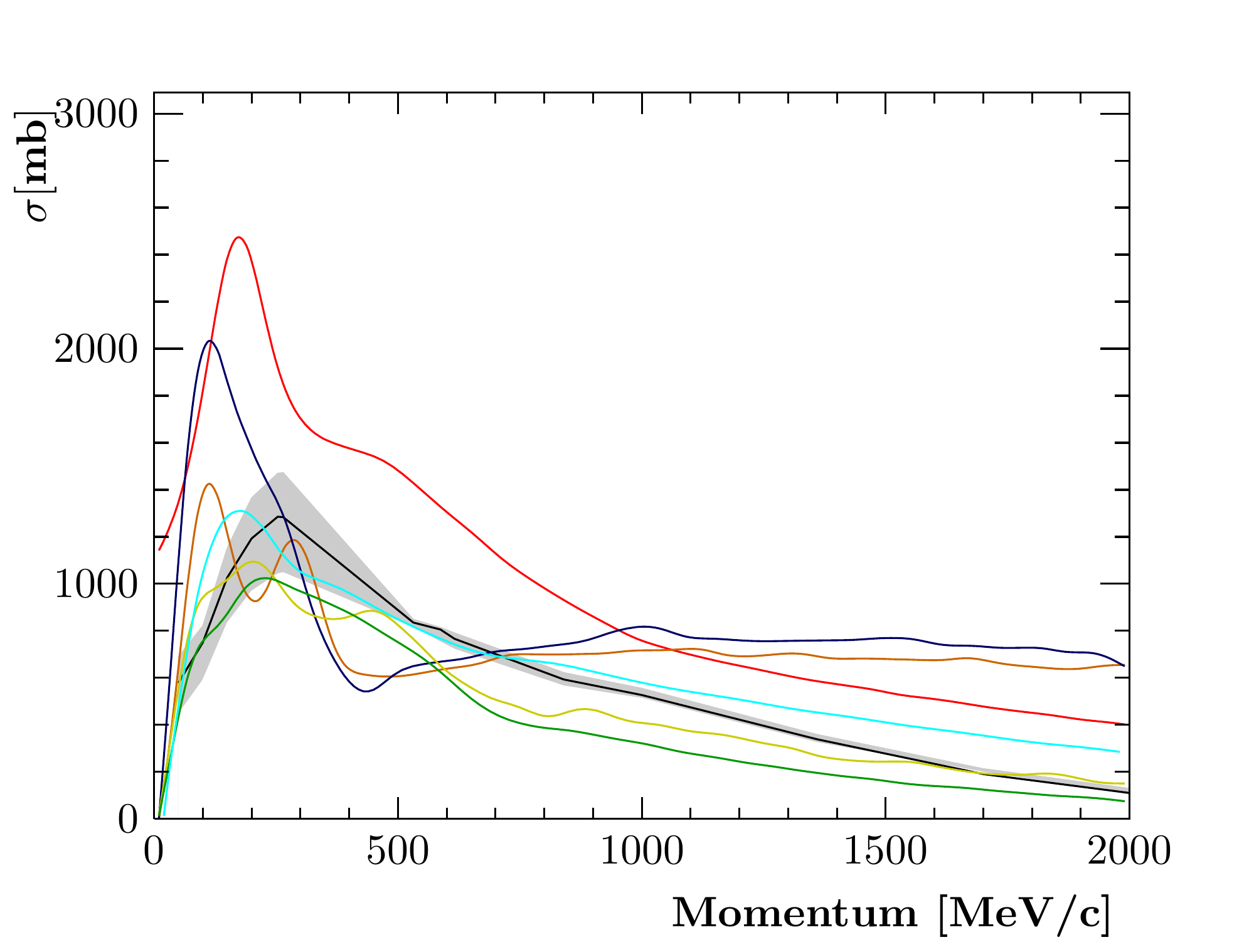}}\\\vspace{-12pt}
  \subfloat[Charge exchange (CX)]{\includegraphics[width=0.33\linewidth]{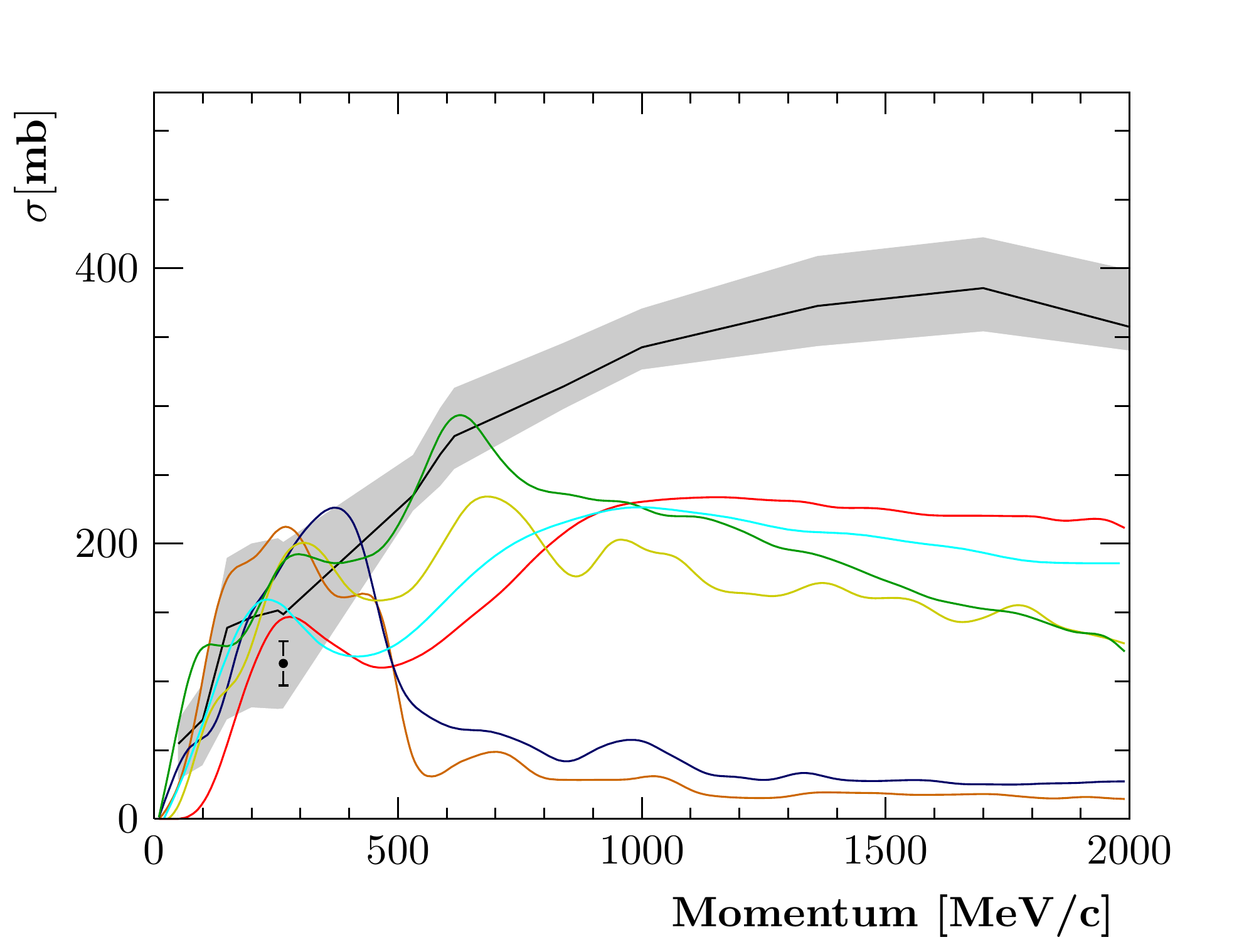}}
  \subfloat[ABS+CX]              {\includegraphics[width=0.33\linewidth]{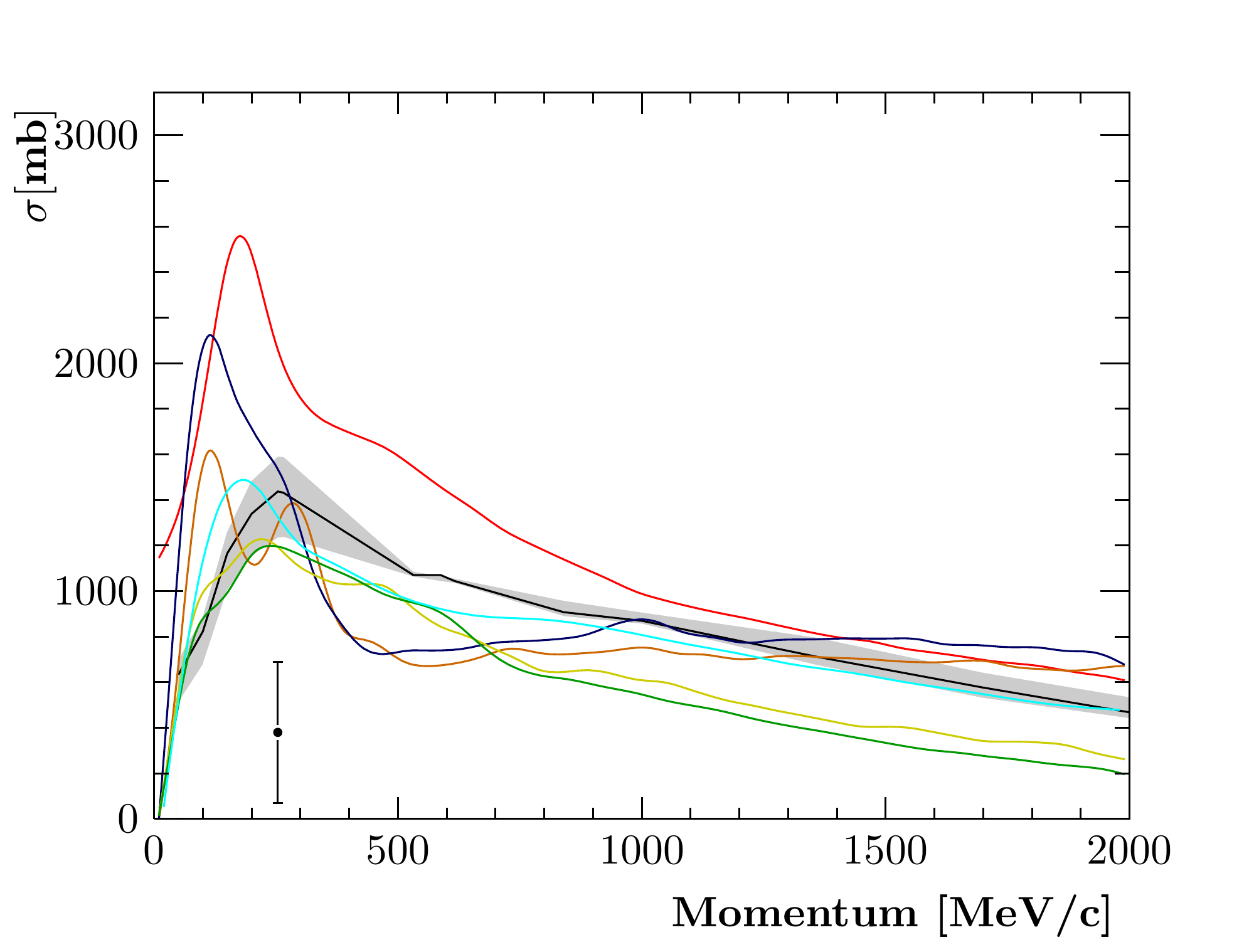}}
  \subfloat                      {\includegraphics[width=0.33\linewidth]{figures/pave.pdf}}
\caption{Comparison of the available $\pi^-$--$^{207}$Pb cross section external data with the NEUT best fit and its $1\sigma$ error band obtained in this work, and other models.}
\label{fig:models-pb-pim}
\end{figure*}

In Section~\ref{sec:models}, a comparison of the tuned NEUT results obtained in Section~\ref{sec:fit-results}, the data used in the fits described in this work (summarized in Table~\ref{tab:piscatdat}), and a variety of alternative intra-nuclear rescattering models were compared for $\pi^{\pm}$--carbon scattering (see Figures~\ref{fig:models-c-pip} and~\ref{fig:models-c-pim}). Here, the similar plots are presented for: oxygen --- Figures~\ref{fig:models-o-pip} and~\ref{fig:models-o-pim}; aluminium --- Figures~\ref{fig:models-al-pip} and~\ref{fig:models-al-pim}; iron --- Figures~\ref{fig:models-fe-pip} and~\ref{fig:models-fe-pim}; copper --- Figures~\ref{fig:models-cu-pip} and~\ref{fig:models-cu-pim}; lead --- Figures~\ref{fig:models-pb-pip} and~\ref{fig:models-pb-pim}.

\begin{figure*}[htbp]
\includegraphics[width=0.9\textwidth]{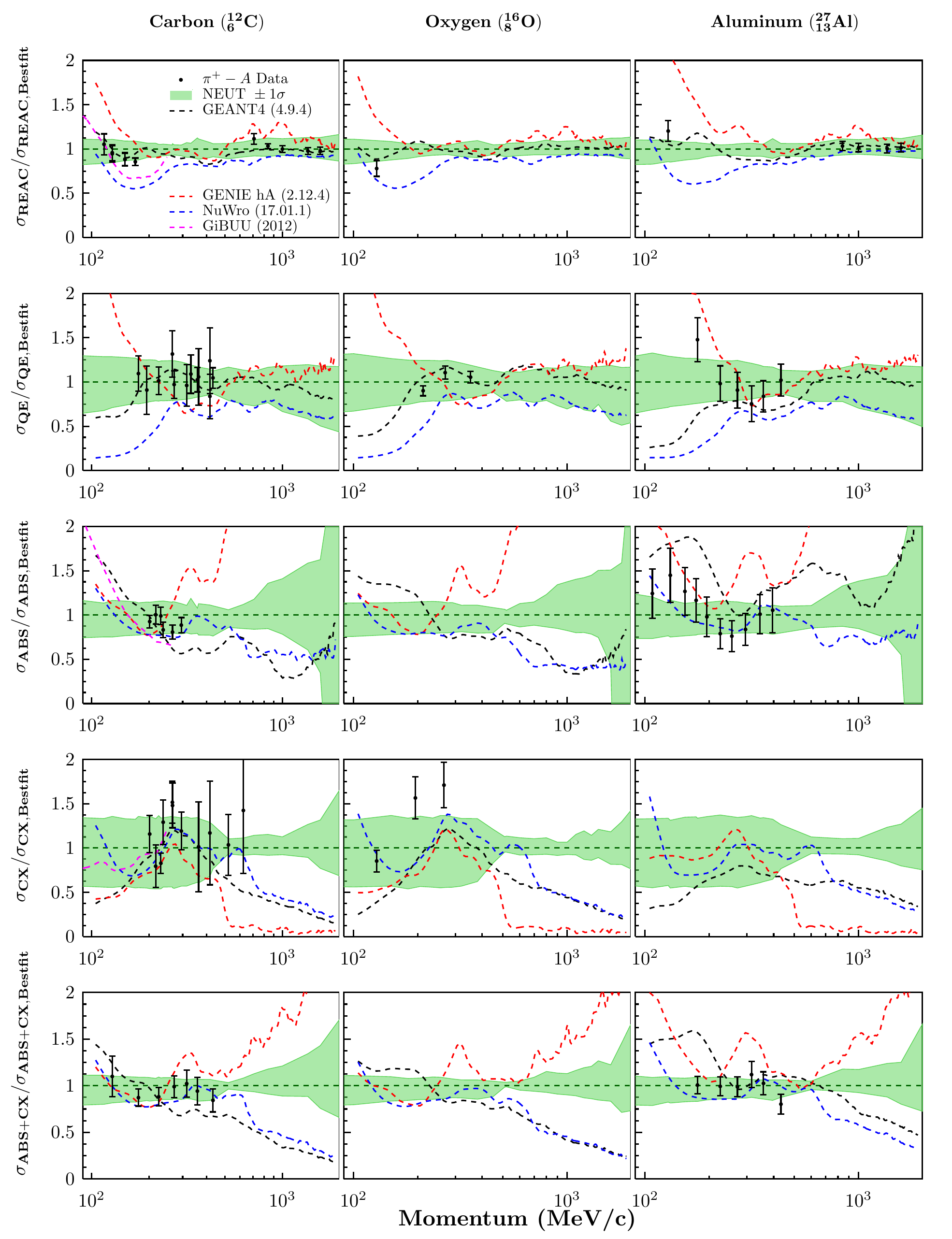}
\caption{Ratios of the \pip--{$^{12}$C, $^{16}$O, $^{27}$Al} data and the GEANT4 (black), GENIE (red), NuWro (blue), and GiBUU (magenta) predictions to the NEUT best fit for the five interaction channels used in the fit. The green band represents the $\pm1\sigma$ band.}
\label{fig:fsifitter-pip-light}
\end{figure*}

\begin{figure*}[htbp]
\centering
\includegraphics[width=0.9\textwidth]{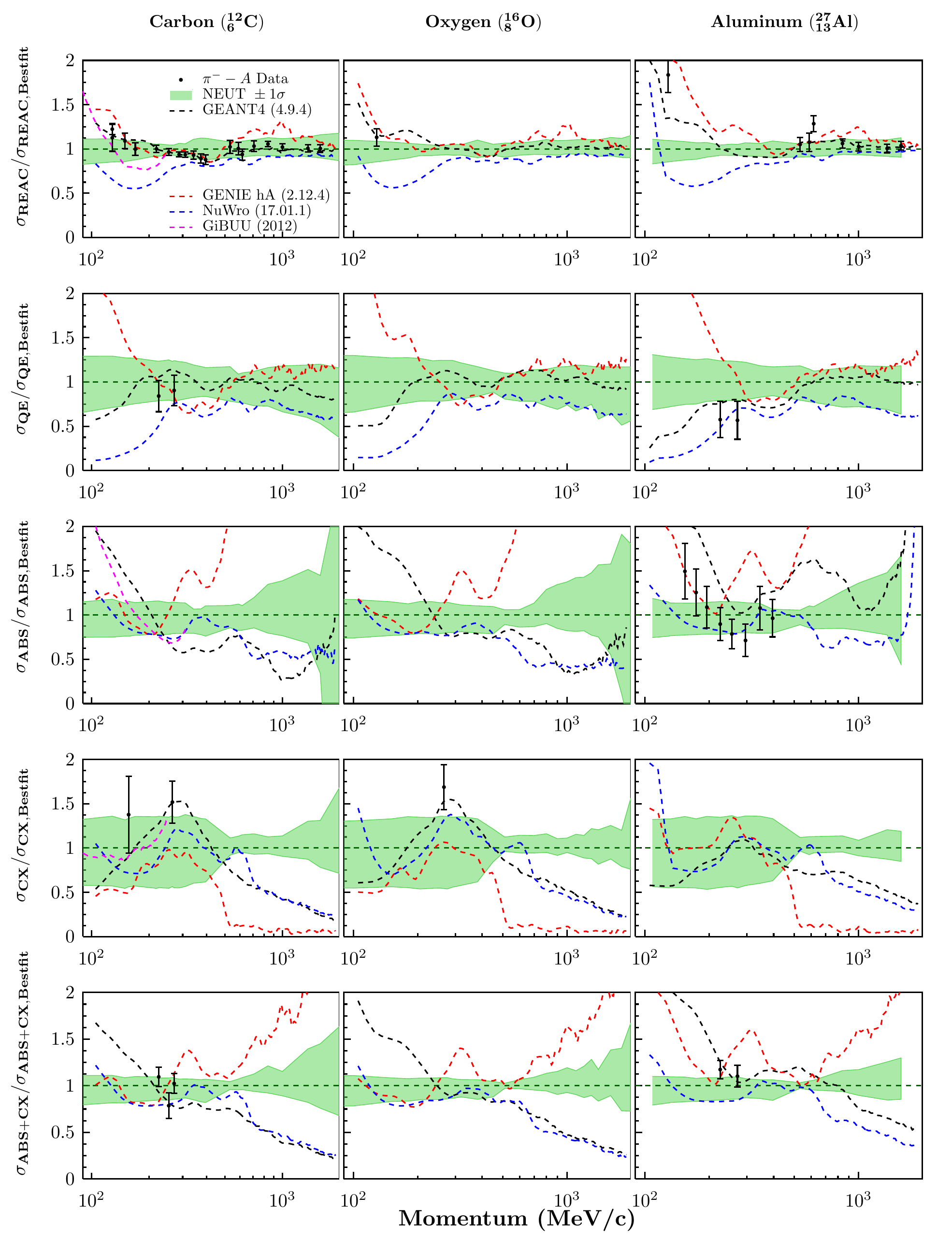}
\caption{Ratios of the \pim--{$^{12}$C, $^{16}$O, $^{27}$Al} data and the GEANT4 (black), GENIE (red), NuWro (blue), and GiBUU (magenta) predictions to the NEUT best fit. The green band represents the $\pm1\sigma$ band.}
\label{fig:fsifitter-pim-light}
\end{figure*}

In addition, ratios of those plots with respect to the best fit NEUT predictions obtained in this work, are presented for reference for \pip--{$^{12}$C,$^{16}$O,$^{27}$Al} (\pim-{$^{12}$C,$^{16}$O,$^{27}$Al}) scattering in Figure~\ref{fig:fsifitter-pip-light} (Figure~\ref{fig:fsifitter-pim-light}), and for \pip--{$^{56}$Fe,$^{63}$Cu,$^{207}$Pb} (\pim-{$^{56}$Fe,$^{63}$Cu,$^{207}$Pb}) scattering in Figure~\ref{fig:fsifitter-pip-light} (Figure~\ref{fig:fsifitter-pim-light}).






\begin{figure*}[htbp]
\includegraphics[width=0.9\textwidth]{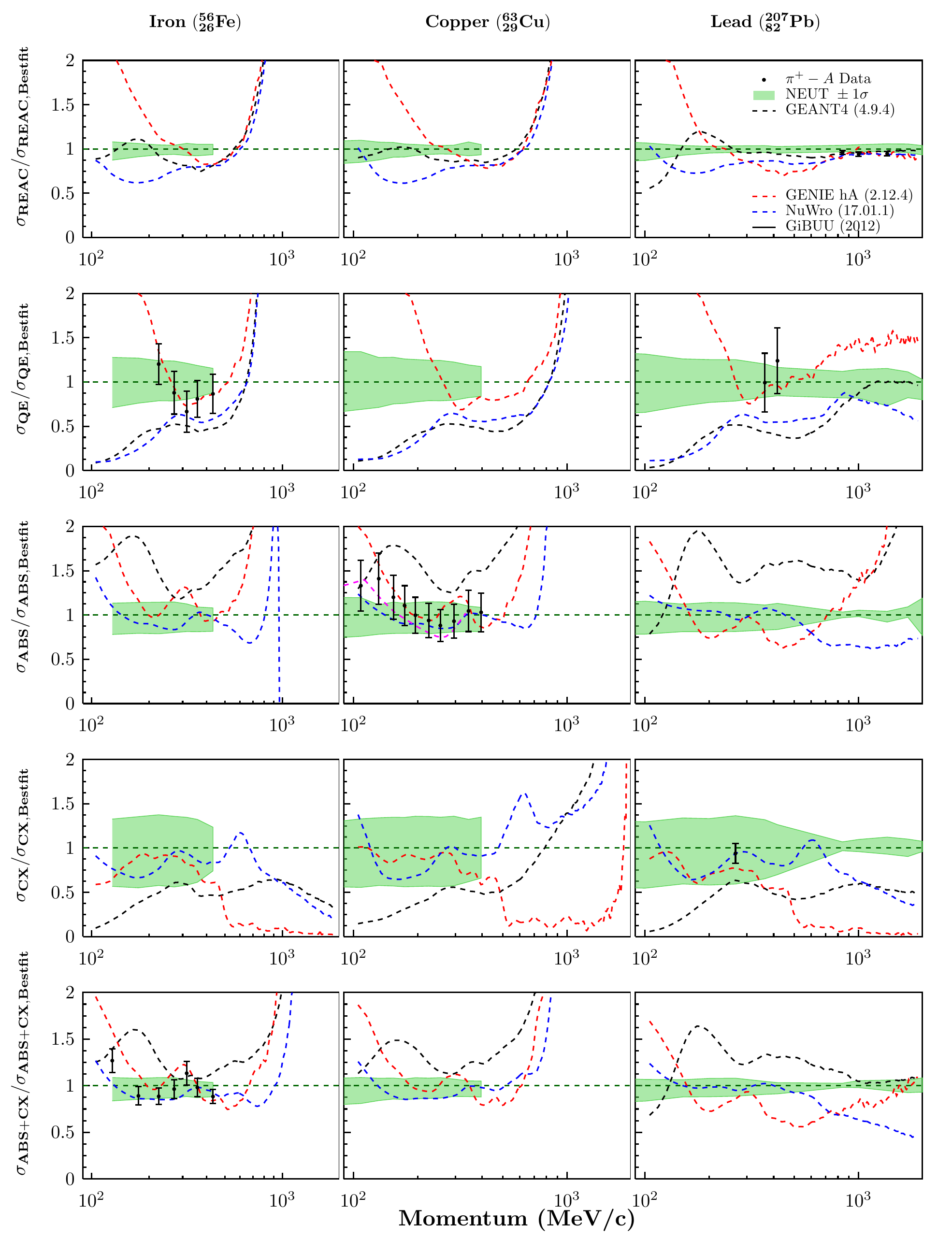}
\caption{Ratios of the \pip--{$^{56}$Fe, $^{63}$Cu, $^{207}$Pb} data and the GEANT4 (black), GENIE (red), NuWro (blue), and GiBUU (magenta) predictions to the NEUT best fit. The green band represents the $\pm1\sigma$ band.}
\label{fig:fsifitter-pip-heavy}
\end{figure*}

\begin{figure*}[htbp]
\centering
\includegraphics[width=0.9\textwidth]{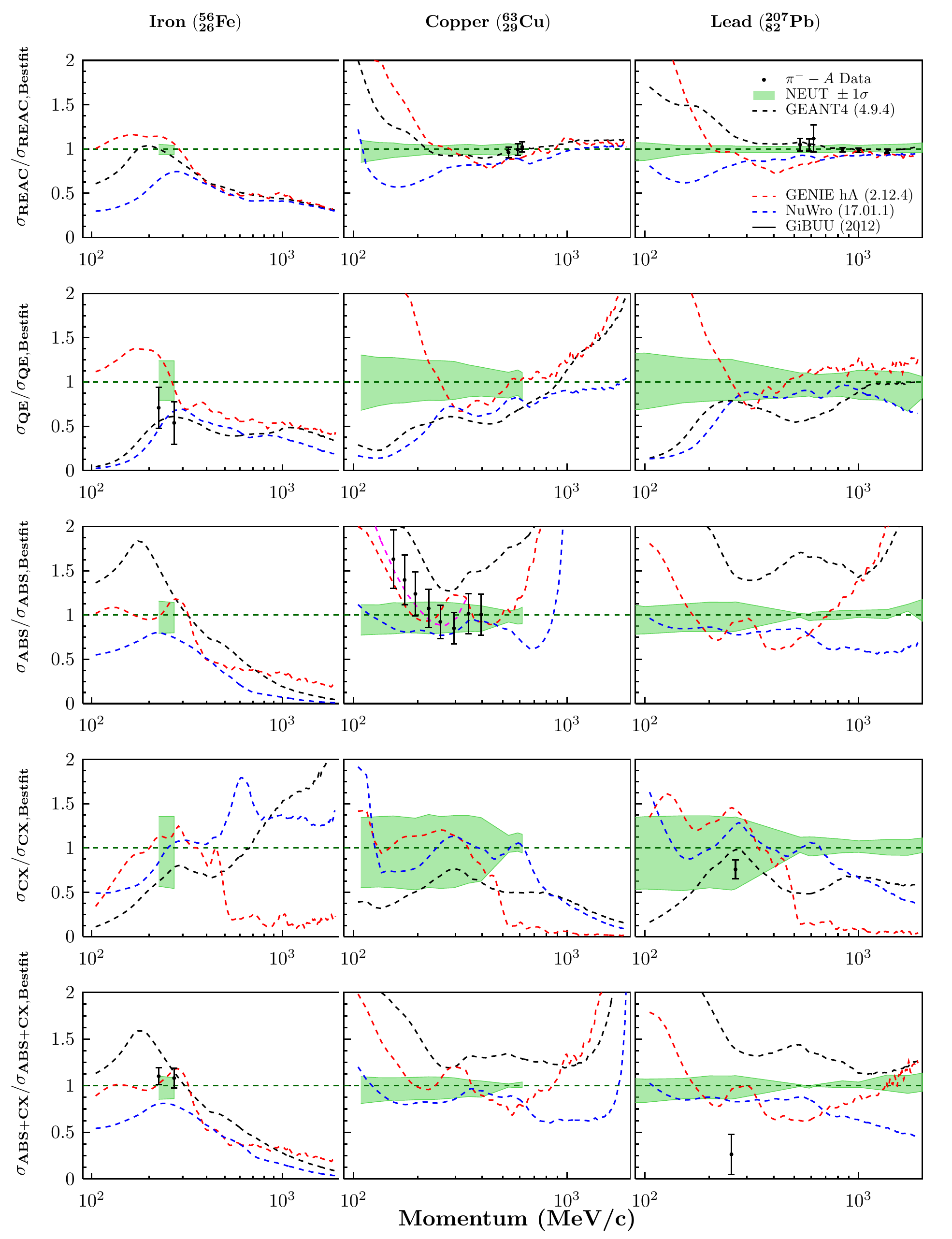}
\caption{Ratios of the \pim--{$^{56}$Fe, $^{63}$Cu, $^{207}$Pb} data and the GEANT4 (black), GENIE (red), NuWro (blue), and GiBUU (magenta) predictions to the NEUT best fit. The green band represents the $\pm1\sigma$ band.}
\label{fig:fsifitter-pim-heavy}
\end{figure*}

\end{document}